\documentclass[twocolumn]{aastex631}

\shorttitle{$D_{\rm frac}$(NH$_3$) in MSF}
\shortauthors{Li et al.}

\begin{document}

\title{The deuterium fractionation of NH$_3$ in massive star-forming regions}

\author[0000-0002-2243-6038]{Yuqiang Li}\thanks{E-mail: liyuqiang@shao.ac.cn}
\affiliation{Shanghai Astronomical Observatory, Chinese Academy of Sciences, No. 80 Nandan Road, Shanghai, 200030, People’s Republic of China}
\affiliation{School of Astronomy and Space Sciences, University of Chinese Academy of Sciences, No. 19A Yuquan Road, Beijing 100049, People’s Republic of China}
\author[0000-0001-6106-1171]{Junzhi Wang}\thanks{E-mail: junzhiwang@gxu.edu.cn}
\affiliation{Guangxi Key Laboratory for Relativistic Astrophysics, Department of Physics, Guangxi University, Nanning 530004, People’s Republic of China}
\author[0000-0003-3520-6191]{Juan Li}\thanks{E-mail: lijuan@shao.ac.cn}
\affiliation{Shanghai Astronomical Observatory, Chinese Academy of Sciences, No. 80 Nandan Road, Shanghai, 200030, People’s Republic of China}
\affiliation{Key Laboratory of Radio Astronomy, Chinese Academy of Sciences, Nanjing 210033, People’s Republic of China}
\author[0009-0001-6483-7366]{Prathap Rayalacheruvu}
\affiliation{School of Earth and Planetary Sciences, National Institute of Science Education and Research, Jatni 752050, Odisha, India}
\affiliation{Homi Bhabha National Institute, Training School Complex, Anushaktinagar, Mumbai 400094, India}
\author[0000-0001-7031-8039]{Liton Majumdar}
\affiliation{School of Earth and Planetary Sciences, National Institute of Science Education and Research, Jatni 752050, Odisha, India}
\affiliation{Homi Bhabha National Institute, Training School Complex, Anushaktinagar, Mumbai 400094, India}
\author[0000-0001-5574-0549]{Yaoting Yan}
\affiliation{Max-Planck-Institut f\"{u}r Radioastronomie, Auf dem H\"{u}gel 69, 53121 Bonn, Germany}
\author[0000-0003-4811-2581]{Donghui Quan}
\affiliation{Research Center for Intelligent Computing Platforms, Zhejiang Laboratory, Hangzhou 311100, People’s Republic of China}
\affiliation{Xinjiang Astronomical Observatory, Chinese Academy of Sciences, 150 Science 1-Street, Urumqi 830011, People’s Republic of China}
\author[0000-0003-2619-9305]{Xing Lu}
\affiliation{Shanghai Astronomical Observatory, Chinese Academy of Sciences, No. 80 Nandan Road, Shanghai, 200030, People’s Republic of China}
\author[0000-0001-9047-846X]{Siqi Zheng}
\affiliation{Shanghai Astronomical Observatory, Chinese Academy of Sciences, No. 80 Nandan Road, Shanghai, 200030, People’s Republic of China}
\affiliation{School of Astronomy and Space Sciences, University of Chinese Academy of Sciences, No. 19A Yuquan Road, Beijing 100049, People’s Republic of China}
\affiliation{Key Laboratory of Radio Astronomy, Chinese Academy of Sciences, Nanjing 210033, People’s Republic of China}
\affiliation{I. Physikalisches Institut, Universit\"{a}t zu Köln, Z\"{u}lpicher Str. 77, 50937 K\"{o}ln, Germany}



\begin{abstract}

Deuteration is sensitive to environmental conditions in star-forming regions.
To investigate NH$_2$D chemistry, we compared the spatial distribution of ortho-NH$_2$D $1_{11}^s-1_{01}^a$, NH$_3$(1,1) and NH$_3$(2,2) in 12 late-stage massive star-forming regions. By averaging several pixels along the spatial slices of ortho-NH$_2$D $1_{11}^s-1_{01}^a$, we obtained the deuterium fractionation of NH$_3$.
In seven targets, the deuterium fractionation of NH$_3$ shows a decreasing trend with increasing rotational temperature. This trend is less clear in the remaining five sources, likely due to limited spatial resolution. However, when considering all 12 sources together, the anticorrelation between NH$_3$ deuterium fractionation and rotational temperature becomes less significant, suggesting that other physical parameters may influence the fractionation. Additionally, we found that the region of highest deuterium fractionation of NH$_3$ is offset from the NH$_3$ peak in each source, likely because the temperature is higher near the NH$_3$ peaks and NH$_2$D may be depleted from the gas phase as the molecular cloud core evolves, {as well as the increased release of CO from grains into the gas phase.}

\end{abstract}

\keywords{stars: massive --- stars: formation --- ISM: molecules --- ISM: abundances}


\section{Introduction}
{The abundances of molecules in which one or more hydrogen atoms have been replaced by deuterium are regarded as good indicators of the evolutionary stages of interstellar molecular clouds
\citep[e.g.][]{2005ApJ...619..379C,2009A&A...493...89E,2011A&A...529L...7F}. Due to their lower zero-point energies compared to their hydrogen-containing counterparts, the formation of deuterated species, such as DCN, DCO$^+$, N$_2$D$^+$ and NH$_2$D, is favored at low temperatures \citep{1982A&A...111...76H}. The ratio of a deuterated species to their hydrogen-containing counterparts, known as deuterium fractionation ($D_{\rm frac}$), can be several orders of magnitude higher than the D/H ratio ($\approx$1.5 $\times$ 10$^{-5}$) in the interstellar medium (ISM) \citep{1995ApJ...451..335L,2003ApJ...587..235O}.} {Consequently, the abundance of these deuterated molecules is often found to be enhanced \citep[e.g.][]{2010A&A...517L...6B,2021MNRAS.502.1104H,2021A&A...650A.172J}. }

The formation of deuterated ammonia, NH$_2$D, NHD$_2$ and ND$_3$, is often considered to be associated with deuterated ions, such as H$_2$D$^+$ and CH$_2$D$^+$, {which convert} ammonia (NH$_3$) in the gas phase \citep[e.g.][]{2001ApJ...553..613R,2008A&A...492..703C,2012A&ARv..20...56C,2014prpl.conf..859C}. {Chemical reactions on the surfaces of dust grains also contribute to the production of deuterated ammonia \citep[e.g.][]{1990ApJ...362L..29T,2008A&A...492..703C,2014prpl.conf..859C,2015A&A...575A..87F}. }Additionally, the abundance of deuterated species could be affected by shocks \citep{2002ApJ...569..322L}. 

NH$_2$D was first detected in Sgr B2 \citep{1978ApJ...219L..43T} and Orion KL \citep{1978ApJ...219L..49R}. {Despite decades of observations and modeling,} our understanding of deuterium chemistry remains limited \citep{2021AJ....161...38O}. Several studies have been carried out to investigate the relationship between $D_{\rm frac}$(NH$_3$) and rotation temperature \citep[e.g.][]{2013ApJ...777...85N,2015A&A...575A..87F,2021A&A...649A..21W}.
With single-pointing observations using single-dish telescopes, no clear trend has been obtained for the relationship between $D_{\rm frac}$(NH$_3$) and $T_{\rm rot}$(2,2;1,1) \citep[e.g.][]{2015A&A...575A..87F,2021A&A...649A..21W}. {For instance,} \cite{2015A&A...575A..87F} reported that no clear correlation between $D_{\rm frac}$(NH$_3$) and ammonia rotation temperature is found with Institut de Radioastronomie Milli${\rm \acute{m}}$etrique (IRAM) and Green Bank Telescope (GBT) observations toward 28 massive star-forming regions in different evolutionary stages. {Similarly,} with IRAM observations toward 992 ATLASGAL (APEX Telescope Large Area Survey) massive clumps, \cite{2021A&A...649A..21W} also reported a flat distribution of $D_{\rm frac}$(NH$_3$) and ammonia rotation temperature. 

However, several interferometric images from case studies provide observational evidence that $D_{\rm frac}$(NH$_3$) is influenced by temperature. \cite{2010A&A...517L...6B} found that the $D_{\rm frac}$(NH$_3$) in young stellar objects (YSOs) was lower than in pre-protostellar cores using IRAM Plateau de Bure Interferometer (PdBI) observations. {Additionally, \cite{2013ApJ...777...85N} suggested that a slightly warmer environment results in lower deuterium fractionation by comparing $D_{\rm frac}$(NH$_3$) in the hot core and compact ridge of Orion KL using Atacama Large Millimeter/submillimeter Array (ALMA) observations.}

Single-pointing observations using single-dish telescopes lack the spatial distribution of $D_{\rm frac}$(NH$_3$), while interferometric observations often focus on one or two targets, which limits the ability to compare the relationship between temperature and $D_{\rm frac}$(NH$_3$) with different physical environments. 
Therefore, {conducting single-dish mapping toward a relatively large sample,} the spatial distribution of NH$_3$ and NH$_2$D could enhance our understanding of the deuterium chemistry of NH$_3$ in massive star-forming regions.

In \cite{2024MNRAS.527.5049L}, we obtained the spatial distribution of ortho-NH$_2$D $1_{11}^s-1_{01}^a$ at 85.926 GHz in 18 Galactic late-stage massive star-forming regions with IRAM 30 m. We found that the spatial distribution of ortho-NH$_2$D $1_{11}^s-1_{01}^a$ at 85.926 GHz is complex and differs from the dense gas tracer H$^{13}$CN 1-0. {To further investigate deuterium fractionation and conduct a more comprehensive study of singly deuterated ammonia (hereafter deuterated ammonia),} we performed a line study of ortho-NH$_2$D $1_{11}^s-1_{01}^a$ at 85.926 GHz, NH$_3$(1,1) at 23.696 GHz and NH$_3$(2,2) at 23.722 GHz toward 12 Galactic late-stage massive star-forming regions in this work. {The data include ortho-NH$_2$D $1_{11}^s-1_{01}^a$ at 85.926 GHz for 12 sources as presented in \cite{2024MNRAS.527.5049L}, as well as  NH$_3$(1,1) and NH$_3$(2,2) data for three sources observed with Effelsberg 100 m. Additionally, NH$_3$(1,1) and NH$_3$(2,2) data for the remaining nine sources were obtained from GBT 100 m archival data.}

In this work, we compared the spatial distributions between ortho-NH$_2$D $1_{11}^s-1_{01}^a$, NH$_3$(1,1) and NH$_3$(2,2) maps in late-stage massive star-forming regions. $D_{\rm frac}$(NH$_3$) is derived for each source and is utilized to analyze its relationship with temperature. The observations are described in Section \ref{observations}, the main results are reported in Section \ref{results}, a discussion is given in Section \ref{discussion}, and a brief summary is given in Section \ref{summary}.


\section{Observations and data reduction}
\label{observations}
\subsection{NH$_3$ data}
\label{NH3 data}
\subsubsection{Effelsberg 100 m observation}
The ammonia inversion transitions, NH$_3$(1,1) and NH$_3$(2,2) were observed toward G081.75+00.59, G109.87+00.77 and G121.29+00.65 during 2021 February and 2023 July with Effelsberg 100 m telescope. The beam size of Effelsberg is 42.5$^{\prime\prime}$ at 23 GHz. The secondary focus receiver S14mm Double Beam RX with a bandwidth of 300 MHz was used during our observation. The Fast Fourier Transform Spectrometer (XFFTS) was used with 65536 channels, providing a frequency resolution of 4.6 kHz and a velocity resolution of 0.058 km s$^{-1}$. The system temperatures were 100–140 K during our observation. In the following analysis, the velocity resolution is smoothed to about 0.6 km s$^{-1}$ with a frequency resolution of 46 kHz at 23 GHz. The on-the-fly (OTF) mapping was used for this observation. A strong continuum source, NGC 7027, was used to calibrate the spectral line flux. The typical rms are about 0.2 K at 0.6 km s$^{-1}$. The details of each source are listed in Table \ref{source}.

\subsubsection{GBT 100 m archival data}
We retrieved the data for NH$_3$(1,1) and NH$_3$(2,2) of nine sources from the GBT archive, and the beam size of GBT is 31.8$^{\prime\prime}$ at 23 GHz. The observations of all GBT archival data have been made using the $K$-band Focal Plane Array (KFPA) receiver.

The NH$_3$(1,1) and NH$_3$(2,2) data of G034.39+00.22, G035.19-00.74 and G075.76+00.33 from \cite{2015MNRAS.452.4029U} were used. The velocity resolution was also smoothed to about 0.6 km s$^{-1}$ and the rms in $T_{\rm mb}$ is about 0.02 K for each source.

The NH$_3$(1,1) and NH$_3$(2,2) data of G023.44-00.18 and G031.28+00.06 from the Radio Ammonia Mid-Plane Survey (RAMPS; \citealt{2018ApJS..237...27H}) were used. The velocity resolution was smoothed to about 0.6 km s$^{-1}$ with about 0.05 K rms in $T_{\rm mb}$ for each source.

In four sources, G015.03-00.67, G035.20-01.73, G081.87+00.78 and G111.54+00.77, NH$_3$(1,1) and NH$_3$(2,2) data from KFPA Examinations of Young STellar Object Natal Environments (KEYSTONE; \citealt{2019ApJ...884....4K}) were used.  The velocity resolution was smoothed to about 0.6 km s$^{-1}$ with 0.03 K rms in $T_{\rm mb}$ for each source.

\subsection{NH$_2$D data}
For sources mentioned in Section \ref{NH3 data}, we used the ortho-NH$_2$D $1_{11}^s-1_{01}^a$ data that were presented in \cite{2024MNRAS.527.5049L}. These ortho-NH$_2$D $1_{11}^s-1_{01}^a$ data were observed by IRAM 30 m with a beam size of about 28.6$^{\prime\prime}$  at 86 GHz. The typical rms are about 0.05 K at 0.6 km s$^{-1}$, corresponding to 195 kHz frequency resolution at 86 GHz. The specific observing parameters are described in \cite{2024MNRAS.527.5049L}.

\subsection{Data reduction}
The lines used consist of ortho-NH$_2$D $1_{11}^s-1_{01}^a$ at 85.926 GHz, NH$_3$(1,1) at 23.696 GHz and NH$_3$(2,2) at 23.722 GHz. Data reduction was conducted with GILDAS software\footnote{\url{http://www.iram.fr/IRAMFR/GILDAS}}. The angular resolution of the IRAM 30 m telescope is about 28.6$^{\prime\prime}$  at 86 GHz, while the angular resolutions of the Effelsberg 100 m telescope and GBT 100 m telescope are about 42.5$^{\prime\prime}$  and 31.8$^{\prime\prime}$, respectively, at 23 GHz.

For G081.75+00.59, G109.87+00.77 and G121.29+00.65, in which the NH$_3$ data were obtained from Effelsberg 100 m observation, the ortho-NH$_2$D $1_{11}^s-1_{01}^a$ data were smoothed to match the Effelsberg resolution of 42.5$^{\prime\prime}$. For these three sources, the ortho-NH$_2$D $1_{11}^s-1_{01}^a$, NH$_3$(1,1) and NH$_3$(2,2) data were regridded to steps of 14$^{\prime\prime}$ (about 1/3 beam size) with a beam size of 42.5$^{\prime\prime}$. 

For the other nine sources, in which NH$_3$ data were obtained from GBT archival data, the ortho-NH$_2$D $1_{11}^s-1_{01}^a$, NH$_3$(1,1) and NH$_3$(2,2) data were regridded to steps of 9$^{\prime\prime}$  (about 1/3 beam size) with a beam size of 30$^{\prime\prime}$ . First-order baselines were used for all spectra. In the following analysis, the beam filling factor is assumed to be unity.

\section{Results}
\label{results}

\subsection{Velocity-integrated intensity maps of ortho-NH$_2$D $1_{11}^s-1_{01}^a$, NH$_3$(1,1) and NH$_3$(2,2)}
\label{Velocity-integrated intensity maps}
\begin{figure}[h]
\gridline{\fig{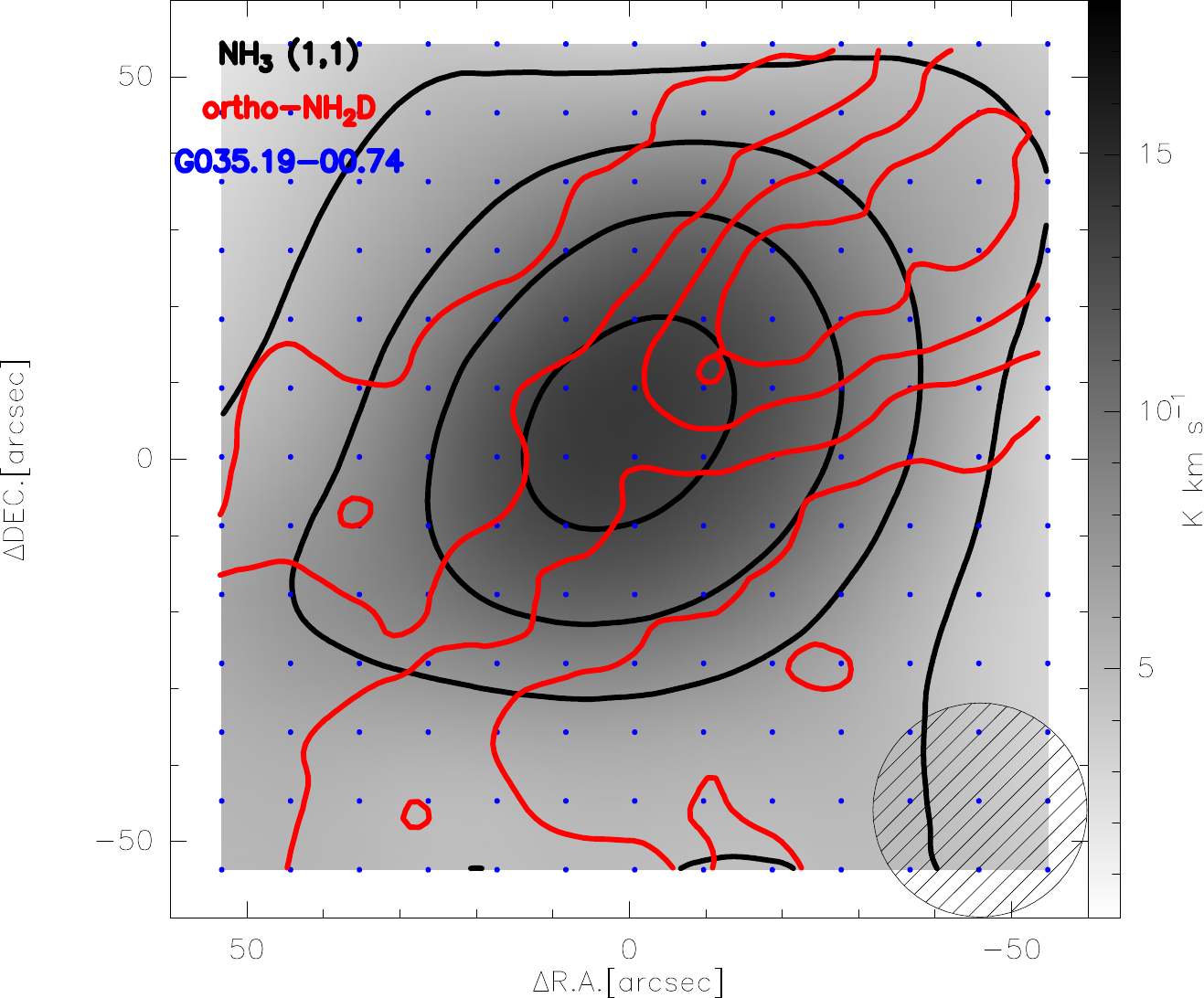}{0.5\textwidth}{(a)}
          \fig{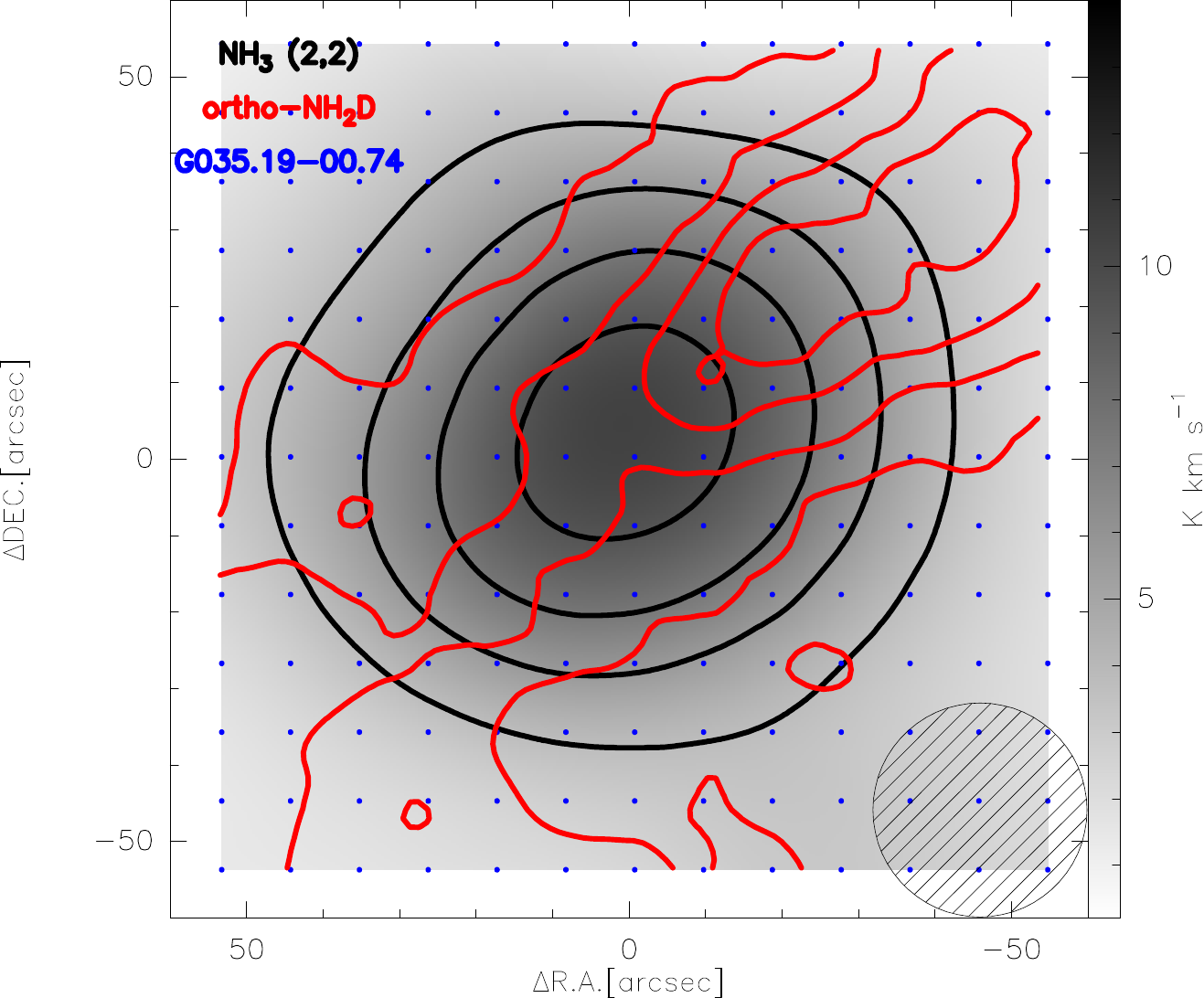}{0.5\textwidth}{(b)}}
\caption{(a) The velocity-integrated intensity of the ortho-NH$_2$D $1_{11}^s-1_{01}^a$ contour (red contour) overlaid on the NH$_3$(1,1) main group velocity-integrated intensity image (gray scale and black contour) in G035.19-00.74. The contour levels start at 5$\sigma$ in steps of 5$\sigma$ for ortho-NH$_2$D $1_{11}^s-1_{01}^a$, while the contour levels start at 90$\sigma$ in steps of 54$\sigma$ for the NH$_3$(1,1) main group. The gray scale starts at 3$\sigma$. (b) Ortho-NH$_2$D $1_{11}^s-1_{01}^a$ at 85.926 GHz velocity-integrated intensity contour (red contour) overlaid on the NH$_3$(2,2) main group velocity-integrated intensity image (gray scale and black contour) in G035.19-00.74. The contour levels start at 5$\sigma$ in steps of 5$\sigma$ for ortho-NH$_2$D $1_{11}^s-1_{01}^a$, while the contour levels start at 54$\sigma$ in steps of 27$\sigma$ for the NH$_3$(2,2) main group. The gray scale starts at 3$\sigma$.}
\label{example_G03519_NH3}
\end{figure}

In cold environment, the five components of ortho-NH$_2$D $1_{11}^s-1_{01}^a$ {can often be} distinguished \cite[e.g.][]{2007A&A...470..221C,2021ApJ...912L...7L,2000ApJ...535..227S}. However, due to the line broadening in late-stage massive star forming regions, three components are blended and only three can be distinguished with single-dish observations \citep[][]{2023MNRAS.526.3673P,2007A&A...467..207P}. Therefore, the fluxes of ortho-NH$_2$D $1_{11}^s-1_{01}^a$ were derived with the velocity-integrated intensity of the emitting channels of its six hyperfine structures instead of Gaussian fitting. {For NH$_3$(1,1) and NH$_3$(2,2),} the main group was used to derive the velocity-integrated intensity. The NH$_3$(1,1) and NH$_3$(2,2) maps present a clear structure for each source. {The ortho-NH$_2$D $1_{11}^s-1_{01}^a$, NH$_3$(1,1) and NH$_3$(2,2) velocity-integrated intensity maps for G035.19-00.74 are shown in Figure \ref{example_G03519_NH3},} while other sources are shown in Figures \ref{G015_NH3}-\ref{G121_NH3}.

From the velocity-integrated intensity maps of ortho-NH$_2$D $1_{11}^s-1_{01}^a$, NH$_3$(1,1) and NH$_3$(2,2), half of the sources (G031.28+00.06, G034.39+00.22, G035.19-00.74, G081.87+00.78, G109.87+02.11 and G111.54+00.77), showed morphological differences in spatial distribution between ortho-NH$_2$D $1_{11}^s-1_{01}^a$ and NH$_3$(1,1) (or NH$_3$(2,2)).  Even though the ortho-NH$_2$D $1_{11}^s-1_{01}^a$ and NH$_3$ emissions can be resolved in  G121.29+00.65,  no clear difference can be found for the spatial distribution of these lines. For the remaining five sources (G015.03-00.67, G023.44-00.18, G035.20-01.73, G075.76+00.33 and G081.75+00.59), the differentiation in distribution between ortho-NH$_2$D $1_{11}^s-1_{01}^a$ and NH$_3$(1,1) (or NH$_3$(2,2)) is less apparent owing to the limited spatial resolution.

\subsection{Temperature and column density}
\label{temperature and column density}
Two ammonia inversion transitions, NH$_3$(1,1) and NH$_3$(2,2), were obtained in 12 sources with Effelsberg observations and {GBT archival data.} Based on the methods of \cite{1983ARA&A..21..239H} and \cite{1992ApJ...388..467M} for calculating NH$_3$ parameters, \cite{2015ApJ...805..171L} provided a Python package\footnote{\url{https://xinglunju.github.io/software.html}} to fit the NH$_3$ spectra and obtained NH$_3$ parameters,  
including the rotation temperature $T_{\rm rot}$(2,2;1,1), which represents the population between NH$_3$(1,1) and NH$_3$(2,2) \citep{1983ARA&A..21..239H}, and the optical depth of NH$_3$(1,1) main group ($\tau$(1,1,m)).

{
At local thermodynamic equilibrium (LTE), the column density can be given by 
\begin{equation}
\label{eq: column density_1}
N_{tot}=\frac{3h}{8\pi^3 S\mu^2\nu}Q(T_{ex})e^{E_u/ kT_{ex}}\int T^{\prime}_{mb}d\nu (cm^{-2}),
\end{equation}

where $h=6.624\times10^{-27} $erg$\cdot$s is the Planck constant,  $\nu$ is the frequency of the transition, $E_u$ is the energy of the upper level, $S$ is the line strength, and $\mu^2$ is the dipole moment. {The values of these parameters for ortho-NH$_2$D $1_{11}^s-1_{01}^a$, NH$_3$(1,1) and NH$_3$(2,2) are listed in Table \ref{parameter}.} $\int T^{\prime}_{mb}d\nu$ is the velocity-integrated intensity with optical depth correction, given by {\cite{1999ApJ...517..209G}:}
\begin{equation}
\label{eq: optical depth correction}
\int T^{\prime}_{mb}d\nu=\int T_{mb}d\nu \frac{\tau}{1-e^{-\tau}},
\end{equation}
where $\int T_{mb}d\nu$ is the velocity-integrated intensity from observation.

The rotation temperature of ammonia, $T_{\rm rot}$(2,2;1,1), is assumed to be {equal to} the excitation temperature of ammonia, $T_{\rm ex}$. The NH$_3$ partition function is determined by considering the contribution of the different metastable levels (from $J,K$=0,0 to $J,K$=6,6), which are given by CDMS\footnote{\url{https://cdms.astro.uni-koeln.de/classic/predictions/catalog/partition_function.html}} \citep{2001A&A...370L..49M,2005JMoSt.742..215M}. Based on the above assumptions, the column density of NH$_3$ can be derived by Eq.\ref{eq: column density_1}.
}

For NH$_2$D, the excitation temperature cannot be derived, {as only one transition was observed.} Therefore, the rotation temperature of ammonia is assumed to be the excitation temperature of NH$_2$D. 
{In regions with relatively low density, the ortho-NH$_2$D $1_{11}^s-1_{01}^a$ may not be thermalized, meaning that $T_{\rm ex}$ is not equal to $T_{\rm rot}$. {According to our checks,} this assumption of $T_{\rm ex}$=$T_{\rm rot}$ results in at most a 15\% error {if the $T_{\rm ex}$ of ortho-NH$_2$D $1_{11}^s-1_{01}^a$ exceeds 10 K.}} 
The ortho/para ratio of NH$_2$D is assumed to be 3 \citep{2000A&A...356.1039T}, {and we assume} that the ortho-NH$_2$D $1_{11}^s-1_{01}^a$ line is optically thin. The NH$_2$D partition function is also determined by considering the contributions of different energy levels (from $J$=0 to $J$=9), {as obtained from CDMS \citep{2001A&A...370L..49M,2005JMoSt.742..215M}.} Using Eq.\ref{eq: column density_1}, the column density of NH$_2$D could be derived.

\begin{figure}[h]
\centering
\includegraphics[width=0.65\textwidth]{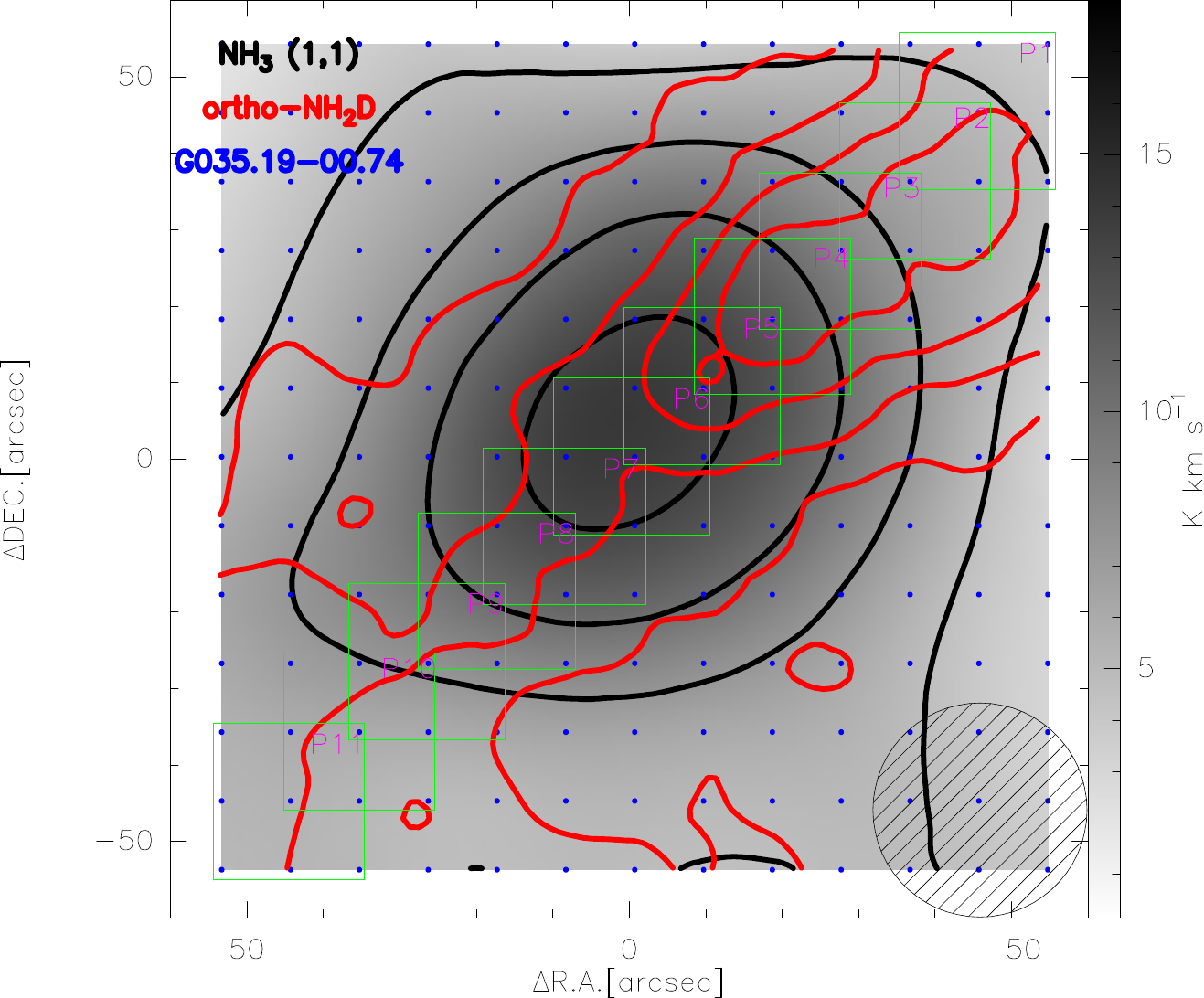}
\caption{The pixels in the box are the average area in G035.19-00.74. The order of the region numbers is the order in Table \ref{D-fraction}.}
\label{example_G03519_NH3_slice}
\end{figure}

To improve the signal-to-noise ratio of NH$_3$(1,1), NH$_3$(2,2) and ortho-NH$_2$D $1_{11}^s-1_{01}^a$, 
several pixels were averaged. {For G081.75+00.59, G109.87+00.77 and G121.29+00.65,} where the NH$_3$ data were obtained from Effelsberg 100 m observations, 4 pixels were averaged. In the other nine sources, 9 pixels were averaged.
{The method of pixel averaging for one of the sources is presented in Figure \ref{example_G03519_NH3_slice},} and other sources are shown in Figure \ref{slice}. In Figure \ref{example_G03519_NH3_slice} (and Figure \ref{slice}), the pixels in the green box are used for averaging. After reducing the spatial resolution, the column density and rotation temperature are listed in Table \ref{D-fraction}. 
In the following analysis, column density and rotation temperature with reduced spatial resolution were used.

\subsection{Deuterium fractionation of NH$_3$}
\label{deuterium fractionation of NH$_3$}
Assuming that the NH$_2$D emission and NH$_3$ emission originate from the same region, the deuterium fractionation of NH$_3$ is derived from the column densities of NH$_3$ and NH$_2$D {listed in Table \ref{D-fraction},} which is calculated after reducing the spatial resolution as mentioned in Section \ref{temperature and column density}.
In the following analysis, $D_{\rm frac}$(NH$_3$) with reduced spatial resolution was used.

\begin{figure}[h]
\centering
\includegraphics[width=0.65\textwidth]{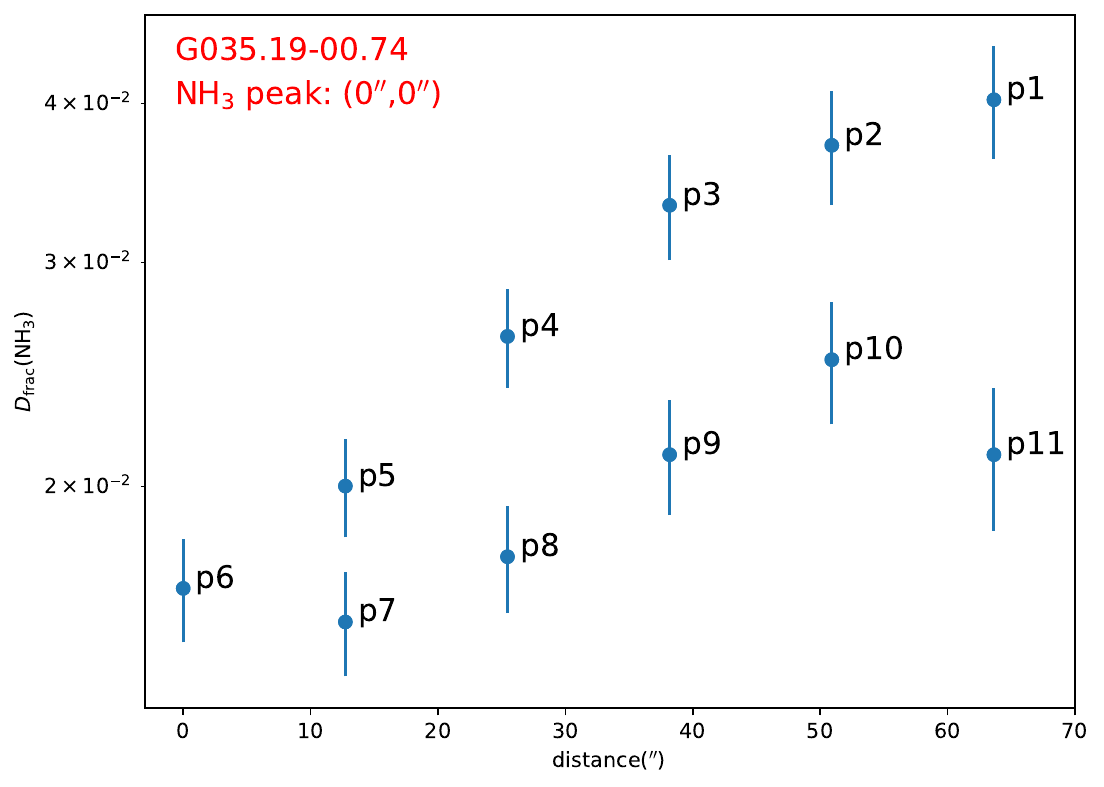}
\caption{The relationship between deuterium fractionation of NH$_3$ and distance of NH$_3$ peak in G035.19-00.74.}
\label{example_G03519_NH3_Dfrac_distance}
\end{figure}

The relationship between deuterium fractionation of NH$_3$ and distance of NH$_3$ peak in G035.19-00.74 is presented in Figure \ref{example_G03519_NH3_Dfrac_distance}, while other sources are shown in Figure \ref{Dfrac_distance}. {Multiple data points at the same distance occur because different locations share the same distance from the NH$_3$ peak.} For the six sources in which the spatial distribution of NH$_3$(1,1) differs from that of ortho-NH$_2$D $1_{11}^s-1_{01}^a$, {the location of the highest deuterium fractionation of NH$_3$ is offset from the NH$_3$ peak.}

The highest deuterium fractionation of NH$_3$ (0.086$\pm$0.007) {in the entire sample is found in G081.87+00.78 with offset (45$^{\prime\prime}$,9$^{\prime\prime}$),} while the NH$_3$ peak is located in offset (0$^{\prime\prime}$,0$^{\prime\prime}$). 
G111.54+00.77 exhibits the largest range of $D_{\rm frac}$(NH$_3$), {varying from 0.014$\pm$0.002 to 0.067$\pm$0.010.} 
{In contrast, for other sources, the variation in $D_{\rm frac}$(NH$_3$) exceeds a factor of two, except G031.28+00.06, G035.20-01.73 and G075.76+00.33, {for which the variation} in $D_{\rm frac}$(NH$_3$) is below a factor of two.} {The ranges and medians of $T_{\rm rot}$ and $D_{\rm frac}$(NH$_3$) for each source are listed in Table \ref{summary_source}.}

\subsection{Individual target detail}
\label{Individual target detail}

\subsubsection{G015.03-00.67}
{G015.03-00.67 (M17) is a molecular cloud complex \citep{1976ApJS...32..603L}  with one of the most luminous giant cometary H\uppercase\expandafter{\romannumeral2} regions in the Galaxy (3.6$\times$10$^6$$L_{\sun}$; \citealt{2009ApJ...696.1278P}). It is the closest (1.98 kpc, \citealt{2011ApJ...733...25X}) giant H\uppercase\expandafter{\romannumeral2} region to the sun. More than 100 OB-type massive young stellar objects have been found in G015.03-00.67 \citep{1991ApJ...368..432L,2009ApJ...696.1278P}. \cite{2023MNRAS.522..503Z} found that the shocked gas is spreading near the boundary of the M17 H\uppercase\expandafter{\romannumeral2} region, likely due to the expansion of ionized gas. }

{In the southwest of M17 with a 4$^{\prime}$$\times$4$^{\prime}$ size, the distributions of NH$_3$(1,1) and NH$_3$(2,2) and ortho-NH$_2$D $1_{11}^s-1_{01}^a$ were obtained (see Figure \ref{G015_NH3}).} A widely distributed NH$_3$(1,1) and NH$_3$(2,2) are detected, while the emission of ortho-NH$_2$D $1_{11}^s-1_{01}^a$ presents a fragmented distribution and weak emission. G015.03-00.67 is the source with the lowest deuterium fractionation in our sample. In this target, the region with the highest deuterium fractionation is located south of G015.03-00.67. The range of $D_{\rm frac}$(NH$_3$) is from 0.0051$\pm$0.0008 to 0.012$\pm$0.002 in G015.03-00.67.

\subsubsection{G023.44-00.18}
{With a strong 6.7 GHz CH$_3$OH maser detected, G023.44-00.18 is thought to have young massive stars inside \citep{1998MNRAS.301..640W}. \cite{2011MNRAS.415L..49R} reported that G023.44-00.18 has two dust cores arranged north and south, based on the Submillimeter Array (SMA) observation. They suggested that these two dust cores are in the pre-UCH\uppercase\expandafter{\romannumeral2} evolutionary stage, with outflow detected in the southern one. Using PdBI observations, \cite{2020A&A...638A.105Z} found several NH$_2$D cores in G023.44-00.18 with high $D_{\rm frac}$(NH$_3$)$\gtrsim$ 0.3, but these cores are offset from the two dust cores. }

Due to the spatial resolution limitation, these NH$_2$D cores cannot be distinguished in the present work (see Figure \ref{G023_NH3}). The ortho-NH$_2$D $1_{11}^s-1_{01}^a$, NH$_3$(1,1) and NH$_3$(2,2) are not detected as asymmetrically distributed spatial structures in the mapping center. {Therefore, a series of similar deuterium fractionations are derived in G023.44-00.18.} {However, the highest $D_{\rm frac}$(NH$_3$) was detected in the northwestern region of G023.44-00.18.} We derived a $D_{\rm frac}$(NH$_3$) range from 0.009$\pm$0.001 to 0.031$\pm$0.006 in this target.

\subsubsection{G031.28+00.06}
{G031.28+00.06 is embedded in the molecular cloud complex W43 \citep{2011A&A...529A..41N}. With Very Large Array (VLA) 6 cm continuum imaging, \cite{1992AJ....103..234F} found a ring emission in G031.28+00.06. Additionally, IRAM 30m observations detected H42$\alpha$ in G031.28+00.06, indicating the presence of an H\uppercase\expandafter{\romannumeral2} region \citep{2024MNRAS.527.5049L}.}

NH$_3$(1,1) and NH$_3$(2,2) exhibit a northeast-to-southwest distribution. But the ortho-NH$_2$D $1_{11}^s-1_{01}^a$ emission presents two structures (see Figure \ref{G031_NH3}) and seems to be separated by the H\uppercase\expandafter{\romannumeral2} region, which is reported in \cite{2024MNRAS.527.5049L}. The center of the spatial distribution of NH$_3$(1,1) and NH$_3$(2,2) is located in the middle of two NH$_2$D structures. The range of $D_{\rm frac}$(NH$_3$) is from 0.009$\pm$0.001 to 0.013$\pm$0.002 in this source.

\subsubsection{G034.39+00.22}
{G034.39+00.22 stretches about 9$\arcmin$ from north to south based on IRAM 30m 1.2 mm continuum observation \citep{2006ApJ...641..389R}. With VLA ammonia observation, \cite{1994ApJS...92..173M} reported that G034.39+00.22 contains a total mass of about 1000 $M_{\sun}$. {Additionally,} using the Nobeyama Radio Observatory (NRO) 45 m telescope observation, \cite{2012ApJ...747..140S} derived the D$_{frac}$(HNC) to be about 0.003. }

We obtain a north-south distribution of NH$_3$(1,1), NH$_3$(2,2) and ortho-NH$_2$D $1_{11}^s-1_{01}^a$ of 2$\arcmin$$\times$2$\arcmin$ size in G034.39+00.22. {The NH$_3$(1,1), NH$_3$(2,2) and ortho-NH$_2$D $1_{11}^s-1_{01}^a$ show distinct structures aligned along the north-south axis, without significant asymmetry in the east-west direction (see Figure \ref{G034_NH3}).} The deuterium fractionation of NH$_3$ in the southern structure is higher than that in the northern structure. In G034.39+00.22, the range of $D_{\rm frac}$(NH$_3$) is from 0.017$\pm$0.002 to 0.035$\pm$0.003.

\subsubsection{G035.19-00.74}
{G035.19-00.74 contains a young B0 star \citep{1984MNRAS.210..173D}. A CO outflow was detected in this source with an orientation of approximately 45$^{\circ}$ along the northeast-southwest direction with James Clerk Maxwell Telescope (JCMT) observations \citep{2003MNRAS.339.1011G}. NH$_3$ mapping with VLA observations revealed the presence of a rotating interstellar disk with a mass of about 150 $M_{\sun}$ {in G035.19-00.74 \citep{1985MNRAS.217..227L}, which is perpendicular to the CO outflow reported by \cite{2003MNRAS.339.1011G}.}}

{In G035.19-00.74, strong ortho-NH$_2$D $1_{11}^s-1_{01}^a$ emission is detected in the southeast and northwest regions, while the NH$_3$ peak is located at the center of the map. Therefore, the high deuterium fractionation of NH$_3$} is located in the southeast and northwest regions, while the deuterium fractionation of NH$_3$ is relatively low in the NH$_3$ peak. We derived a $D_{\rm frac}$(NH$_3$) range from 0.016$\pm$0.001 to 0.040$\pm$0.004 in this source.

\subsubsection{G035.20-01.73}
{G035.20-01.73 (W 48) contains a 2$\arcmin$$\times$2$\arcmin$ H\uppercase\expandafter{\romannumeral2} region \citep{1989ApJS...69..831W,2005ApJ...626..253R}. With Herschel and Berkeley Illinois Maryland Association (BIMA) interferometer observations, {\cite{2014MNRAS.440..427R} reported an east-west evolutionary gradient within this UCH\uppercase\expandafter{\romannumeral2} region. For G035.20-01.73, it was found that the 3.5 mm dust emission peak does not coincide with the brightest NH$_2$D peak with PdBI observations \citep{2011A&A...530A.118P}. }} 

The east-west distributions of ortho-NH$_2$D $1_{11}^s-1_{01}^a$, NH$_3$(1,1) and NH$_3$(2,2) are {present} in G035.20-01.73 (see Figure \ref{G03520_NH3}). The ortho-NH$_2$D $1_{11}^s-1_{01}^a$ emission is located in the west of G035.20-01.73 and shows a similar distribution to the result in \cite{2011A&A...530A.118P}. In contrast, the emission of ortho-NH$_2$D $1_{11}^s-1_{01}^a$, NH$_3$(1,1) and NH$_3$(2,2) is weaker in the east of G035.20-01.73, where the UCH\uppercase\expandafter{\romannumeral2} region is located \citep{2014MNRAS.440..427R}. There is also an east-west gradient in the deuterium fractionation of NH$_3$, increasing farther away from the UCH\uppercase\expandafter{\romannumeral2} region. The range of $D_{\rm frac}$(NH$_3$) is from 0.024$\pm$0.003 to 0.041$\pm$0.004 in G035.20-01.73.

\subsubsection{G075.76+00.33}
{G075.76+00.33 is located in the giant molecular cloud ON2 \citep{1973A&A....29..309M}. An UCHII region in the north of G075.76+00.33 was reported by \cite{1989ApJS...69..831W}. }

The northeast-to-southwest distributions of ortho-NH$_2$D $1_{11}^s-1_{01}^a$, NH$_3$(1,1) and  NH$_3$(2,2) are presented in Figure \ref{G075_NH3}. {However, the ortho-NH$_2$D $1_{11}^s-1_{01}^a$, NH$_3$(1,1) and  NH$_3$(2,2) emissions are not associated with the  UCH\uppercase\expandafter{\romannumeral2} region as reported in \citep{1989ApJS...69..831W}.} This target exhibits the narrowest range of $D_{\rm frac}$(NH$_3$) in our sample, ranging from 0.018$\pm$0.2 to 0.023$\pm$0.003.

\subsubsection{G081.75+00.59}
{G081.75+00.59 (DR 21) is a giant star-forming complex in the Cygnus X molecular cloud complex \citep{1978ApJ...223..840D,2009ApJ...694.1056K} with about 25$\arcmin$-length ridge based on Herschel observations\citep{2007A&A...476.1243M,2012A&A...543L...3H}. It contains several YSOs and high-mass protostars \citep[e.g.][]{1986ApJ...300..737H,1991ApJ...378..576M,2007A&A...476.1243M}. With IRAM observation, \cite{2023pcsf.conf..253C} reported an extended distribution of ortho-NH$_2$D $1_{11}^s-1_{01}^a$, DCN 1-0, DNC 1-0 and DCO$^+$ 1-0 in Cygnus X. }

We obtain a similar ortho-NH$_2$D $1_{11}^s-1_{01}^a$ distribution to that reported in \cite{2023pcsf.conf..253C} (see Figure \ref{G08175_NH3}). In G081.75+00.59, the abundance of NH$_3$ increases from north to south, whereas the NH$_2$D abundance decreases along the same direction. We derived a $D_{\rm frac}$(NH$_3$) range from 0.022$\pm$0.002 to 0.068$\pm$0.008 in G081.75+00.59.

\subsubsection{G081.87+00.78}
{North of G081.75+00.59, G081.87+00.78 (W75N) is {another massive star-forming region} in the Cygnus X molecular cloud complex \citep{1978ApJ...223..840D,2006A&A...445..971P}. This target is reported {to host a large-scale, high-velocity molecular outflow \citep{2003ApJ...584..882S} and contains a B-type YSO \citep{2004ApJ...601..952S}. }}

The ortho-NH$_2$D $1_{11}^s-1_{01}^a$, NH$_3$(1,1) and NH$_3$(2,2) present a southeast-to-northwest distribution in G081.87+00.78, while ortho-NH$_2$D $1_{11}^s-1_{01}^a$ {is predominantly located} to the east of NH$_3$(1,1) and NH$_3$(2,2) (see figure \ref{G08187_NH3}). G081.87+00.78 has the highest deuterium fractionation of NH$_3$ (0.086$\pm$0.007) in our sample. Besides, the range of $D_{\rm frac}$(NH$_3$) is from 0.036$\pm$0.003 to 0.086$\pm$0.007 in this source.

\subsubsection{G109.87+02.11}
{With several YSOs detected, G109.87+02.11 (Cep A) is an active massive star-forming region \citep{1984ApJ...276..204H}. HW2, the brightest radio source in G109.87+02.11, is associated with a zero-age main-sequence (ZAMS) star of early B spectral type \citep{1984ApJ...276..204H}. With IRAM 30 m observations, \cite{2017MNRAS.466..248L} detected strong ortho-NH$_2$D emission in the northeast of G109.87+02.11, suggesting that the abundance of NH$_2$D may be associated with shock.}

As is shown in Figure \ref{G109_NH3}, strong ortho-NH$_2$D $1_{11}^s-1_{01}^a$ emission is detected in the northeast of G109.87+02.11, while NH$_3$(1,1) and  NH$_3$(2,2) are weak, {indicating high deuterium fractionation in this region.} We derived a $D_{\rm frac}$(NH$_3$) range from 0.011$\pm$0.002 to 0.055$\pm$0.023 in this target.  

\subsubsection{G111.54+00.77}
{G111.54+00.77 (NGC 7538) is a massive star-forming region located in a giant molecular cloud complex \citep{2000ApJ...537..221U} {and is associated with a H\uppercase\expandafter{\romannumeral2} region \citep[e.g.][]{2016ApJ...824..125L,2017MNRAS.467.2943S,2022A&A...659A..77B}.} With Herschel observations, several high-mass dense clump candidates were identified in G111.54+00.77, which are expected to produce several intermediate-to-high-mass stars in the future \citep{2013ApJ...773..102F}. The mass of gas and dust in this target is estimated to be several thousand solar masses \citep{2005ApJ...625..891R}.}

In G111.54+00.77, {widespread distributions of ortho-NH$_2$D $1_{11}^s-1_{01}^a$, NH$_3$(1,1) and NH$_3$(2,2) have been detected (see figure \ref{G111_NH3}). NH$_3$(1,1) and NH$_3$(2,2) are predominantly detected in the southern region, whereas ortho-NH$_2$D $1_{11}^s-1_{01}^a$  is more concentrated in the southwest, indicating a high deuterium fractionation in the southwest of G111.54+00.77.} This target exhibits the widest range of $D_{\rm frac}$(NH$_3$) in our sample, ranging from 0.014$\pm$0.002 to 0.067$\pm$0.010.

\subsubsection{G121.29+00.65}
{With the detection of 6.7 GHz CH$_3$OH,} G121.29+00.65 (L1287) is forming a massive star \citep{2010A&A...511A...2R}. {A bipolar outflow oriented in the northeast-southwest direction} was observed through SMA observations \citep{2019A&A...621A.140J}. 

The distributions of ortho-NH$_2$D $1_{11}^s-1_{01}^a$, NH$_3$(1,1) and NH$_3$(2,2) are shown in Figure \ref{G121_NH3}. Both NH$_2$D and NH$_3$ {exhibit distributions that are perpendicular to the bipolar outflow as reported by \citep{2019A&A...621A.140J}.} The ortho-NH$_2$D $1_{11}^s-1_{01}^a$, NH$_3$(1,1) and NH$_3$(2,2) show a northwest-to-southeast distribution. Similar to G035.19-00.74, the high deuterium fractionation in G121.29+00.65 is detected in the southeastern and northwestern regions, while it is relatively low in the mapping center, where the NH$_3$ peak is located. The range of $D_{\rm frac}$(NH$_3$) is from 0.020$\pm$0.003 to 0.041$\pm$0.014 in this target.

\section{Discussion}
\label{discussion}

\subsection{Environmental impact of the enrichment of deuterated ammonia}
\label{Environmental impact of the enrichment of deuterated ammonia}
\begin{figure}
\centering
\gridline{\fig{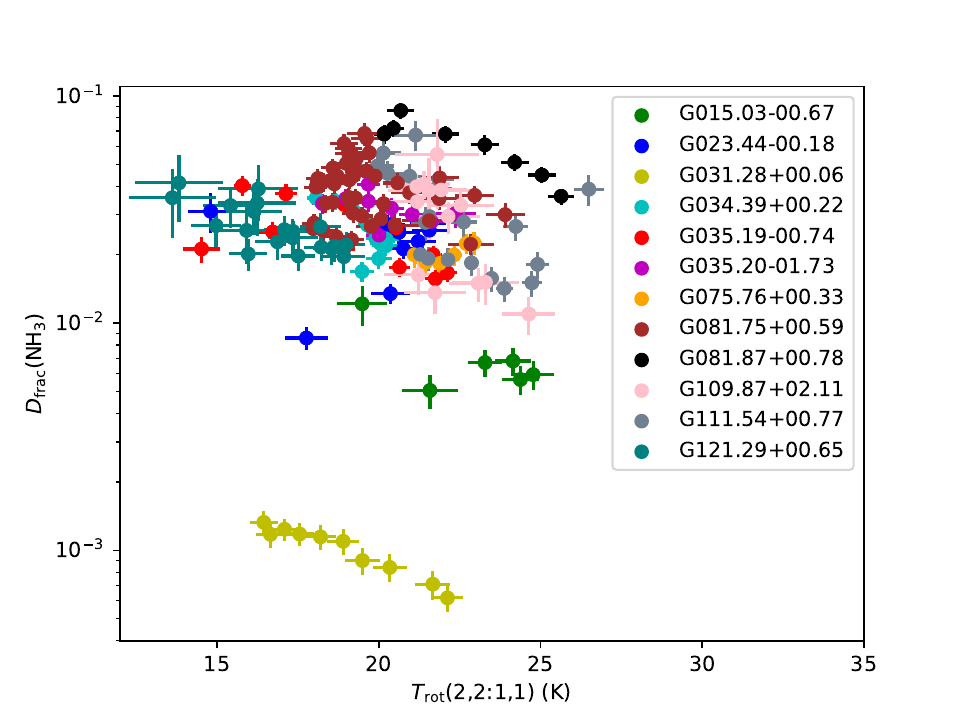}{0.33\textwidth}{(a)}
          \fig{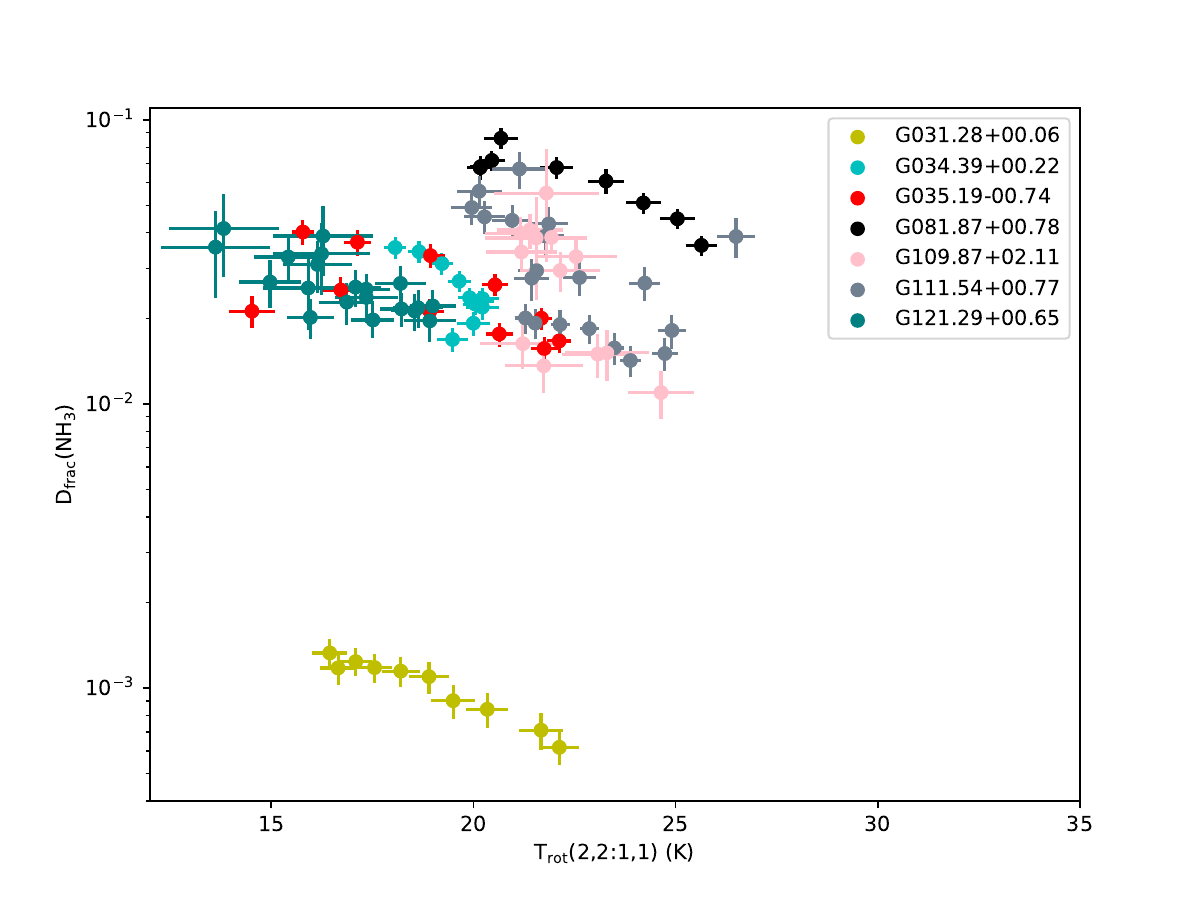}{0.33\textwidth}{(b)}
          \fig{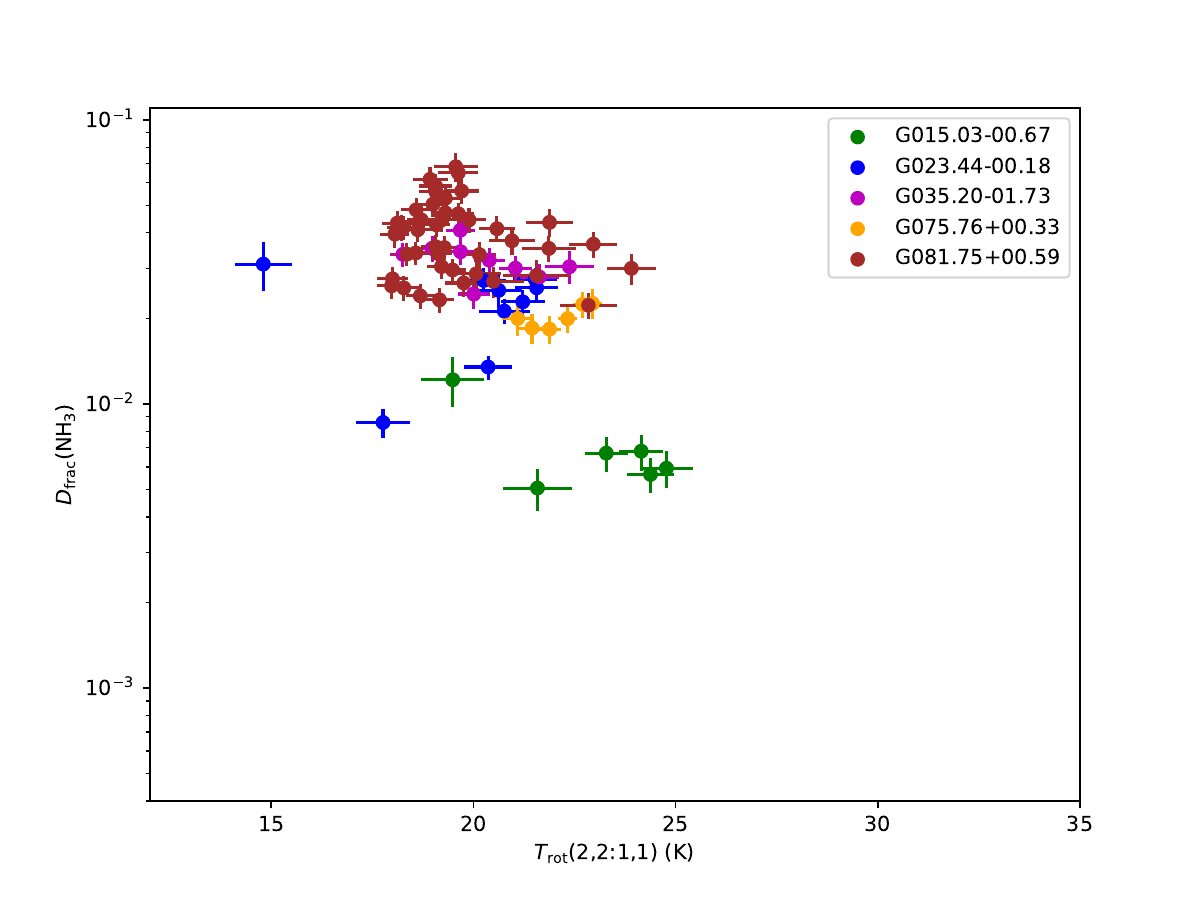}{0.33\textwidth}{(b)}
          }
\caption{The deuterium fractionation of ammonia compared with the ammonia rotation temperature. Each source is represented by a different color. (a) $D_{\rm frac}$(NH$_3$) compared with $T_{\rm rot}$(2,2;1,1) over the whole sample. (b) $D_{\rm frac}$(NH$_3$) compared with $T_{\rm rot}$(2,2;1,1) for seven sources that present clear anticorrelation between $D_{\rm frac}$(NH$_3$) and $T_{\rm rot}$(2,2;1,1). (c) $D_{\rm frac}$(NH$_3$) compared with $T_{\rm rot}$(2,2;1,1) for five sources that do not present clear correlation between $D_{\rm frac}$(NH$_3$) and $T_{\rm rot}$(2,2;1,1).}
\label{D_fraction temperature}
\end{figure}

Seven targets, G031.28+00.06, G034.39+00.22, G035.19-00.74, G081.87+00.78, G109.87+02.11, G111.54+00.77 and G121.29+00.65, exhibit an anticorrelation between $D_{\rm frac}$(NH$_3$) and $T_{\rm rot}$(2,2;1,1) within each source (see Figure \ref{D_fraction temperature}). This anticorrelation indicates that $D_{\rm frac}$(NH$_3$) is influenced by temperature, with $D_{\rm frac}$(NH$_3$) decreasing as temperature increases in these seven targets. Such anticorrelation was reported in case studies with interferometric observations  \citep{2010A&A...517L...6B,2013ApJ...777...85N}. Therefore, the temperature could indeed be an important physical parameter to influence the deuterium fractionation of NH$_3$. For the other five targets, G015.03-00.67, G023.44-00.18, G035.20-01.73, G075.76+00.33 and G081.75+00.59, the relationship between $D_{\rm frac}$(NH$_3$) and $T_{\rm rot}$(2,2;1,1) could not be analyzed owing to limitations in spatial resolution and sensitivity. {Our 12 targets are classified based on the presence or absence of clear anticorrelation between $D_{\rm frac}$(NH$_3$) and $T_{\rm rot}$(2,2;1,1), as listed in Table \ref{summary_source}.}

{However,} no clear trend of the relation between  $D_{\rm frac}$(NH$_3$) and $T_{\rm rot}$(2,2;1,1) is evident when analyzing all data across 12 targets together (see Figure \ref{D_fraction temperature}). 
{Furthermore, there is no clear trend between $D_{\rm frac}$(NH$_3$) and $T_{\rm rot}$(2,2;1,1) exhibited in either one of the groups even when considered separately, regardless of the presence or absence of a clear anticorrelation between $D_{\rm frac}$(NH$_3$) and $T_{\rm rot}$(2,2;1,1) (see also Figure \ref{D_fraction temperature}).}
Similar results were reported by \cite{2023MNRAS.526.3673P}, where no correlation between $D_{\rm frac}$(NH$_3$) and $T_{\rm rot}$(2,2;1,1) was found when considering five high-mass star-forming regions as a whole, based on IRAM 30 m and Effelsberg 100 m observations.
Such a result is somehow also consistent with that from single-pointing observations using single-dish telescopes \citep{2015A&A...575A..87F, 2021A&A...649A..21W}.  {Physical parameters, such as ionization rate, density and evolutionary stage, vary among sources,} which could contribute to the lack of a clear relationship between $D_{\rm frac}$(NH$_3$) and temperature when analyzing a large sample. 

{H$_2$D$^+$ plays a key role in deuterium fractionation processes \citep[e.g.][]{1992A&A...258..479P,2009A&A...493...89E,2012A&ARv..20...56C,2014prpl.conf..859C}.}
{In cold molecular gas, 
the chemical model shows that the abundance ratio of H$_2$D$^+$ to H$^+_3$ drops rapidly when the gas kinetic temperature is above $\sim$ 15 K  \citep{2008A&A...492..703C},} leading to reduced deuterium fractionation as temperature rises. However, $D_{\rm frac}$(NH$_3$) may be influenced not only by temperature but also by other physical parameters. To understand how $D_{\rm frac}$(NH$_3$) is influenced by other physical parameters, we constructed a chemical model that considers different physical parameters, including temperature, density, cosmic-ray ionization rates and evolutionary timescales.

\subsection{Chemical model}
\label{Chemical model}
The chemical model used in this work is the DNAUTILUS gas-grain chemical model, an updated version of the NAUTILUS model \citep{2014MNRAS.440.3557R,2015MNRAS.447.4004R,2015MNRAS.453L..48W} described in \cite{2017MNRAS.466.4470M}. This code solves for the time dependent abundances of molecules, in both two-phase mode, which treats the entire grain as chemically homogeneous, and three-phase mode, which treats the grain surface and bulk as chemically distinct. This chemical model considers the gas- and grain-phase chemical species from the KIDA network\footnote{\url{https://kida.astrochem-tools.org}}\citep{2015ApJS..217...20W}, including multiply deuterated isotopologues. The chemical network consists of 15 elements, all of which are listed with their initial abundances in Table \ref{tab:initial_abundances}. Only H$_{2}$ and HD exist in molecular form, while the other elements remain in atomic form. Elements such as  C, S, Si, Fe, Na, Mg, Cl and P, whose ionization potential is less than 13.6 eV, appear as singly positive ions. {Though the model can handle the spin chemistry of H$_2$, H$_2^+$, H$_3^+$, and their corresponding deuterated isotopologues, the present network includes deuteration up to 14 atoms without considering the spin of any H- or D-bearing species. This simplification is necessary for simplicity and computational efficiency given the large parameter space that we choose to explore in this work.} The network includes 1606 gas species, 1472 grain surface species, and 1472 grain-mantle species linked by 83715 gas-phase reactions, 10,967 reactions on grain surfaces, and 9431 reactions in the grain mantles.  More details are described in \cite{2017MNRAS.466.4470M}.

To obtain the deuterium fractionation of NH$_3$ varying with temperature, the model was run with dust and gas temperatures ranging from 6 K to 60 K, in the steps of 2 K. To account for different physical environments, the model was run with varying proton densities ({$n_{\mathrm{H}}$}) and cosmic-ray ionization rates ($\zeta$). Specifically, {the proton densities were set to 10$^4$ cm$^{-3}$, 10$^5$ cm$^{-3}$, 10$^6$ cm$^{-3}$ and 10$^7$ cm$^{-3}$, and the cosmic-ray ionization rates to 3$\times$10$^{-18}$ s$^{-1}$, 1.5$\times$10$^{-17}$ s$^{-1}$, 3$\times$10$^{-17}$ s$^{-1}$, 6$\times$10$^{-17}$ s$^{-1}$ and 3$\times$10$^{-16}$ s$^{-1}$}. A proton density of {n$_{\mathrm{H}}$}=10$^6$ cm$^{-3}$ is typical for hot cores \citep[e.g.][]{1998ARA&A..36..317V,2000prpl.conf..299K}, and $\zeta$=3$\times$10$^{-17}$ s$^{-1}$ is the average cosmic-ray ionization rate measured in massive star-forming regions by \cite{2000A&A...358L..79V}. A visual extinction of 10 mag was used in each simulation. {According to our tests, visual extinction does not significantly impact the simulation results.} The physical parameters used in our simulation are listed in Table \ref{tab:para1}.

\begin{figure}
\centering
\includegraphics[width=0.65\textwidth]{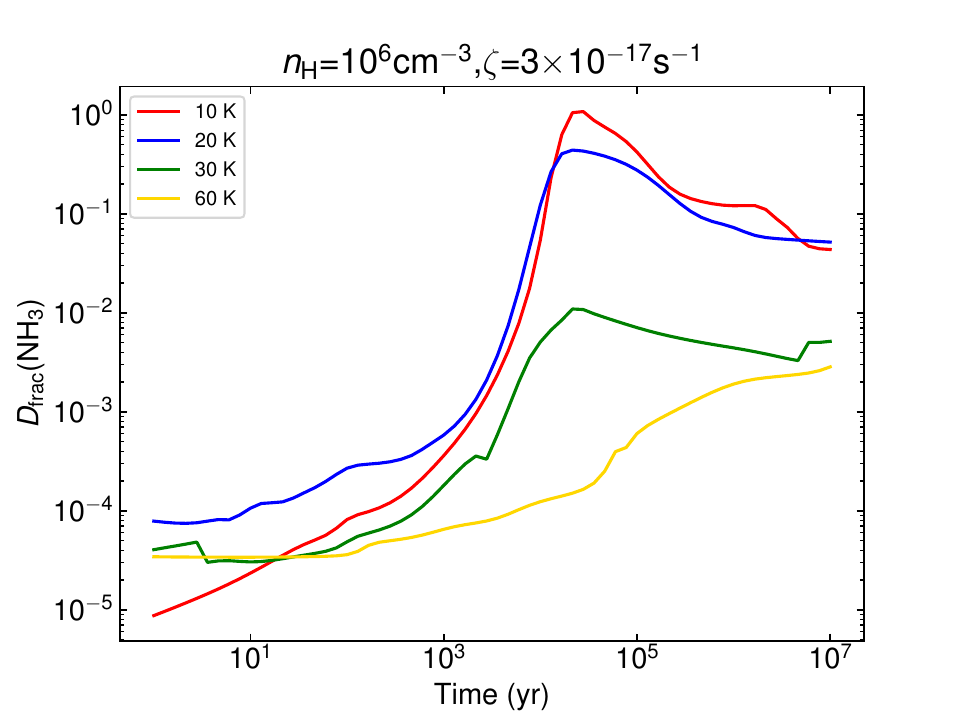}
\caption{Deuterium fractionation of NH$_3$ as a function of timescale simulated by the gas-grain chemical model. The timescales for the red, blue, green and yellow lines are 10, 20, 30 and 60 K, respectively.}
\label{model_time_scale}
\end{figure}

\begin{figure}
\centering
\includegraphics[width=\textwidth]{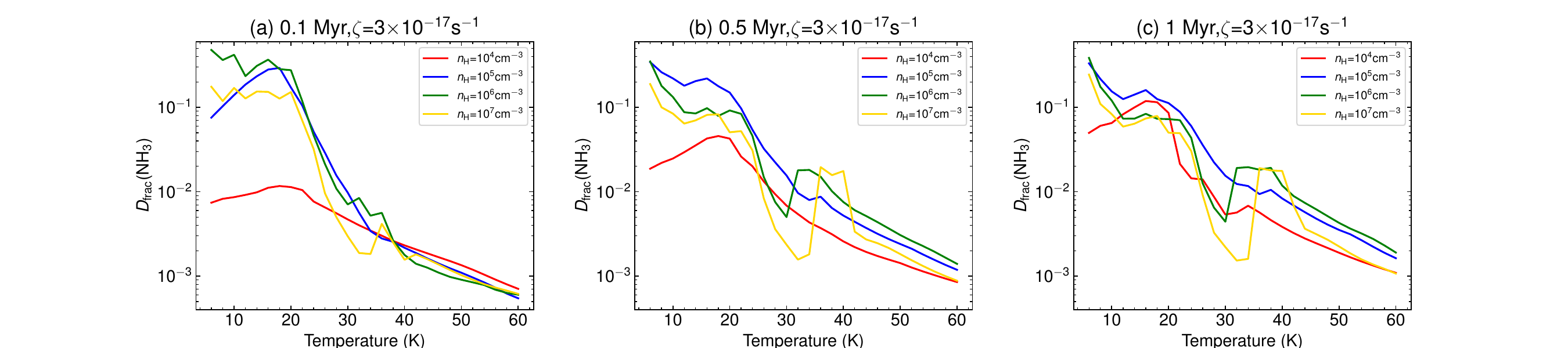}
\caption{Deuterium fractionation of NH$_3$ as a function of temperature for different densities simulated by the gas-grain chemical model. The density of H$_2$ for the red, blue, green and yellow lines is 10$^4$, 10$^5$, 10$^6$ and 10$^7$ cm$^{-3}$, respectively. The timescales for the left, middle and right panels are 0.1, 0.5 and 1 Myr, respectively. The cosmic-ray ionization rate is set to $\zeta$=3$\times$10$^{-17}$ s$^{-1}$ for each model.}
\label{model_density}
\end{figure}

\begin{figure}
\centering
\includegraphics[width=\textwidth]{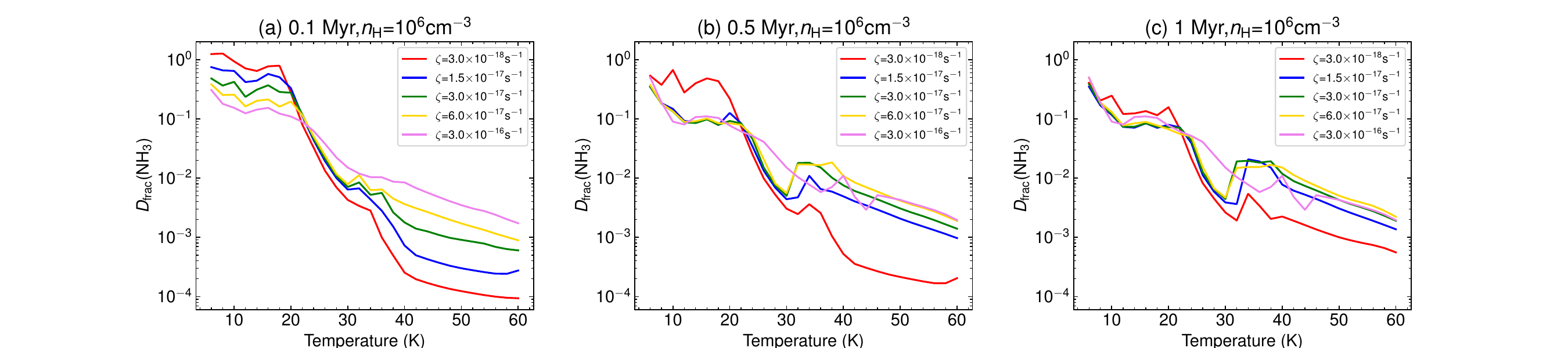}
\caption{Deuterium fractionation of NH$_3$ as a function of temperature for different cosmic-ray ionization rates simulated by the gas-grain chemical model. The cosmic-ray ionization rate for the red, blue, green, yellow and violet lines is 3$\times$10$^{-18}$, 1.5$\times$10$^{-17}$, 3$\times$10$^{-17}$, 6$\times$10$^{-17}$ and 3$\times$10$^{-16}$ s$^{-1}$ respectively. The timescales for the left, middle and right panels are 0.1r, 0.5 and 1 Myr, respectively. The cosmic-ray ionization rate is set to $n_{\rm H_2}$=10$^6$ cm$^{-3}$ for each model.}
\label{model_ionization}
\end{figure}

The simulation results are present in Figure \ref{model_time}-\ref{model_ionization}. In Figure \ref{model_time}, with a fixed proton density {$n_{\mathrm{H}}$}=10$^6$ cm$^{-3}$) and cosmic-ray ionization rate ($\zeta$=3$\times$10$^{-17}$ s$^{-1}$), there is only a slight difference between the different timescales (0.1, 0.5 and 1 Myr) regarding the relationship between the deuterium fractionation of NH$_3$ and temperature. $D_{\rm frac}$(NH$_3$) generally decreases as temperature increases, though this trend is not strictly monotonic. A ``bulge" in $D_{\rm frac}$(NH$_3$) is seen between 30 and 40 K before continuing its downward trend with higher temperatures. Some major production and destruction pathways of NH$_3$ and NH$_2$D, {in which the reactions have a contribution of at least $\textgreater$ 10\%,} are listed in table \ref{tab:major_reactions}. For all temperatures $T$ up to 60 K considered in the simulation, the electron recombination reaction (equation \ref{eq:NH3prod}) remains a significant pathway for the production of NH$_3$ and NH$_2$D:

\begin{equation}\label{eq:NH3prod}
\text{NH}_4^+/\text{NH}_3\text{D}^+ + \text{e}^- \rightarrow \text{H} + \text{NH}_3/\text{NH}_2\text{D}.
\end{equation}

However, the destruction pathways of NH$_{3}$ and NH$_{2}$D evolve as temperature increases, influencing the NH$_{2}$D/NH$_{3}$ ratios. {The pathways are primarily dominated by various ion-molecule reactions, as shown in Table 6. At lower temperatures ($T$ $\lesssim$ 24 K), the pathways involving H$_{3}$$^{+}$ and H$^{+}$ are the major ones. However, as the temperature increases ($T$ $\gtrsim$ 24 K), other competing reactions involving HCO$^{+}$ and H$_{3}$O$^{+}$ become more favorable, with minor contributions from N$_{2}$H$^{+}$, S$^{+}$, and the depletion of NH$_{3}$ on grain surfaces.} 

With a fixed proton density {$n_{\mathrm{H}}$}=10$^6$ cm$^{-3}$) and cosmic-ray ionization rate ($\zeta$=3$\times$10$^{-17}$ s$^{-1}$), the simulation results between $D_{\rm frac}$(NH$_3$) and timescale are presented in Figure \ref{model_time_scale}. $D_{\rm frac}$(NH$_3$) evolves over time, initially showing an increasing trend before beginning to decrease at about 10$^4$ yr. 

{In Figure \ref{model_density}, with a fixed cosmic-ray ionization rate ($\zeta$=3$\times$10$^{-17}$ s$^{-1}$), no clear trend is found among different densities in the relationship between deuterium fractionation of NH$_3$ and temperature. Except for temperatures above 30 K at 0.1 Myr in our simulations, different densities lead to {variations of a factor of several or even dozens} in deuterium fractionation of NH$_3$. Furthermore, at 0.5 and 1 Myr, no clear relationship between density and deuterium fractionation of NH$_3$ is evident. However, at 0.1 Myr, the simulation results indicate a high deuterium fractionation of NH$_3$ at high densities and low temperatures.}

With a fixed proton density {n$_{\mathrm{H}}$}=10$^6$ cm$^{-3}$), there is a clear difference among different cosmic-ray ionization rates in the relationship between deuterium fractionation of NH$_3$ and temperature (see Figure \ref{model_ionization}). {Except for temperature at 20-26K, different cosmic-ray ionization rates can lead to {variations of a factor of several or even dozens} in deuterium fractionation of NH$_3$. At the timescale of 0.1 Myr, the deuterium fractionation of NH$_3$ increases as the ionization rate decreases when temperatures are below 20 K, but it increases with higher cosmic-ray ionization rates when temperatures exceed 30 K. However, at the timescales of 0.5 Myr and 1 Myr, the relationship between deuterium fractionation of NH$_3$ and cosmic-ray ionization rate becomes more complex, making it difficult to determine the exact influence of cosmic-ray ionization rate on deuterium fractionation of NH$_3$.}

\subsection{Comparison between chemical model and observation}
\label{Comparison between models and observation}
\begin{figure}
\centering
\includegraphics[width=0.65\textwidth]{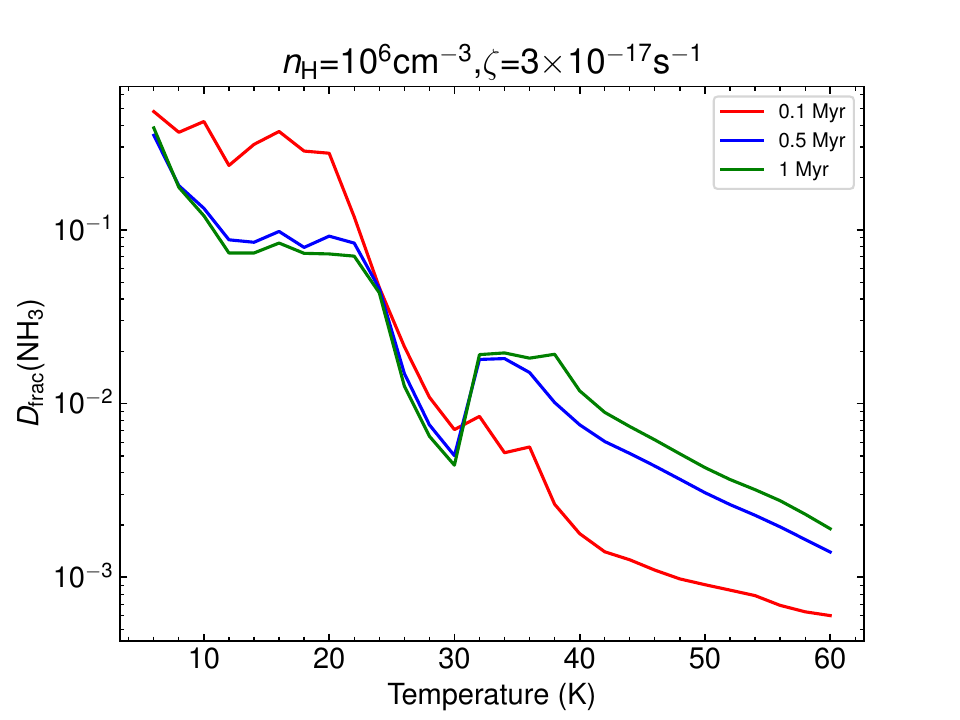}
\caption{Deuterium fractionation of NH$_3$ as a function of temperature simulated by the gas-grain chemical model. The timescales for the red, blue and green lines are 0.1, 0.5 and 1 Myr, respectively.}
\label{model_time}
\end{figure}

Most targets present a trend of increasing $D_{\rm frac}$(NH$_3$) with decreasing temperature, while this trend becomes unclear when all 12 targets are considered as a whole. According to our chemical model, when using consistent values for density, cosmic-ray ionization rates and timescale, $D_{\rm frac}$(NH$_3$) is high at low temperature ($T$$\lesssim$20 K) and low at high temperature ($T$$\gtrsim$50 K) (see Figure \ref{model_time}). This suggests an anticorrelation between $D_{\rm frac}$(NH$_3$) and temperature, explaining why most targets in this study, along with interferometric observations \citep[e.g.][]{2010A&A...517L...6B,2013ApJ...777...85N} show a trend of increasing $D_{\rm frac}$(NH$_3$) with decreasing temperature.

{
Our targets are located in different physical environments.
As described in Section \ref{Individual target detail}, although all targets are late-stage massive star-forming regions, there are slight differences in their evolutionary stages. For instance, G015.03-00.67 is a giant H\uppercase\expandafter{\romannumeral2} region \citep{1976ApJS...32..603L} and is associated with over 100 OB-type massive young stellar objects \citep{1991ApJ...368..432L,2009ApJ...696.1278P}, whereas G121.29+00.65 is forming a massive star \citep{2010A&A...511A...2R}. With IRAM 30m observations, dense gas tracer H$^{13}$CN 1-0 was mapped for our sources \citep{2024MNRAS.527.5049L}, but the velocity-integrated intensity of H$^{13}$CN 1-0 differs among the sources. {This suggests that the sources may have different densities.}

At the same temperature, our chemical model shows that different timescales, proton densities, or cosmic-ray ionization rates can lead to varying $D_{\rm frac}$(NH$_3$) (see Figure \ref{model_time}-\ref{model_ionization}). This indicates that different physical environments can lead to different $D_{\rm frac}$(NH$_3$). 
Consequently, when all 12 targets are considered collectively, there is no clear relationship between $D_{\rm frac}$(NH$_3$) and temperature.
}

In brief, $D_{\rm frac}$(NH$_3$) generally tends to decrease with increasing temperature. 
{But $D_{\rm frac}$(NH$_3$) is influenced not only by temperature but also by other physical parameters.}

{Limited by the evolutionary stage of the sample, we lack information on the early evolutionary stages. Additionally, our observation lacks other spectral lines that are unable to {estimate} the effect of different cosmic-ray ionization rates on $D_{\rm frac}$(NH$_3$). Combined with the early-stage massive star-forming regions and ionization degree tracer (such as HCO$^+$, N$_2$H$^+$ and their deuterated counterparts; \citealt{1998ApJ...499..234C,2002P&SS...50.1133C}), the future work could provide further insights into how $D_{\rm frac}$(NH$_3$) is influenced by different physical environments.}

\begin{figure}[h]
\centering
\includegraphics[width=0.65\textwidth]{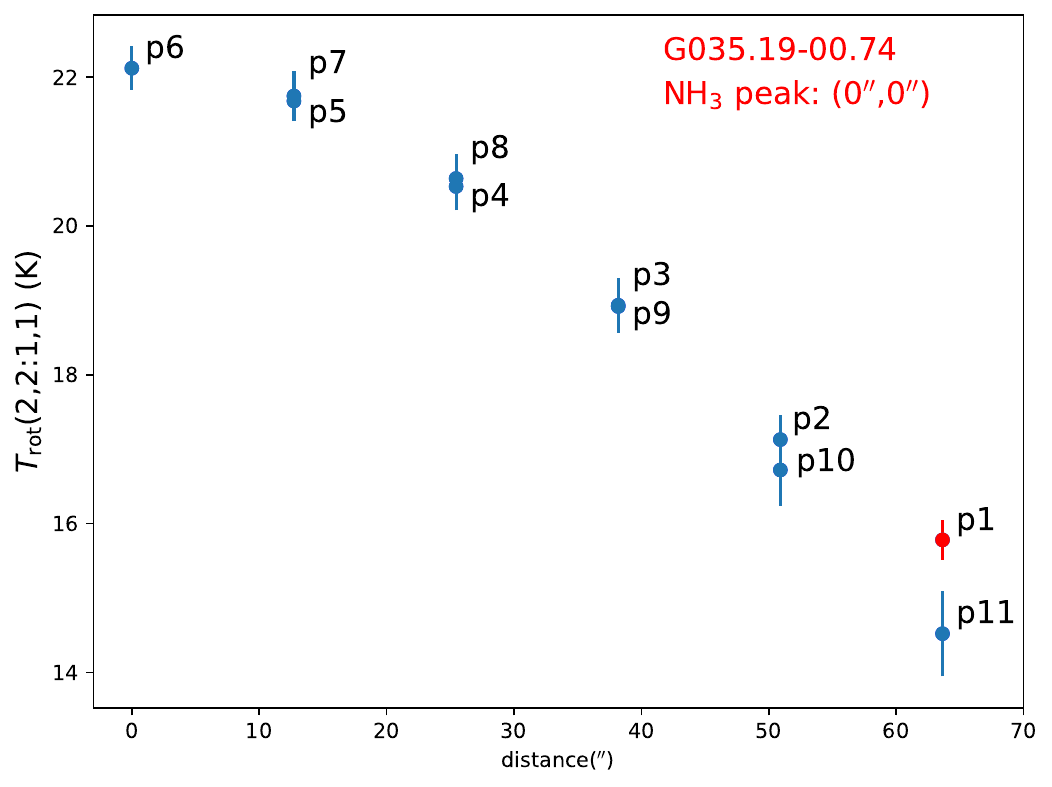}
\caption{The relationship between rotation temperature of NH$_3$ and distance of NH$_3$ peak in G035.19-00.74. The red marker is the position with the highest deuterium fractionation of NH$_3$.}
\label{example_G03519_NH3_Dfrac_distance_temperature}
\end{figure}

{
\subsection{Nonoverlap between NH$_3$ and $D_{\rm frac}$(NH$_3$) peaks}
\label{NH3 and Dfrac peaks}
The location of the highest deuterium fractionation of NH$_3$ deviates from the NH$_3$ peak in almost all sources, regardless of whether there is a difference in morphology between the spatial distribution of ortho-NH$_2$D $1_{11}^s-1_{01}^a$ and NH$_3$(1,1) (or NH$_3$(2,2)) or not. Although the location of the highest deuterium fractionation of NH$_3$ deviates from the NH$_3$ peak by less than 20$\arcsec$ in G035.20-01.73 and G075.76+00.33, the NH$_3$ and $D_{\rm frac}$(NH$_3$) peaks still may not be colocated. 

As shown in Figure \ref{example_G03519_NH3_Dfrac_distance_temperature}, the temperature decreases with increasing distance from the NH$_3$ peak in G035.19-00.74, while other sources are shown in Figure \ref{distance_temperature}. Only one source, G075.76+00.33 presents that the highest value of $D_{\rm frac}$(NH$_3$) and the highest rotation temperature are situated in the same position. The reason may be that the differentiation in the distribution of ortho-NH$_2$D $1_{11}^s-1_{01}^a$ and NH$_3$(1,1) (or NH$_3$(2,2)) is less apparent owing to the limited spatial resolution. In other sources, the highest deuterium fractionation of NH$_3$ occurs at relatively low temperatures. Our simulation shows that $D_{\rm frac}$(NH$_3$) decreases with increasing temperature, suggesting that the spatial offset between the NH$_3$ peaks and $D_{\rm frac}$(NH$_3$) peaks may result from the highest deuterium fractionation of NH$_3$ occurring in cooler regions.
}

According to the chemical model of \cite{2015A&A...581A.122S}, which considers the variation of physical conditions with distance from the core center, {$D_{\rm frac}$(NH$_3$) increases} as one moves away from the core, reaching a maximum at a few thousand au before gradually declining. This may be caused by the depletion of NH$_2$D from the gas phase as the molecular cloud evolves, leading to a decrease in $D_{\rm frac}$(NH$_3$) near the core region \citep{2013A&A...554A..92S,2015A&A...578A..55S}. 
{As the molecular cloud evolves, CO can be released from grains. The increased CO abundance consumes H$_3^+$, preventing the production of H$_2$D$^+$ and subsequently hindering the formation of NH$_2$D \citep[e.g.][]{2005A&A...438..585R,2012A&ARv..20...56C,2013A&A...554A..92S,2015A&A...578A..55S,2015A&A...581A.122S}. This may be one of the reasons for the low $D_{\rm frac}$(NH$_3$) near the core.}

Assuming that the NH$_3$ peak corresponds to the molecular cloud core in our sample, our results show a trend similar to that reported by \cite{2015A&A...581A.122S}. However, the distance between the region of highest $D_{\rm frac}$(NH$_3$) and the molecular cloud core in our results is an order of magnitude greater than in their simulations. This difference may be because single-dish telescope observations focus on extended spatial structures, while the simulations by \cite{2015A&A...581A.122S} focus on small-scale spatial structures. 
{At the scale of molecular clouds, the single-dish telescope beam encompasses multiple protostellar cores and young stars, which may be at different evolutionary stages. 
In different evolutionary regions of molecular clouds, the extent of NH$_2$D depletion and CO release from grains into the gas phase may vary.}
Therefore, despite this difference in scale, we suggest that the variation of $D_{\rm frac}$(NH$_3$) with distance from the molecular cloud core in our results on $\sim$parsec scales is similar to the spatial distribution patterns reported in \cite{2015A&A...581A.122S}.
{The spatial offset between the NH$_3$ and $D_{\rm frac}$(NH$_3$) peaks may also be due to the depletion of NH$_2$D in the gas phase as the molecular cloud core evolution, as well as the increased release of CO from grains into the gas phase.}

{
In brief, the $D_{\rm frac}$(NH$_3$) peak occurring outside of the NH$_3$ peak may be attributed to several factors. One factor is the easier formation of NH$_2$D at low temperatures. Another factor is the gradual depletion of NH$_2$D from the gas phase as the molecular cloud evolves. Additionally, the increasing release of CO from grains into the gas phase inhibits the production of NH$_2$D.
}

%

\section{Conclusions}
\label{summary}
Using mapping data of ortho-NH$_2$D $1_{11}^s-1_{01}^a$ at 85.926 GHz, NH$_3$(1,1) at 23.696 GHz and NH$_3$(2,2) at 23.722 GHz toward a sample of 12 late stage massive star-forming regions, we analyzed the spatial distribution of $D_{\rm frac}$(NH$_3$) and investigated the relationship between $D_{\rm frac}$(NH$_3$) and $T_{\rm rot}$(2,2;1,1) in each source. Our main results include the following:

\begin{enumerate}
	\item In our sample, the highest deuterium fractionation of NH$_3$ is detected in G081.87+00.78 with an offset of (45$^{\prime\prime}$,9$^{\prime\prime}$) with respect to the NH$_3$ peak. The widest range of $D_{\rm frac}$(NH$_3$) is found in G111.54+00.77, with values ranging from 0.014$\pm$0.002 to 0.067$\pm$0.010. 
	\item The region of highest deuterium fractionation of NH$_3$ deviates from the NH$_3$ peak for the seven sources with different morphology in the spatial distribution of NH$_3$ and ortho-NH$_2$D $1_{11}^s-1_{01}^a$ emission.
	\item We obtained the NH$_3$ rotation temperature in the range of 13-27 K. A clear trend of increasing $D_{\rm frac}$(NH$_3$) with decreasing $T_{\rm rot}$(2,2;1,1) is found in seven sources, but no clear trend of the relation between $D_{\rm frac}$(NH$_3$) and $T_{\rm rot}$(2,2;1,1) can be found when considering all targets as a whole.
	\item Employing the gas-grain chemical model, we found a decrease in $D_{\rm frac}$(NH$_3$) with increasing temperature. However, $D_{\rm frac}$(NH$_3$) may also be influenced by other physical parameters.
	\item {The relatively lower temperature regions are not situated in the NH$_3$ peak.  Additionally, molecular cloud evolves may lead to the depletion of NH$_2$D in the gas phase and an increased release of CO from grains into the gas phase. These factors contribute to the spatial offset between the NH$_3$ peak and the location of the highest $D_{\rm frac}$(NH$_3$).} 
\end{enumerate}   

\begin{acknowledgements}
This work is supported by the National Key R$\&$D Program of China (No. 2022YFA1603101)  and the National Natural Science Foundation of China (NSFC, Grant No. 12173067). This study is based on observations carried out under project numbers 042-19, 147-19, and 127-20 with the IRAM 30 m telescope. IRAM is supported by INSU/CNRS (France), MPG (Germany) and IGN (Spain). This work is based on observations with the 100 m telescope of the MPIfR (Max-Planck-Institut f\"{u}r Radioastronomie) at Effelsberg. We would like to extend our sincere gratitude to James Urquhart for his invaluable contribution to this work. His generous provision of NH$_3$ data significantly enriched the research presented here and was instrumental in achieving the depth and comprehensiveness of our findings. This research used the Canadian Advanced Network For Astronomy Research (CANFAR) operated in partnership by the Canadian Astronomy Data Centre and The Digital Research Alliance of Canada with support from the National Research Council of Canada the Canadian Space Agency, CANARIE and the Canadian Foundation for Innovation. This publication makes use of molecular line data from the Radio Ammonia Mid-Plane Survey (RAMPS). RAMPS is supported by the National Science Foundation under grant AST-1616635. Yaoting Yan and Siqi Zheng would like to thank the China Scholarship Council (CSC) for support. We sincerely thank the reviewers for their valuable suggestions, which have greatly improved the quality of this paper.
\end{acknowledgements}

%

\vspace{5mm}



\bibliography{sample631}{}
\bibliographystyle{aasjournal}

\onecolumngrid 

\begin{deluxetable*}{ccccccccc}
\tablecaption{Source List}
\label{source}
\tablewidth{0pt}
\tablehead{
\colhead{Source Name} & \colhead{R.A. (J2000)} & \colhead{Decl. (J2000)} &
 \colhead{$D_{\rm GC}$} & \colhead{$D_{\sun}$} & \colhead{$v_{\rm LSR}$}  & \colhead{Size} & \colhead{Ammonia Data Origin} \\
\colhead{} & \colhead{(hh:mm:ss)} & \colhead{(dd:mm:ss)} &
 \colhead{(kpc)} & \colhead{(kpc)} & \colhead{(km s$^{-1}$)} & \colhead{} & \colhead{}
}
\startdata
G015.03-00.67  & 18:20:22.01  & -16:12:11.30    &  6.4 & 2.0 & 22    & 4$'\times4'$ & \cite{2019ApJ...884....4K}\\
G023.44-00.18  & 18:34:39.29  & -08:31:25.40    &  3.7 & 5.9 & 97    & 2$'\times2'$ &\cite{2018ApJS..237...27H} \\
G031.28+00.06  & 18:48:12.39  & -01:26:30.70   &  5.2 & 4.3 & 109   & 2$'\times2'$ &\cite{2018ApJS..237...27H} \\
G034.39+00.22  & 18:53:19.00  & +01:24:50.80    &  7.1 & 1.6 & 57    & 2$'\times2'$ &\cite{2015MNRAS.452.4029U}\\
G035.19-00.74  & 18:58:13.05  & +01:40:35.70   &  6.6  & 2.2 & 30    & 2$'\times2'$ &\cite{2015MNRAS.452.4029U}\\
G035.20-01.73  & 19:01:45.54  & +01:13:32.50   &  5.9 &  3.3 & 42    & 3$'\times2'$ &\cite{2019ApJ...884....4K}\\
G075.76+00.33  & 20:21:41.09  & +37:25:29.30   &  8.2 & 3.5 & -9    & 2$'\times2'$ &\cite{2015MNRAS.452.4029U}\\
G081.75+00.59  & 20:39:01.99  & +42:24:59.30   &  8.2 & 1.5 & -3    & 2$'\times3'$ &Effelsberg observation\\
G081.87+00.78  & 20:38:36.43  & +42:37:34.80   &  8.2  &  1.3 & 7     & 2$'\times2'$ &\cite{2019ApJ...884....4K}\\
G109.87+02.11  & 22:56:18.10  & +62:01:49.50   &  8.6 & 0.7 &-7    & 3$'\times2'$ &Effelsberg observation\\
G111.54+00.77  & 23:13:45.36  & +61:28:10.60    &  9.6 & 2.6 & -57   & 2$'\times4'$ &\cite{2019ApJ...884....4K}\\
G121.29+00.65  & 00:36:47.35  & +63:29:02.20   &  8.8 & 0.9 & -23   & 2$'\times2'$ &Effelsberg observation\\
\enddata
\end{deluxetable*}

\begin{deluxetable*}{cccccc}[h]
\tablecaption{Parameters of NH$_2$D and NH$_3$}
\label{parameter}
\tablewidth{0pt}
\tablehead{
\colhead{Transitions} & \colhead{$\nu$} & \colhead{$\log_{10}$A$_{ul}$} &
 \colhead{$S\mu^2$} & \colhead{$g_{up}$} & \colhead{$E_u$} \\
\colhead{} & \colhead{(MHz)} & \colhead{(s$^{-1}$)} &
 \colhead{(D$^2$)} & \colhead{} & \colhead{(K)}
}
\startdata
ortho-NH$_2$D $1_{11}^s-1_{01}^a$ & 85926.2780 & -5.1067 & 28.598 & 27 & 20.7 \\
para-NH$_3$(1,1) $F$=2-2 & 23694.4955 & -7.3774 & 1.625 & 6   & 23.4 \\
para-NH$_3$(1,1) $F$=1-1 & 23694.4949 & -6.8999 & 8.133 & 10 & 23.4 \\
para-NH$_3$(2,2) $F$=2-2 & 23722.6321 & -6.8074 & 10.03 & 10 & 64.9 \\
para-NH$_3$(2,2) $F$=3-3 & 23722.6335 & -6.6996 & 18.00 & 14 & 64.9 \\
para-NH$_3$(2,2) $F$=1-1 & 23722.6342 & -6.7745 & 6.490 & 6   & 64.9 \\
\enddata
\tablecomments{The transitions of NH$_2$D and NH$_3$ used for this work. We considered the $F$=0-1, $F$=2-1, $F$=2-2, $F$=1-1, $F$=1-2 and $F$=1-0 transitions of ortho-NH$_2$D $1_{11}^s-1_{01}^a$, and the $F$=1-1 and $F$=2-2 transitions of NH$_3$(1,1) when calculating the column densities. Therefore, the parameters of ortho-NH$_2$D $1_{11}^s-1_{01}^a$ are considered together as six transitions in this table, while the parameters of NH$_3$(1,1) and NH$_3$(2,2) are listed independently. Column (1): chemical formula and transition quantum number. Column (2): rest frequency. Column (3): emission coefficient. Column (4): $S$ is the line strength and $\mu^2$ is the dipole moment. Column (5): Upper state degeneracy. Column (6): Upper state energy level. These parameters are taken from CDMS.}
\end{deluxetable*}

\begin{deluxetable*}{cccccc}[h]
\tablecaption{Summary of $T_{\rm rot}$(2,2;1,1) and $D_{\rm frac}$(NH$_3$)}
\label{summary_source}
\tablewidth{0pt}
\tablehead{
\colhead{Source Name} & \colhead{$T_{\rm rot}$ (Range)} & \colhead{$T_{\rm rot}$ (Median)} &
 \colhead{$D_{\rm frac}$(NH$_3$)$\times$100 (Range)} & \colhead{$D_{\rm frac}$(NH$_3$)$\times$100 (Median)} & \colhead{$T_{\rm rot}$ vs $D_{\rm frac}$(NH$_3$)} \\
\colhead{} & \colhead{(K)} & \colhead{(K)} &
 \colhead{} & \colhead{} & \colhead{}
}
\startdata
G015.03-00.67 & 19.5 - 24.8 & 23.7 & 0.51 - 1.2 & 0.68 & no \\ 
G023.44-00.18 & 14.8 - 21.6 & 20.4 & 0.9 - 3.1 & 1.3 & no \\ 
G031.28+00.06 & 16.4 - 22.1 & 17.1 & 0.062 - 0.13 & 0.12 & anticorrelation \\ 
G034.39+00.22 & 18.1 - 20.2 & 20.0 & 1.7 - 3.5 & 2.2 & anticorrelation \\ 
G035.19-00.74 & 14.5 - 22.1 & 22.1 & 1.6 - 4.0 & 1.7 & anticorrelation \\ 
G035.20-01.73 & 18.3 - 22.4 & 19.7 & 2.4 - 4.1 & 3.4 & no \\ 
G075.76+00.33 & 21.1 - 22.9 & 22.1 & 1.8 - 2.3 & 1.9 & no \\ 
G081.75+00.59 & 18.0 - 23.9 & 18.6 & 2.2 - 6.8 & 3.4 & no \\ 
G081.87+00.78 & 20.2 - 25.6 & 22.1 & 3.6 - 8.6 & 6.8 & anticorrelation \\ 
G109.87+02.11 & 21.2 - 24.6 & 21.2 & 1.1 - 5.5 & 4.0 & anticorrelation \\ 
G111.54+00.77 & 20.0 - 26.5 & 22.2 & 1.4 - 6.7 & 3.5 & anticorrelation \\ 
G121.29+00.65 & 13.6 - 19.0 & 18.4 & 2.0 - 4.1 & 2.1 & anticorrelation \\ 
\enddata
\tablecomments{ Column (6): whether $T_{\rm rot}$(2,2;1,1) and $D_{\rm frac}$(NH$_3$) present anticorrelation.}
\end{deluxetable*}


\begin{table*}[ht]
\centering
\caption{Initial elemental abundances used in our chemical model \label{tab:initial_abundances}}
\begin{tabular}{ll}
\hline
\hline
{Element} & {Abundance Relative to H} \\
\hline
H$_2$ &   $0.50$ \\
He    &   $9.00\times 10^{-2}$ \\
N     &   $6.20\times 10^{-5}$ \\
O     &   $2.40\times 10^{-4}$ \\
C$^+$ &   $1.70\times 10^{-4}$ \\
S$^+$ &   $8.00\times 10^{-8}$ \\
Si$^+$&  {$8.00\times 10^{-9}$} \\
Fe$^+$&   $3.00\times 10^{-9}$ \\
Na$^+$&   $2.00\times 10^{-9}$ \\
Mg$^+$&   $7.00\times 10^{-9}$ \\
P$^+$ &   $2.00\times 10^{-10}$ \\
Cl$^+$&   $1.00\times 10^{-9}$ \\
F     &   $6.68\times 10^{-9}$ \\
HD    &   $1.60\times 10^{-5}$ \\
\hline
\end{tabular}
\end{table*}

\begin{table*}[ht]
\centering
\caption{Physical parameters used in our chemical model \label{tab:para1}}
\begin{tabular}{lll}
\hline
\hline
{Parameter} & {Value} \\
\hline
$T$ (K) & 6, 60 (in steps of 2) \\
n$_{\rm {H}}$ (cm$^{-3}$) & $10^4$, $10^5$, $10^6$, $10^7$ \\
$\zeta$ (s$^{-1}$) & $3\times10^{-18}$, $1.5\times10^{-17}$, $3\times10^{-17}$, $6\times10^{-17}$, $3\times10^{-16}$ \\
\hline
\end{tabular}
\end{table*}

\begin{table*}[ht]
\centering
\caption{Major Production and Destruction Reactions of NH$_3$ and NH$_2$D at {$n_{\mathrm{H}}$} = $10^{6}$ cm$^{-3}$ and $\zeta$ = 3$\times$10$^{-17}$ s$^{-1}$}
\label{tab:major_reactions}
\begin{tabular}{cccc}
\hline
\hline
{$T$ (K)} & {NH$_3$} & {NH$_2$D} \\
\hline
$T < 24$ & 
\begin{tabular}[c]{@{}c@{}}NH$_4^+$ + e$^-$ $\rightarrow$ H + NH$_3$ \\[24pt]
NH$_3$ + H$^+$ $\rightarrow$ H + NH$_3^+$ \\[6pt]
NH$_3$ + H$_3^+$ $\rightarrow$ H$_2$ + NH$_4^+$ \\[6pt]
\end{tabular}
& 
\begin{tabular}[c]{@{}c@{}}NH$_3$D$^+$ + e$^-$ $\rightarrow$ H + NH$_2$D \\[6pt]
NH$_2$D$_2^+$ + e$^-$ $\rightarrow$ D + NH$_2$D \\[18pt]
NH$_2$D + H$^+$ $\rightarrow$ H + NH$_2$D$^+$ \\[6pt]
NH$_2$D + H$_3^+$ $\rightarrow$ H$_2$ + NH$_3$D$^+$ \\[6pt]
NH$_2$D + H$_3^+$ $\rightarrow$ HD + NH$_4^+$ \\ [6pt]
\end{tabular} \\
\hline
$T > 24$ & 
\begin{tabular}[c]{@{}c@{}}
NH$_4^+$ + e$^-$ $\rightarrow$ H + NH$_3$ \\[18pt]
NH$_3$ $\rightarrow$ ice-NH$_{3}$ \\[6pt]
NH$_3$ + HCO$^+$ $\rightarrow$ CO + NH$_4^+$ \\[6pt]
NH$_3$ + H$_3$O$^+$ $\rightarrow$ H$_2$O + NH$_4^+$ \\[6pt]
NH$_3$ + N$_2$H$^+$ $\rightarrow$ N$_2$ + NH$_4^+$ \\[6pt]
NH$_3$ + S$^+$ $\rightarrow$ S + NH$_3^+$ \\[6pt]
\end{tabular}
& 
\begin{tabular}[c]{@{}c@{}}NH$_3$D$^+$ + e$^-$ $\rightarrow$ H + NH$_2$D \\[18pt]
NH$_2$D $\rightarrow$ ice-NH$_2$D \\[6pt]
NH$_2$D + HCO$^+$ $\rightarrow$ CO + NH$_3$D$^+$ \\[6pt]
NH$_2$D + N$_2$H$^+$ $\rightarrow$ N$_2$ + NH$_3$D$^+$ \\[6pt]
NH$_2$D + H$_3$O$^+$ $\rightarrow$ H$_2$O + NH$_3$D$^+$ \\[6pt]
\end{tabular} \\
\hline
\end{tabular}
\end{table*}



\onecolumngrid 

\clearpage
\appendix

\section{NH$_3$ spectra fitting result}
From each source, we have selected typical NH$_3$(1,1) and NH$_3$(2,2) spectra to show. The spectra of each source NH$_3$(1,1) and NH$_3$(2,2) are from the same position. The results of the spectral line fitting for NH$_3$ and the corresponding physical parameters obtained from the fitting are illustrated in Figure \ref{NH3_spectra}, using the code by \citep{2015ApJ...805..171L}.

\setcounter{figure}{0}
\renewcommand{\thefigure}{A\arabic{figure}}
\begin{figure}[h]
\centering
\includegraphics[width=0.3\textwidth]{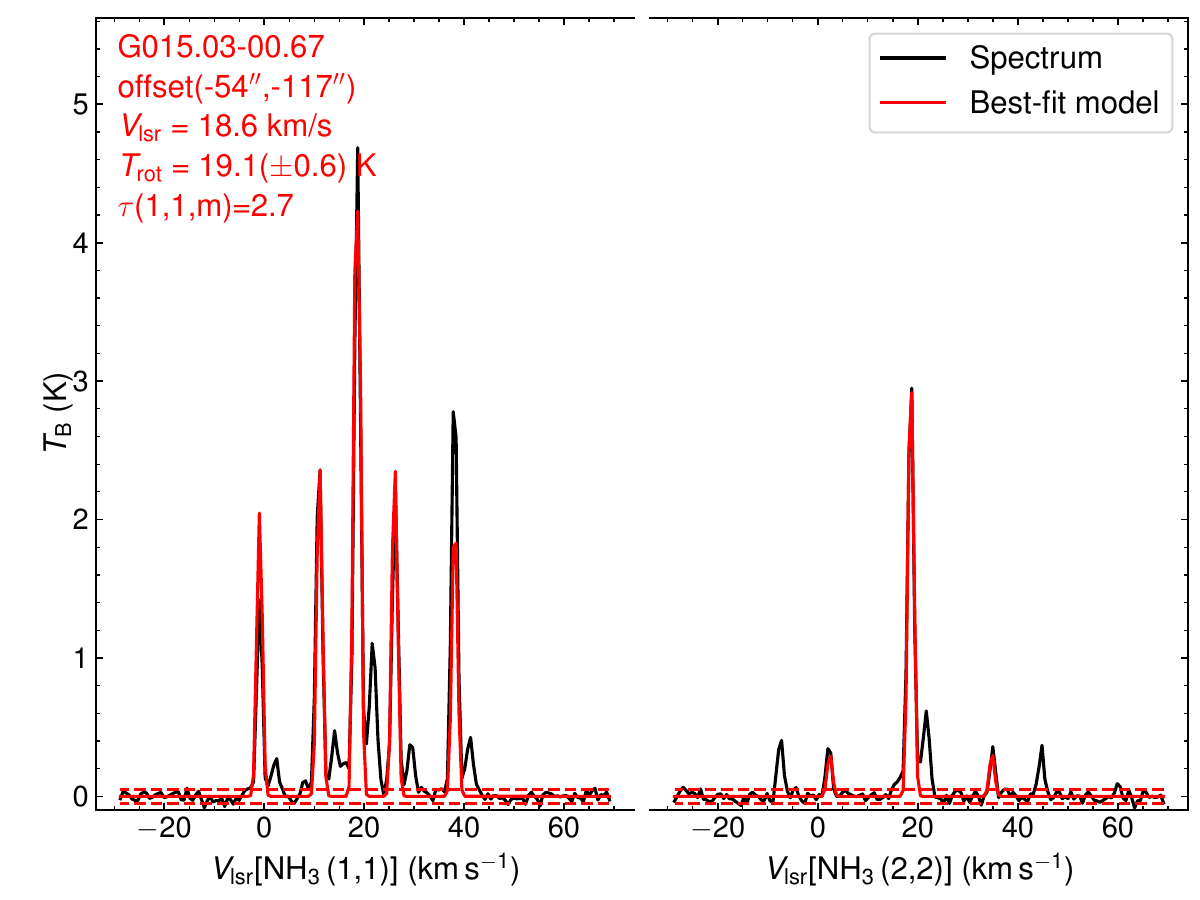}\includegraphics[width=0.3\textwidth]{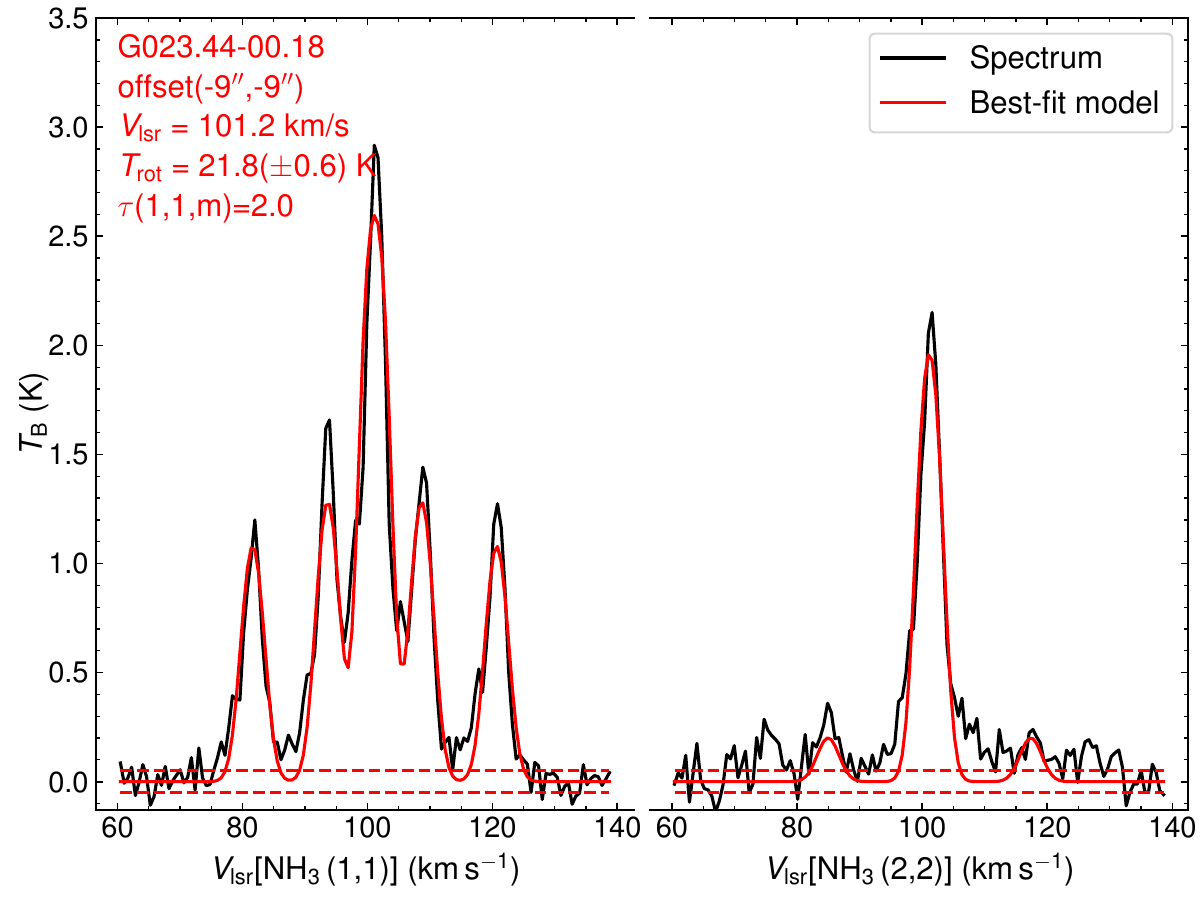}\includegraphics[width=0.3\textwidth]{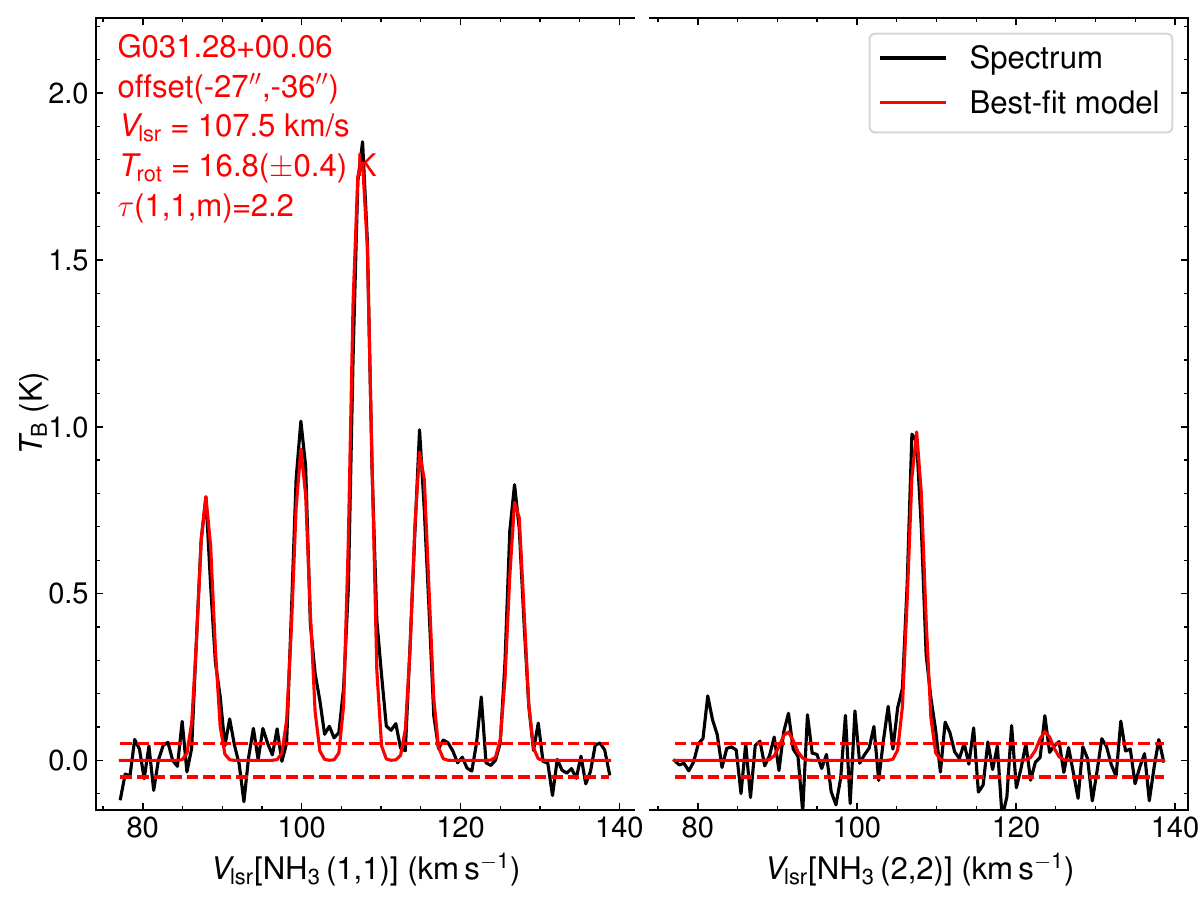}\\
\includegraphics[width=0.3\textwidth]{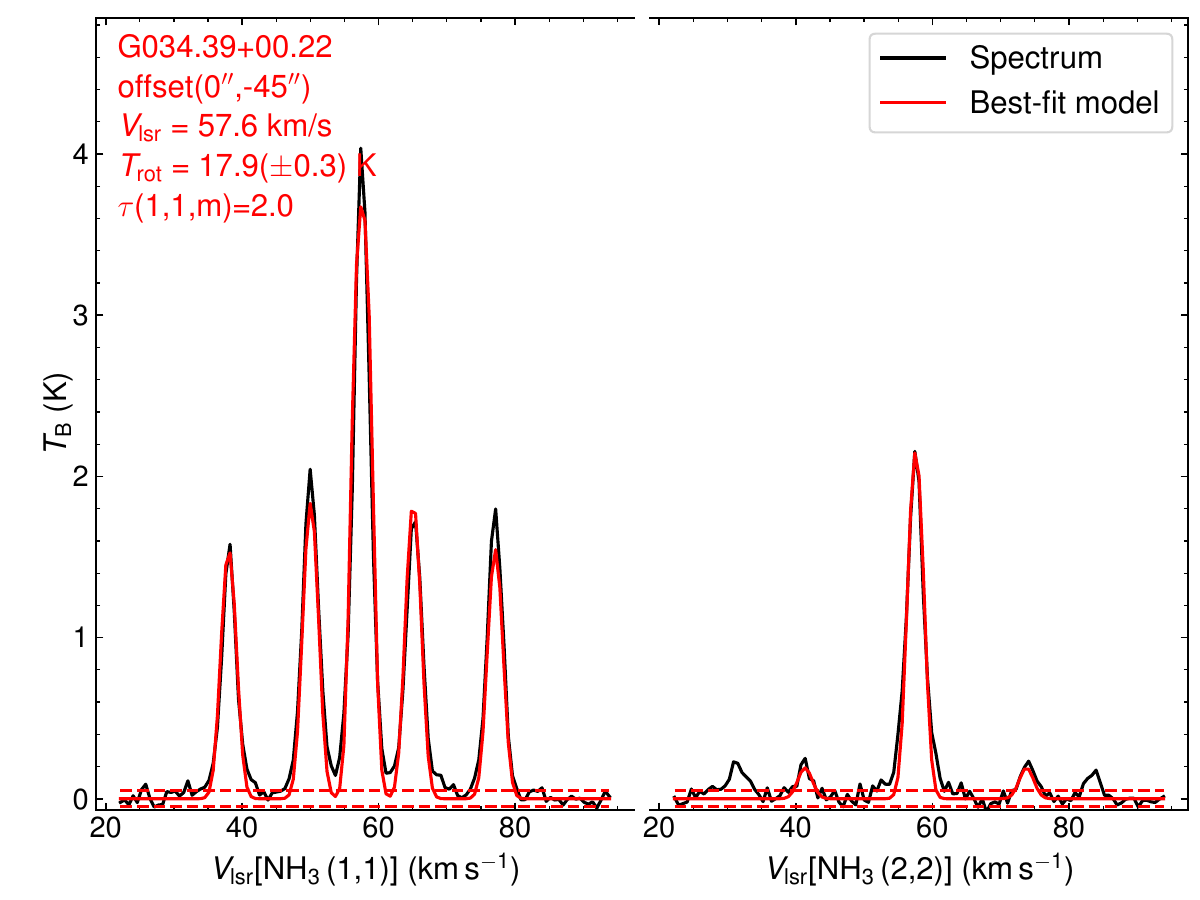}\includegraphics[width=0.3\textwidth]{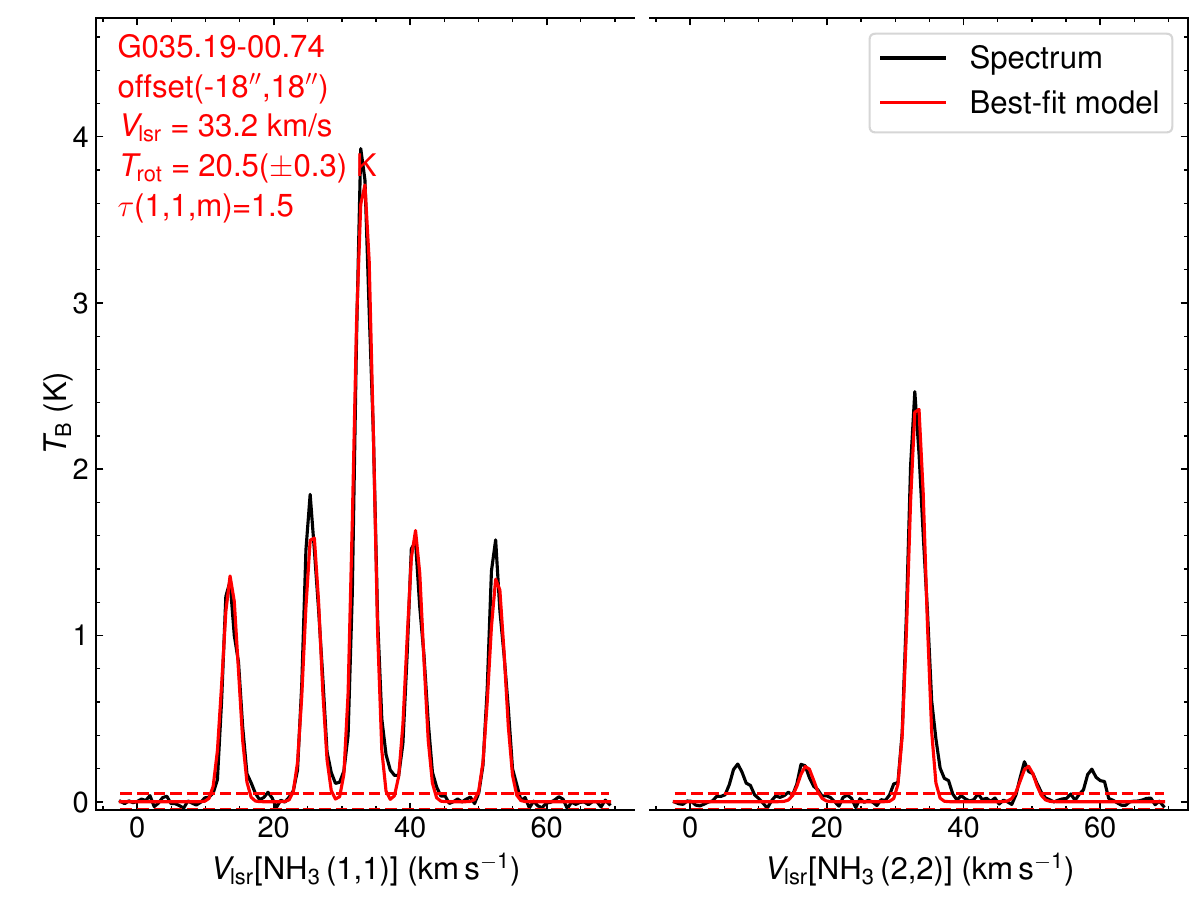}\includegraphics[width=0.3\textwidth]{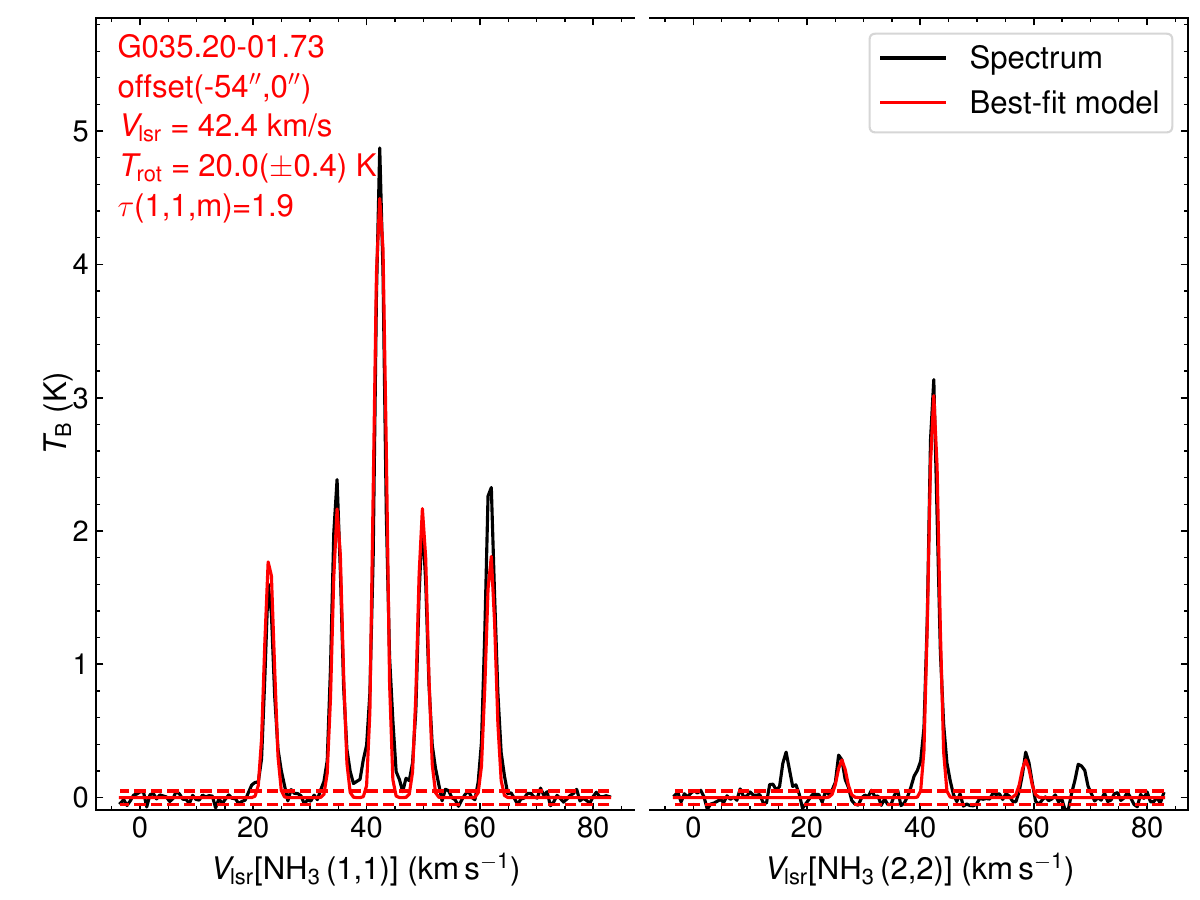}\\
\includegraphics[width=0.3\textwidth]{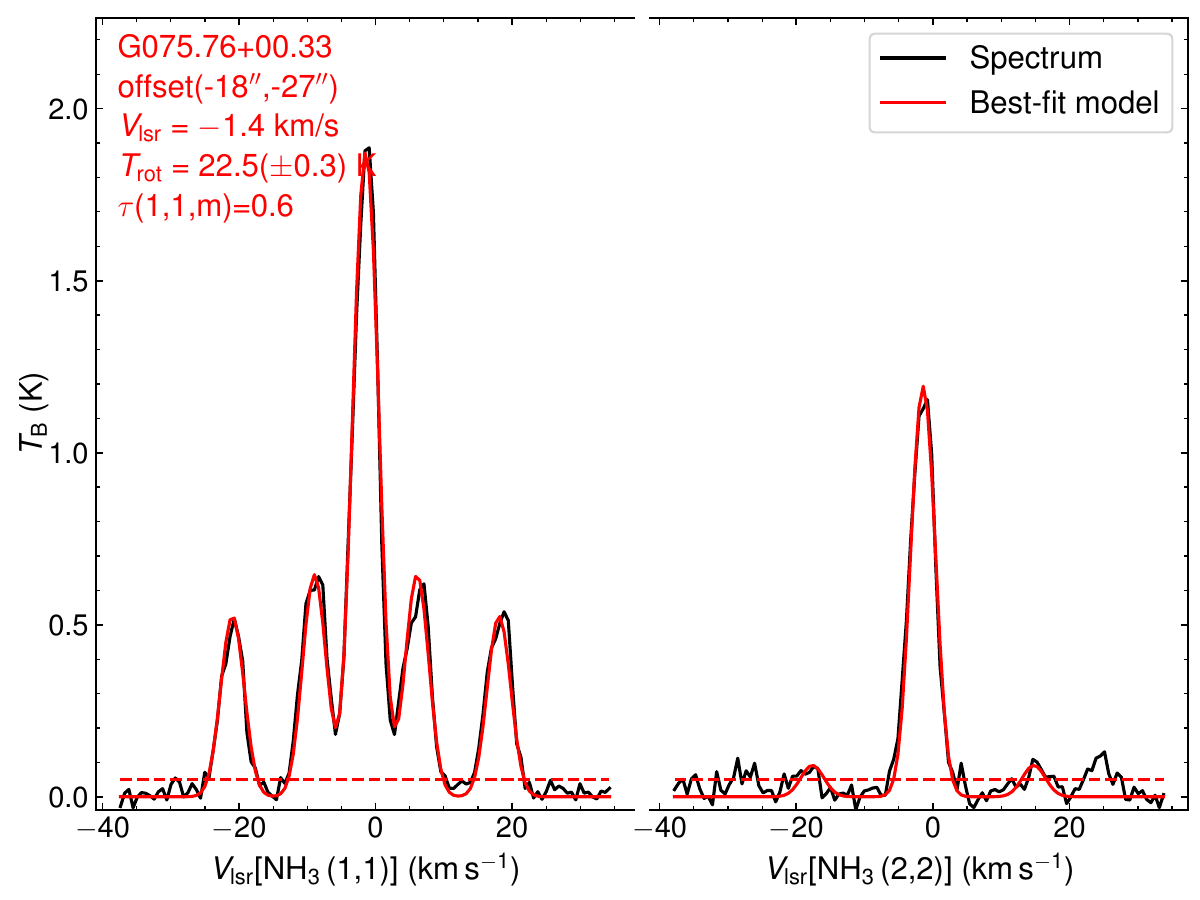}\includegraphics[width=0.3\textwidth]{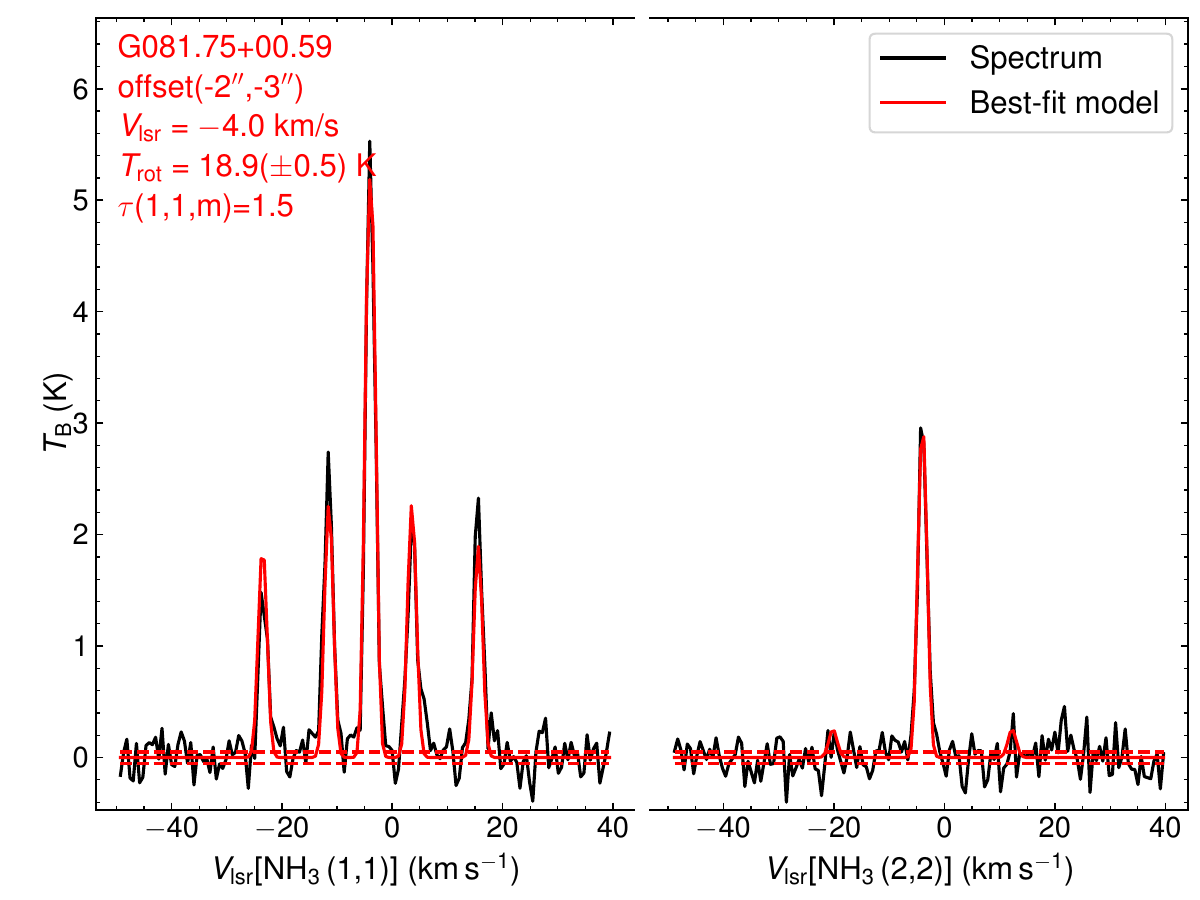}\includegraphics[width=0.3\textwidth]{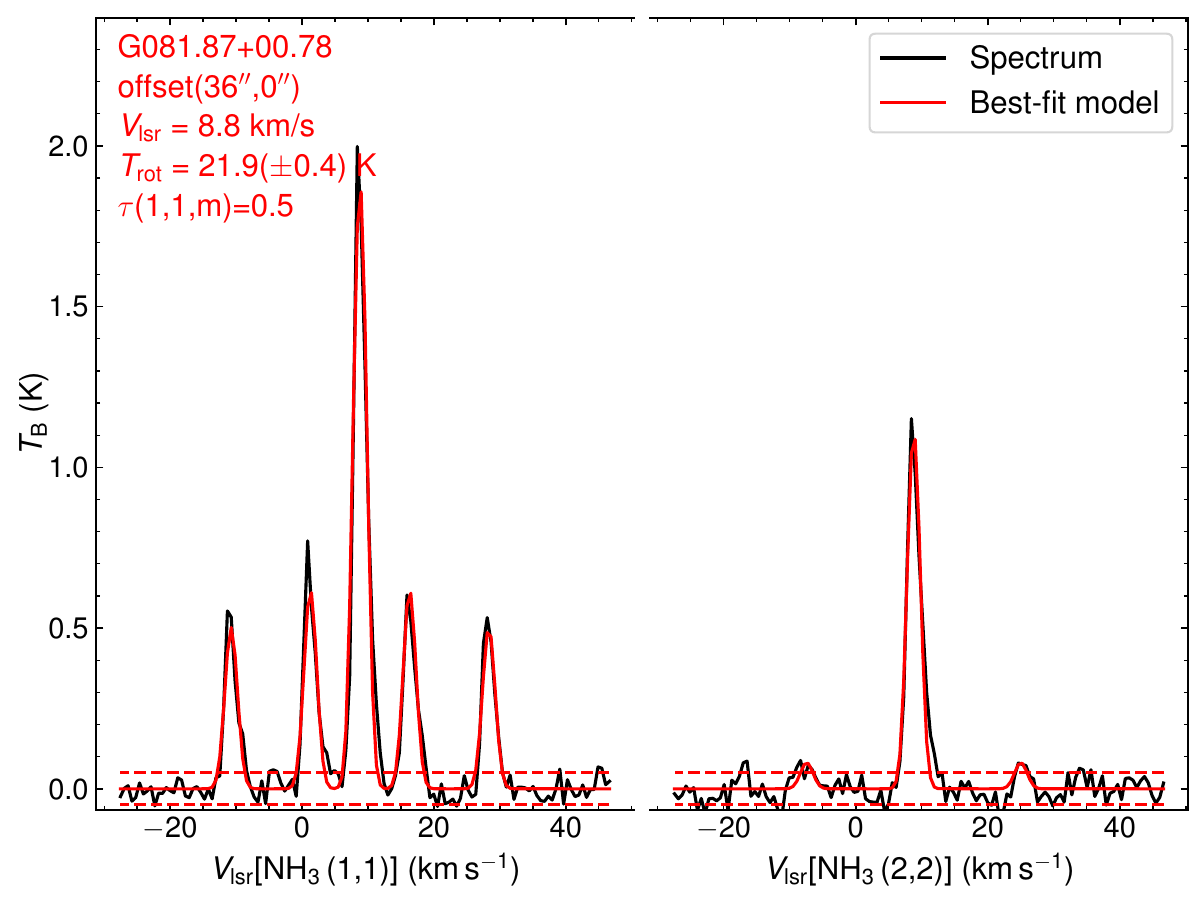}\\
\includegraphics[width=0.3\textwidth]{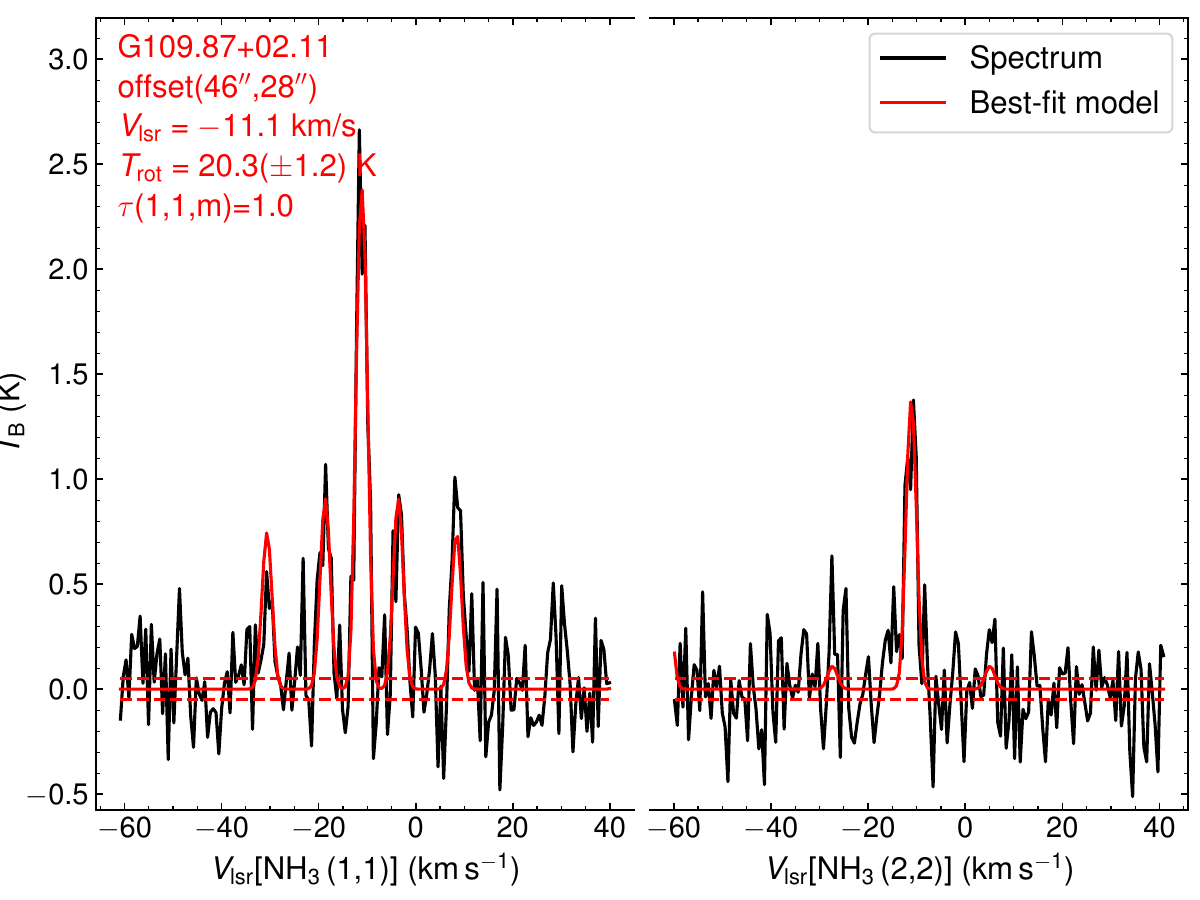}\includegraphics[width=0.3\textwidth]{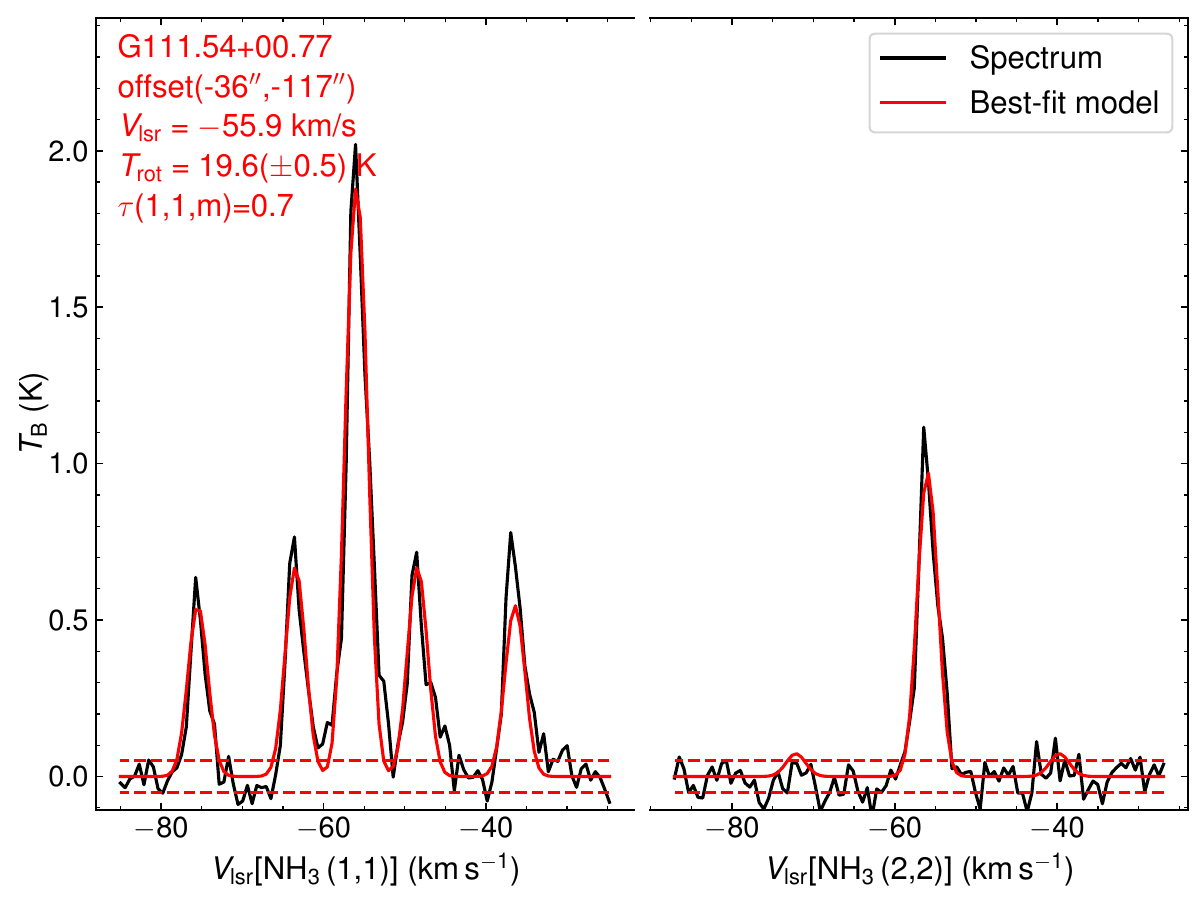}\includegraphics[width=0.3\textwidth]{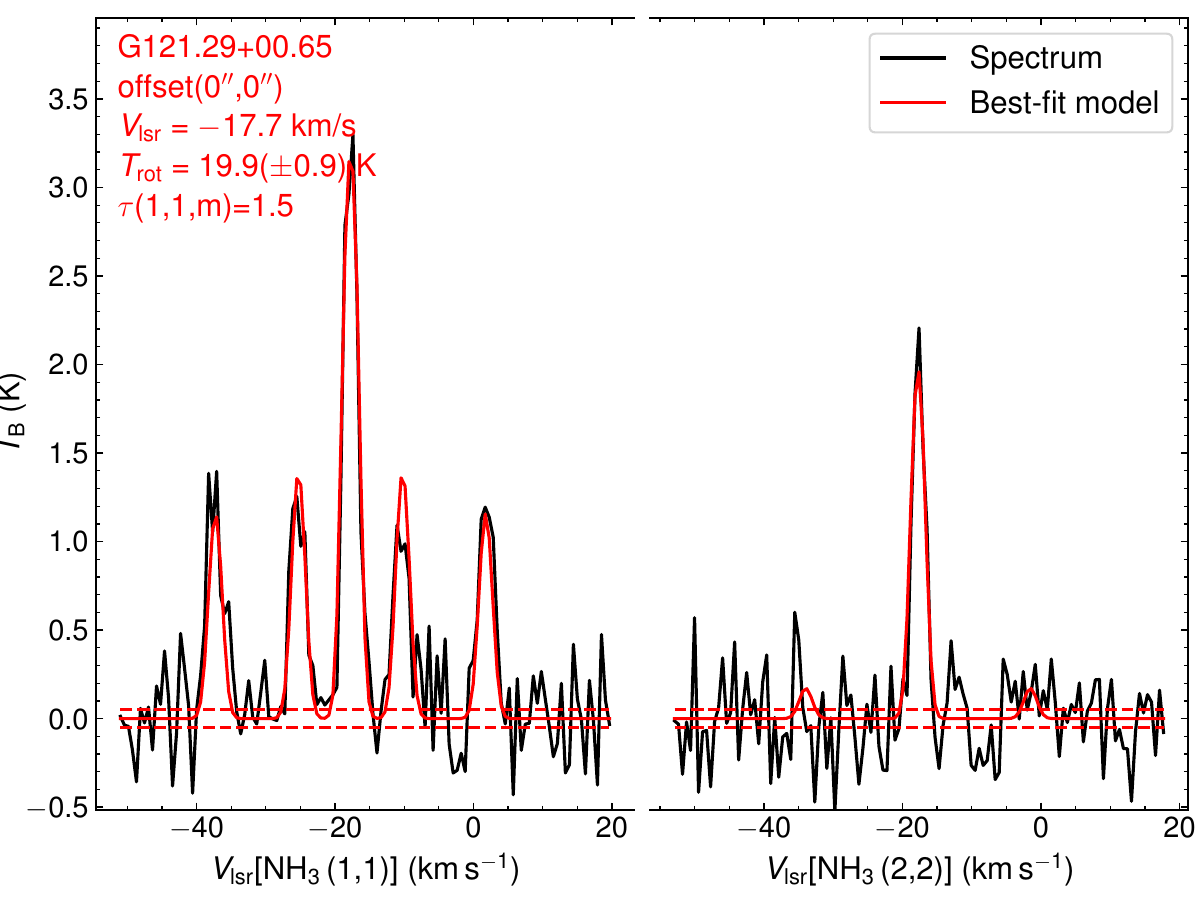}
\caption{The observed spectra of NH$_3$(1,1) and NH$_3$(2,2) are in black, while the best fit from the code of \cite{2015ApJ...805..171L} is in red.}
\label{NH3_spectra}
\end{figure}

\newpage

\section{The distribution of NH$_3$(1,1), NH$_3$(2,2) and ortho-NH$_2$D $1_{11}^s-1_{01}^a$}
\setcounter{figure}{0}
\renewcommand{\thefigure}{B\arabic{figure}}

\begin{figure}[h]
\gridline{\fig{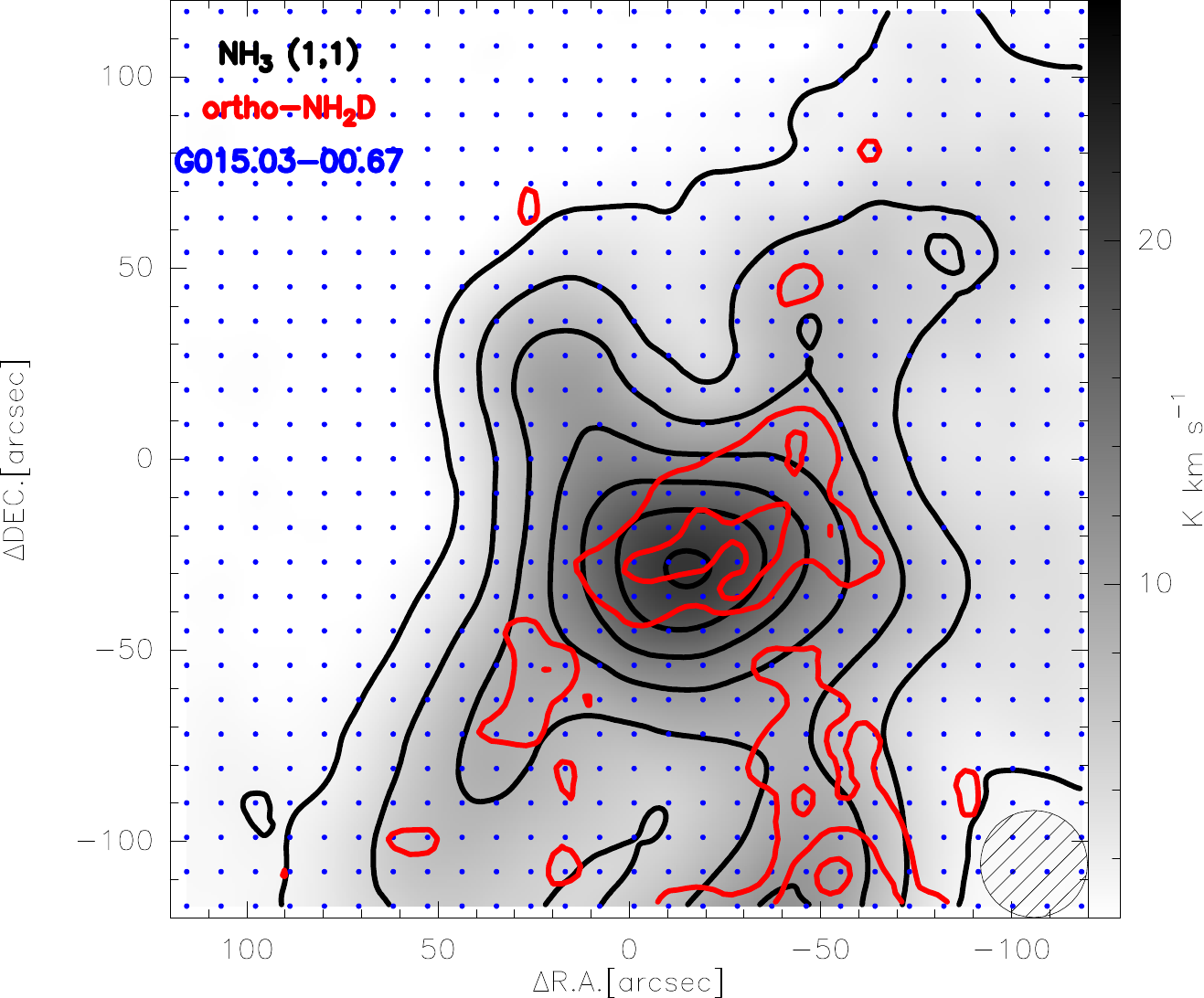}{0.4\textwidth}{(a)}
          \fig{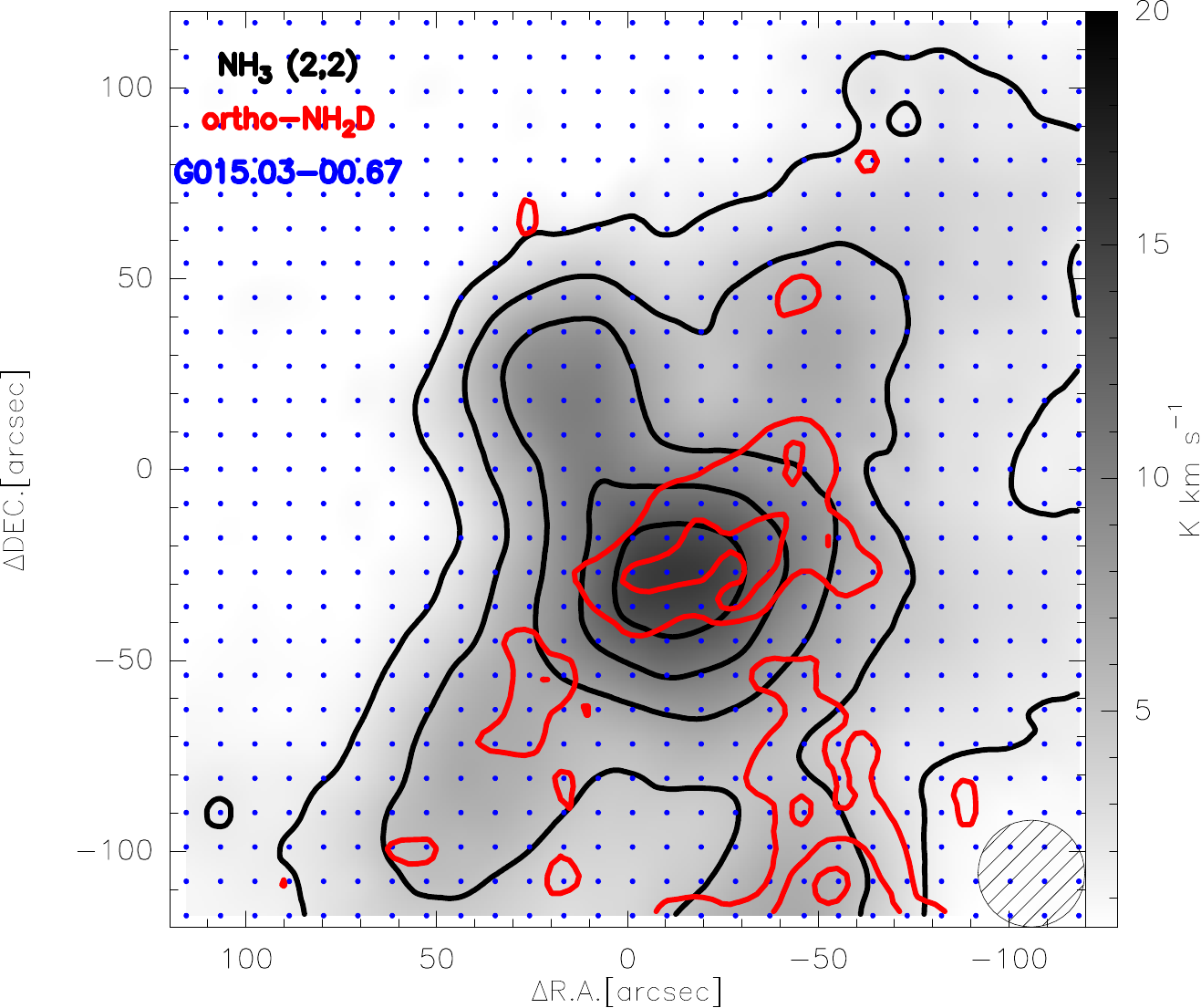}{0.4\textwidth}{(b)}}
\caption{(a) The velocity-integrated intensity of the ortho-NH$_2$D $1_{11}^s-1_{01}^a$ contour (red contour) overlaid on the NH$_3$(1,1) main group velocity-integrated intensity image (gray scale and black contour) in G015.03-00.67. The contour levels start at 3$\sigma$ in steps of 2$\sigma$ for NH$_2$D $1_{11}^s-1_{01}^a$, while the contour levels start at 15$\sigma$ in steps of 36$\sigma$ for the NH$_3$(1,1) main group. The gray scale starts at 3$\sigma$. (b) The velocity-integrated intensity of the ortho-NH$_2$D $1_{11}^s-1_{01}^a$ contour (red contour) overlaid on the NH$_3$(2,2) main group velocity-integrated intensity image (Gray scale and black contour) in G015.03-00.67. The contour levels start at 3$\sigma$ in steps of 2$\sigma$ for the ortho-NH$_2$D $1_{11}^s-1_{01}^a$, while the contour levels start at 15$\sigma$ in steps of 24$\sigma$ for NH$_3$(2,2) main group. The gray scale starts at 3$\sigma$.}
\label{G015_NH3}
\end{figure}

\begin{figure}[h]
\gridline{\fig{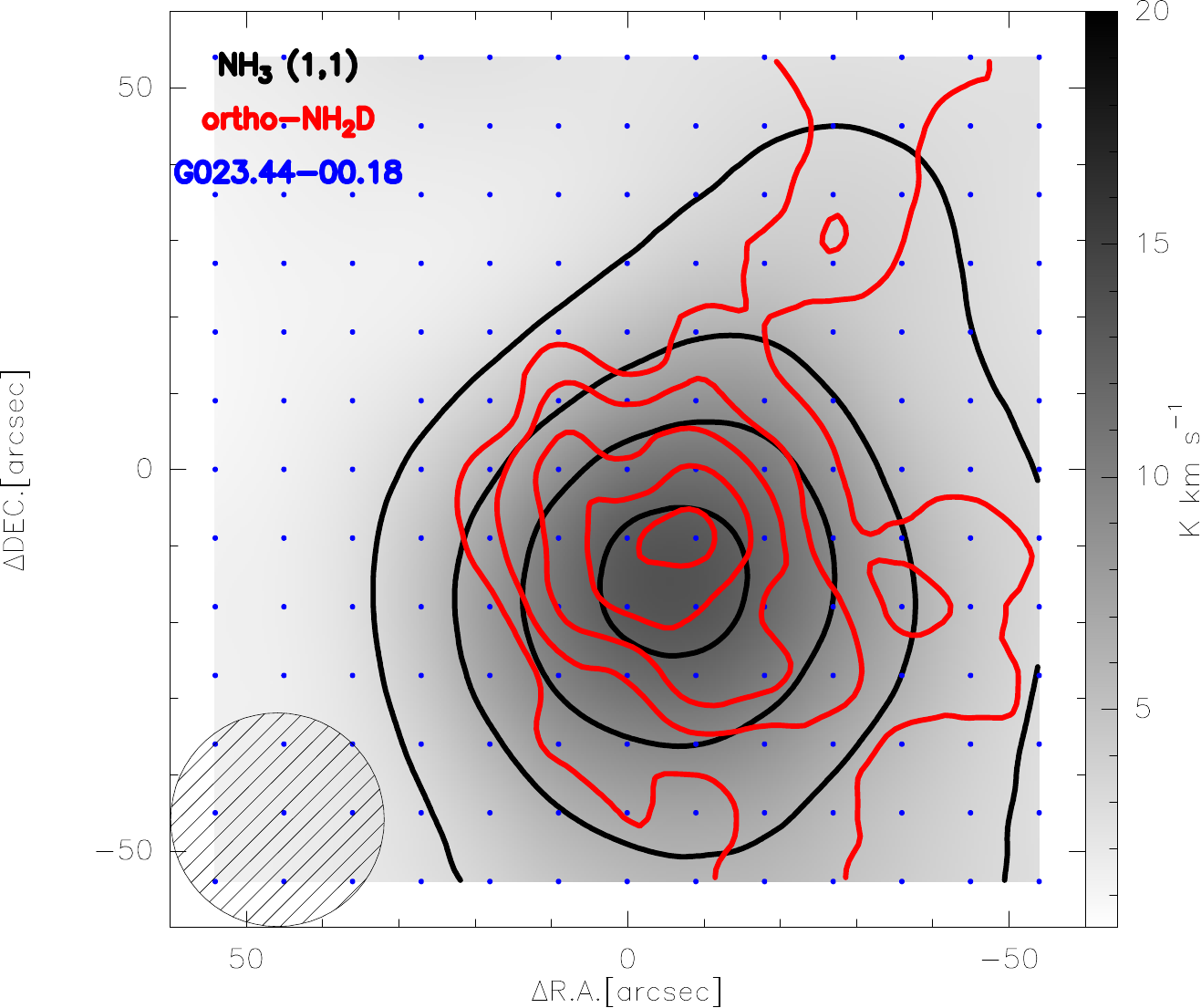}{0.5\textwidth}{(a)}
          \fig{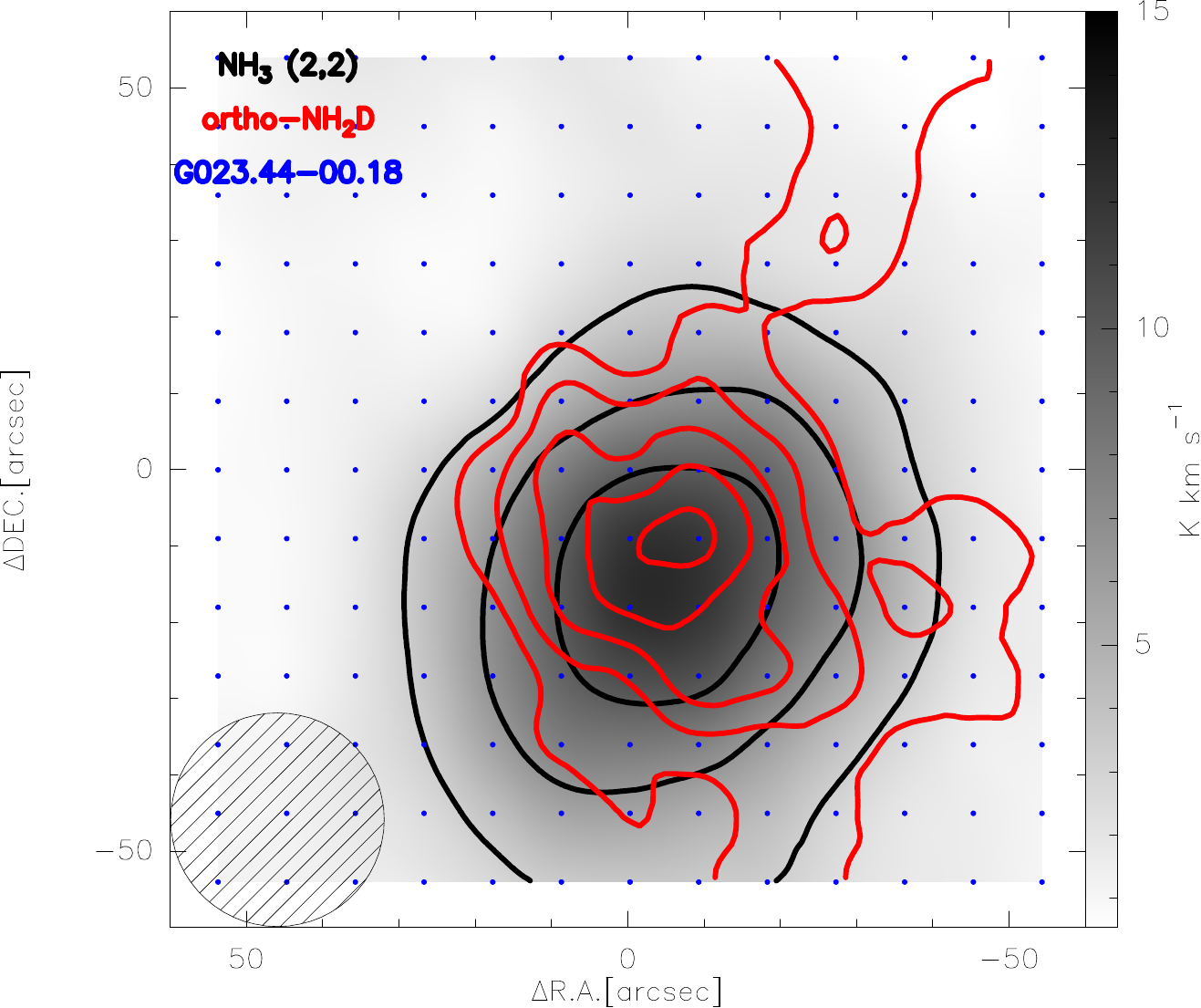}{0.5\textwidth}{(b)}}
\caption{Same as Figure \ref{G015_NH3}, but for source G023.44-00.18. (a) The red contour levels start at 5$\sigma$ in steps of 3$\sigma$, the black contour levels start at 30$\sigma$ in steps of 30$\sigma$, and the gray scale starts at 3$\sigma$. (b) The red contour levels start at 5$\sigma$ in steps of 3$\sigma$, the black contour levels start at 18$\sigma$ in steps of 18$\sigma$, and the gray scale starts at 3$\sigma$.
}
\label{G023_NH3}
\end{figure}

\begin{figure}[h]
\gridline{\fig{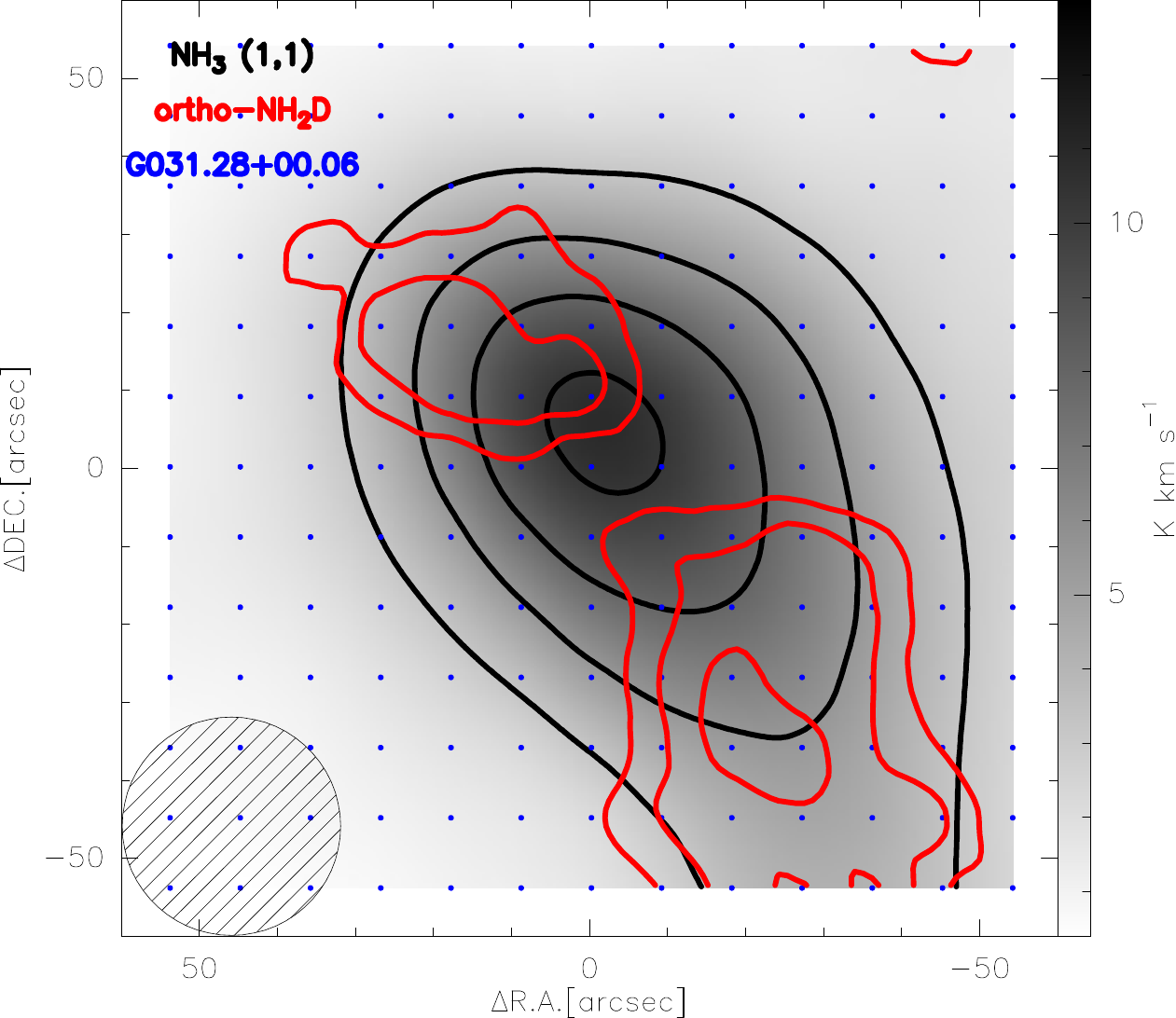}{0.5\textwidth}{(a)}
          \fig{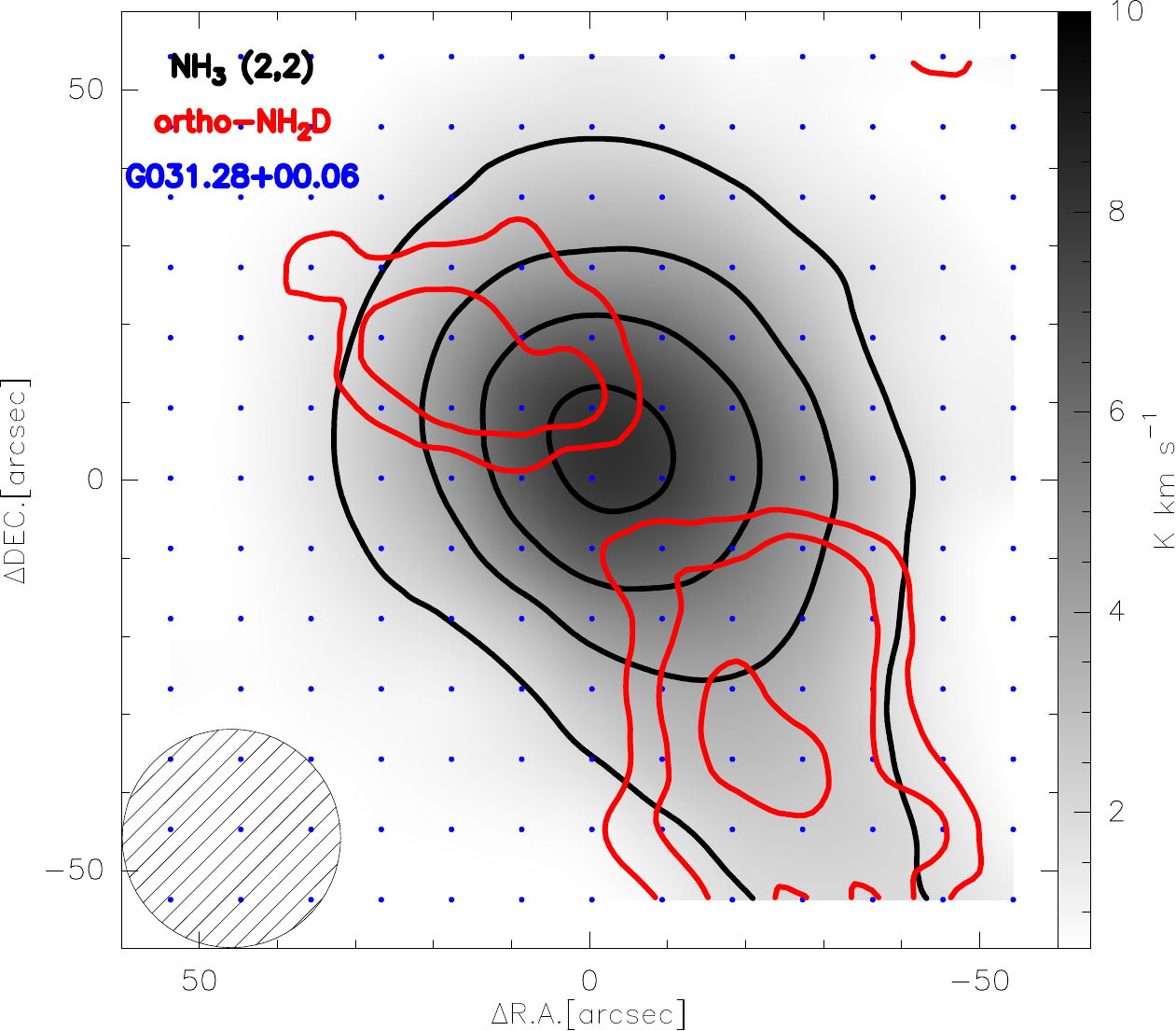}{0.5\textwidth}{(b)}}
\caption{Same as Figure \ref{G015_NH3}, but for source G031.28+00.06. (a) The red contour levels start at 3$\sigma$ in steps of 2$\sigma$, the black contour levels start at 24$\sigma$ in steps of 18$\sigma$, and the gray scale starts at 3$\sigma$. (b) The red contour levels start at 3$\sigma$ in steps of 2$\sigma$, the black contour levels start at 9$\sigma$ in steps of 9$\sigma$, and the gray scale starts at 3$\sigma$.
}
\label{G031_NH3}
\end{figure}

\begin{figure}[h]
\gridline{\fig{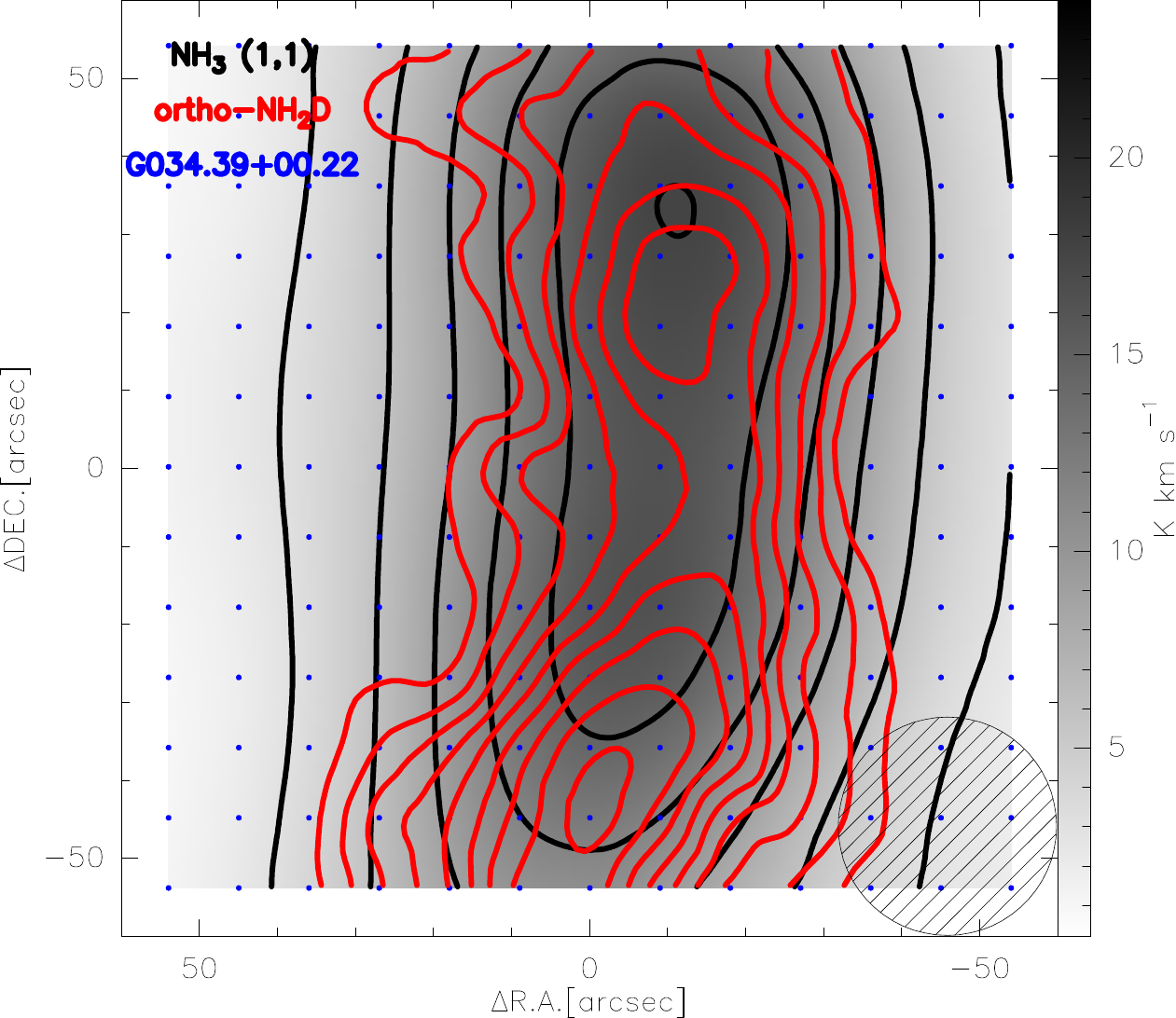}{0.5\textwidth}{(a)}
          \fig{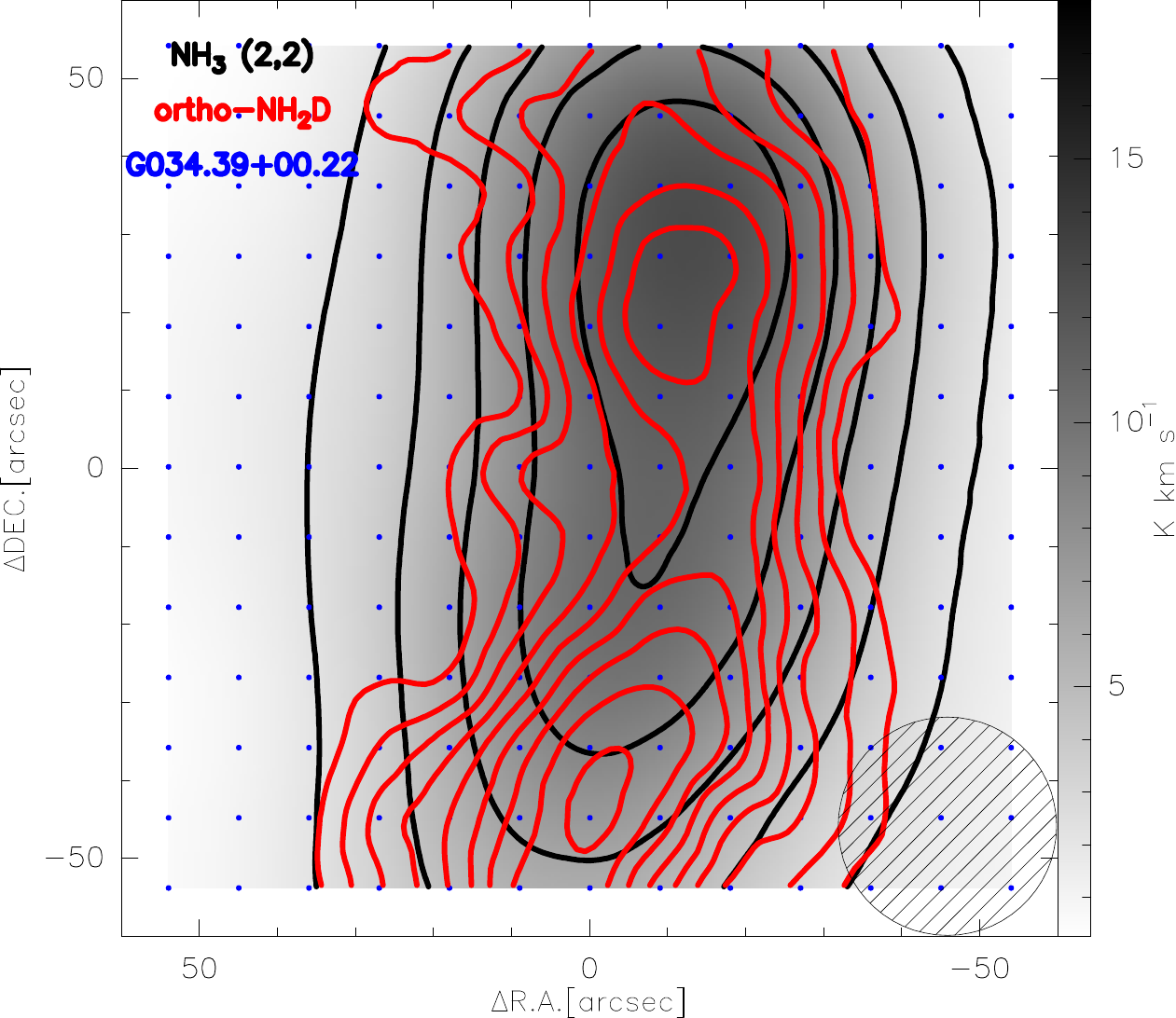}{0.5\textwidth}{(b)}}
\caption{Same as Figure \ref{G015_NH3}, but for source G034.39+00.22. (a) The red contour levels start at 5$\sigma$ in steps of 2$\sigma$, the black contour levels start at 42$\sigma$ in steps of 42$\sigma$, and the gray scale starts at 3$\sigma$. (b) The red contour levels start at 5$\sigma$ in steps of 2$\sigma$, the black contour levels start at 24$\sigma$ in steps of 24$\sigma$, and the gray scale starts at 3$\sigma$.
}
\label{G034_NH3}
\end{figure}

\begin{figure}[h]
\gridline{\fig{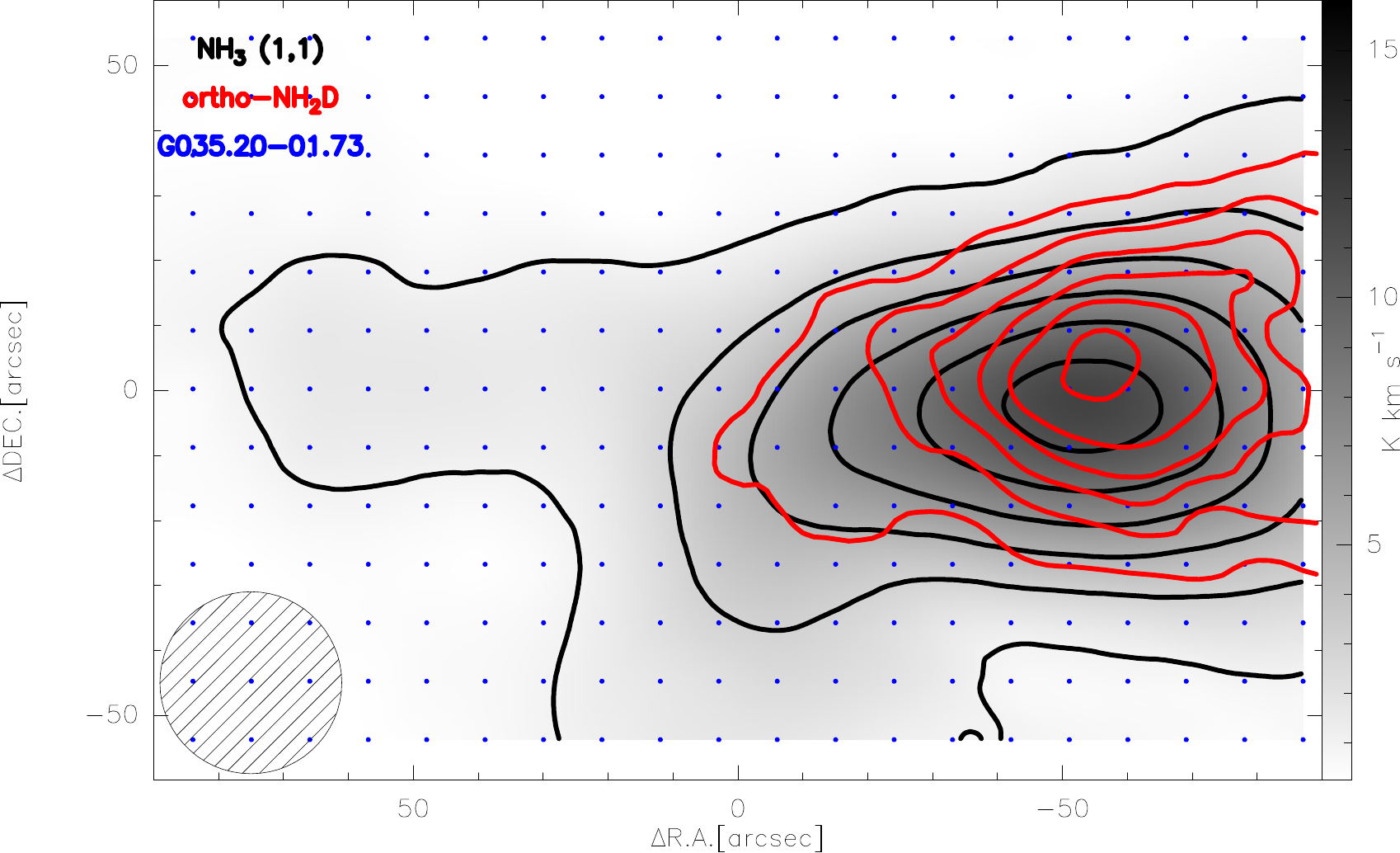}{0.5\textwidth}{(a)}
          \fig{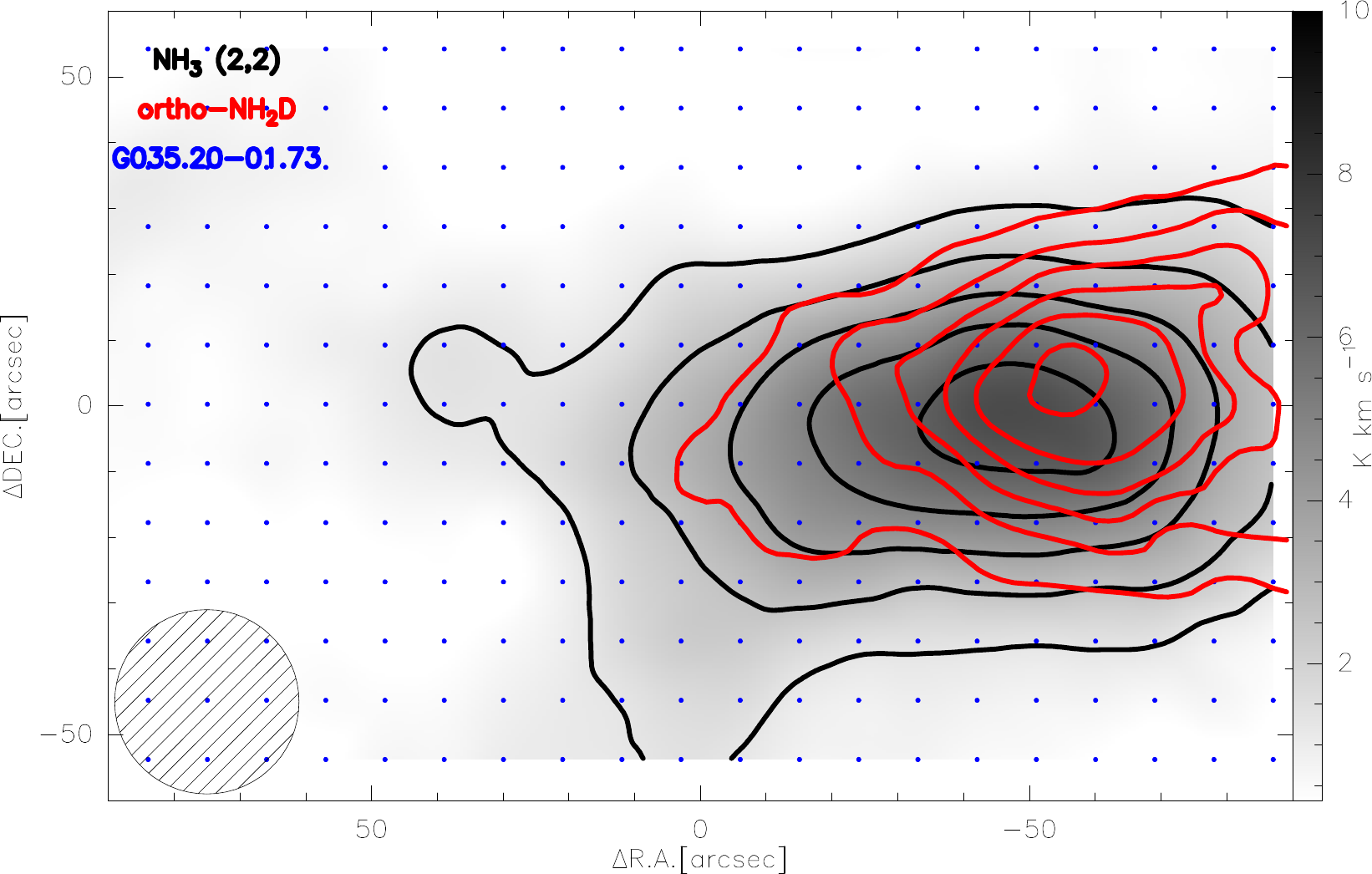}{0.48\textwidth}{(b)}}
\caption{Same as Figure \ref{G015_NH3}, but for source G035.20-01.73. (a) The red contour levels start at 5$\sigma$ in steps of 2$\sigma$, the black contour levels start at 24$\sigma$ in steps of 33$\sigma$, and the gray scale starts at 3$\sigma$. (b) The red contour levels start at 5$\sigma$ in steps of 2$\sigma$, the black contour levels start at 18$\sigma$ in steps of 24$\sigma$, and the gray scale starts at 3$\sigma$.
}
\label{G03520_NH3}
\end{figure}

\begin{figure}[h]
\gridline{\fig{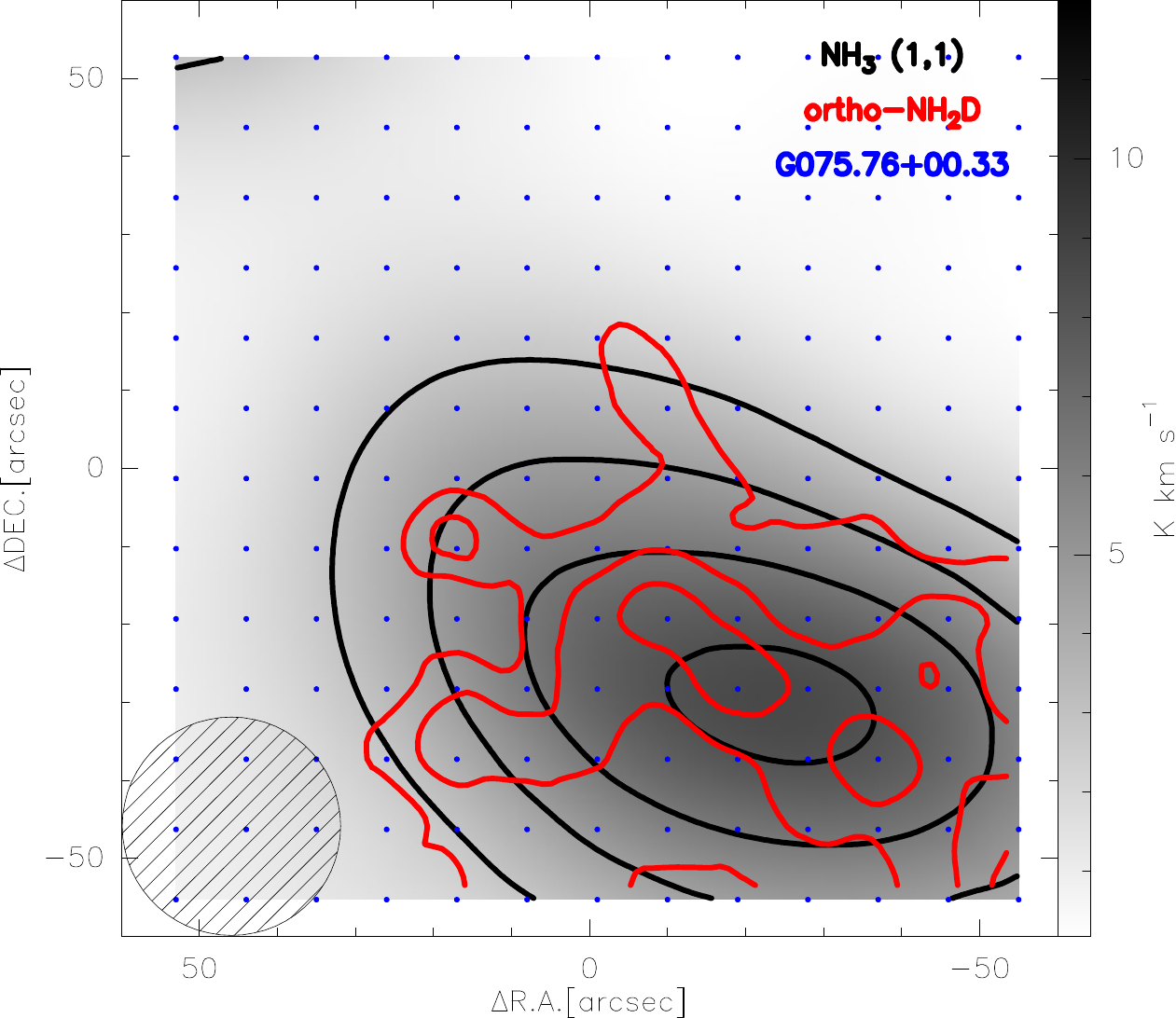}{0.5\textwidth}{(a)}
          \fig{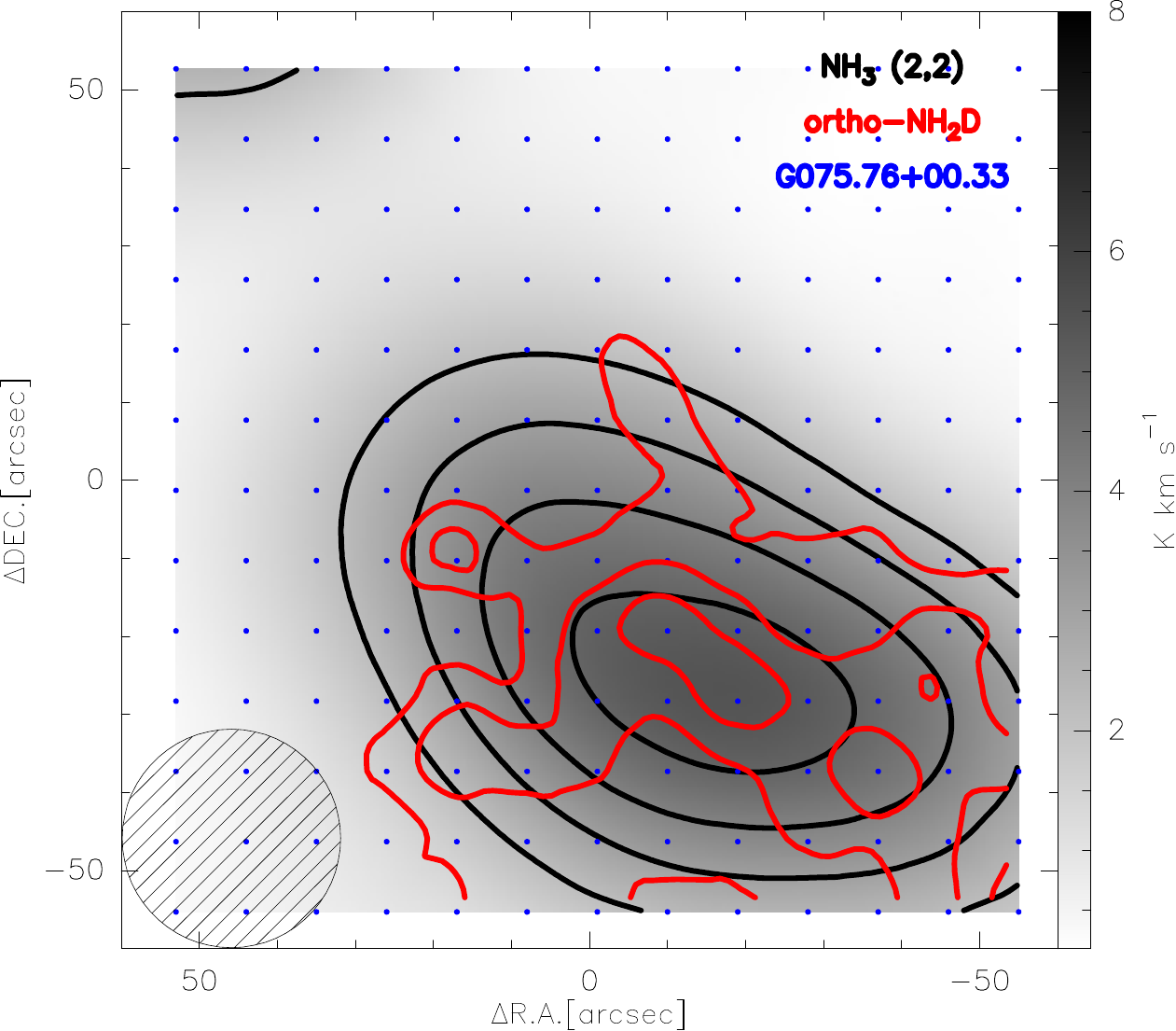}{0.5\textwidth}{(b)}}
\caption{Same as Figure \ref{G015_NH3}, but for source G075.76+00.33. (a) The red contour levels start at 3$\sigma$ in steps of 2$\sigma$, the black contour levels start at 48$\sigma$ in steps of 30$\sigma$, and the gray scale starts at 3$\sigma$. (b) The red contour levels start at 3$\sigma$ in steps of 2$\sigma$, the black contour levels start at 36$\sigma$ in steps of 15$\sigma$, and the gray scale starts at 3$\sigma$.
}
\label{G075_NH3}
\end{figure}

\begin{figure}[h]
\gridline{\fig{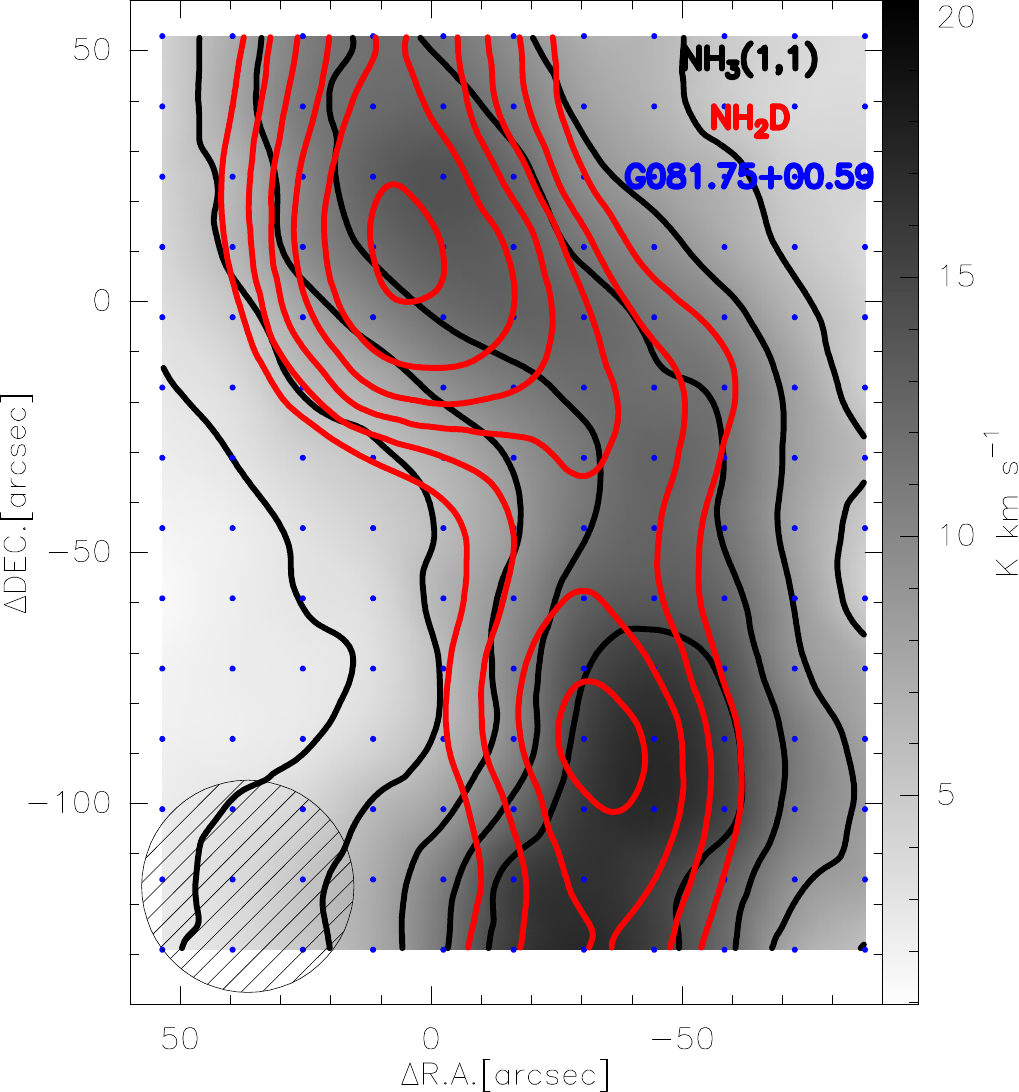}{0.5\textwidth}{(a)}
          \fig{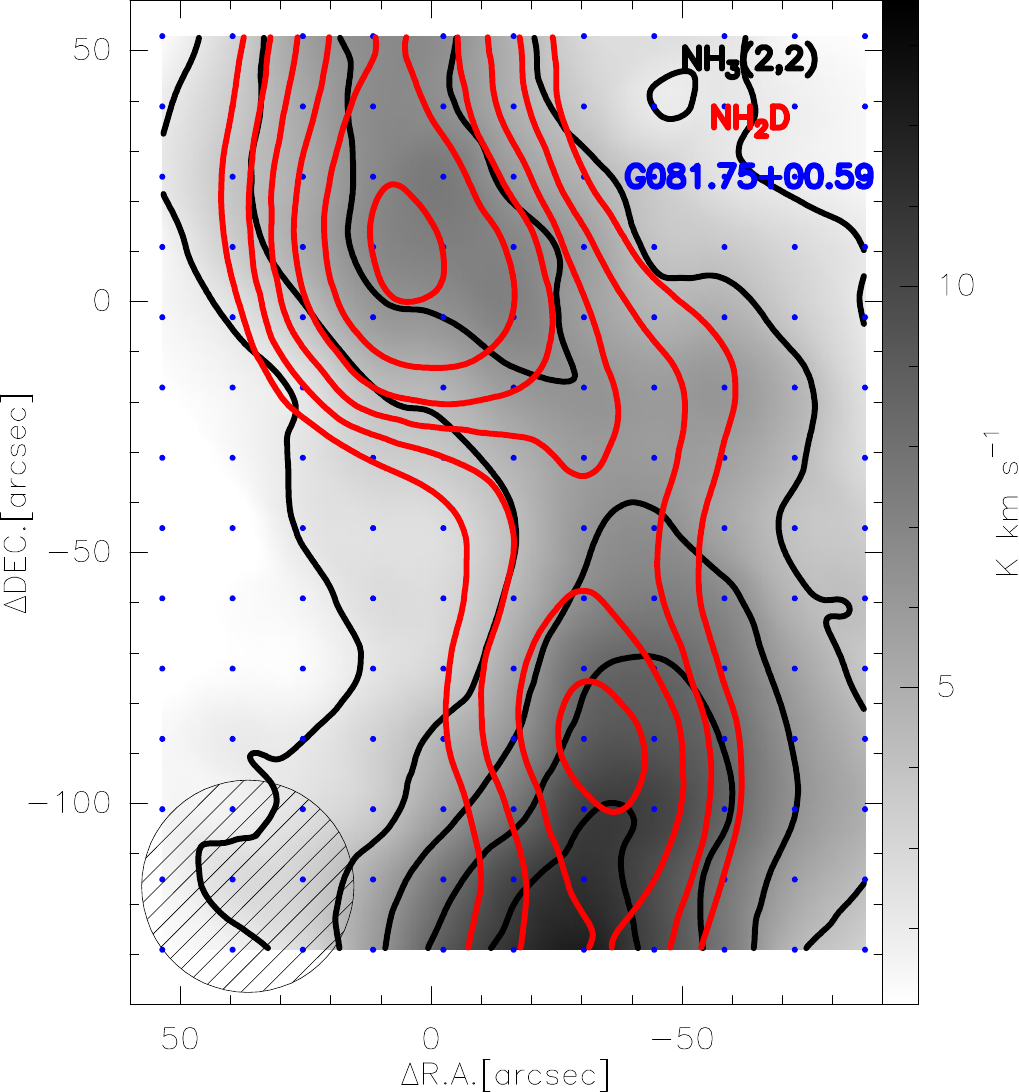}{0.5\textwidth}{(b)}}
\caption{Same as Figure \ref{G015_NH3}, but for source G081.75+00.59. (a) The red contour levels start at 36$\sigma$ in steps of 10$\sigma$, the black contour levels start at 9$\sigma$ in steps of 9$\sigma$, and the gray scale starts at 3$\sigma$. (b) The red contour levels start at 36$\sigma$ in steps of 10$\sigma$, the black contour levels start at 6$\sigma$ in steps of 9$\sigma$, and the gray scale starts at 3$\sigma$.
}
\label{G08175_NH3}
\end{figure}

\begin{figure}[h]
\gridline{\fig{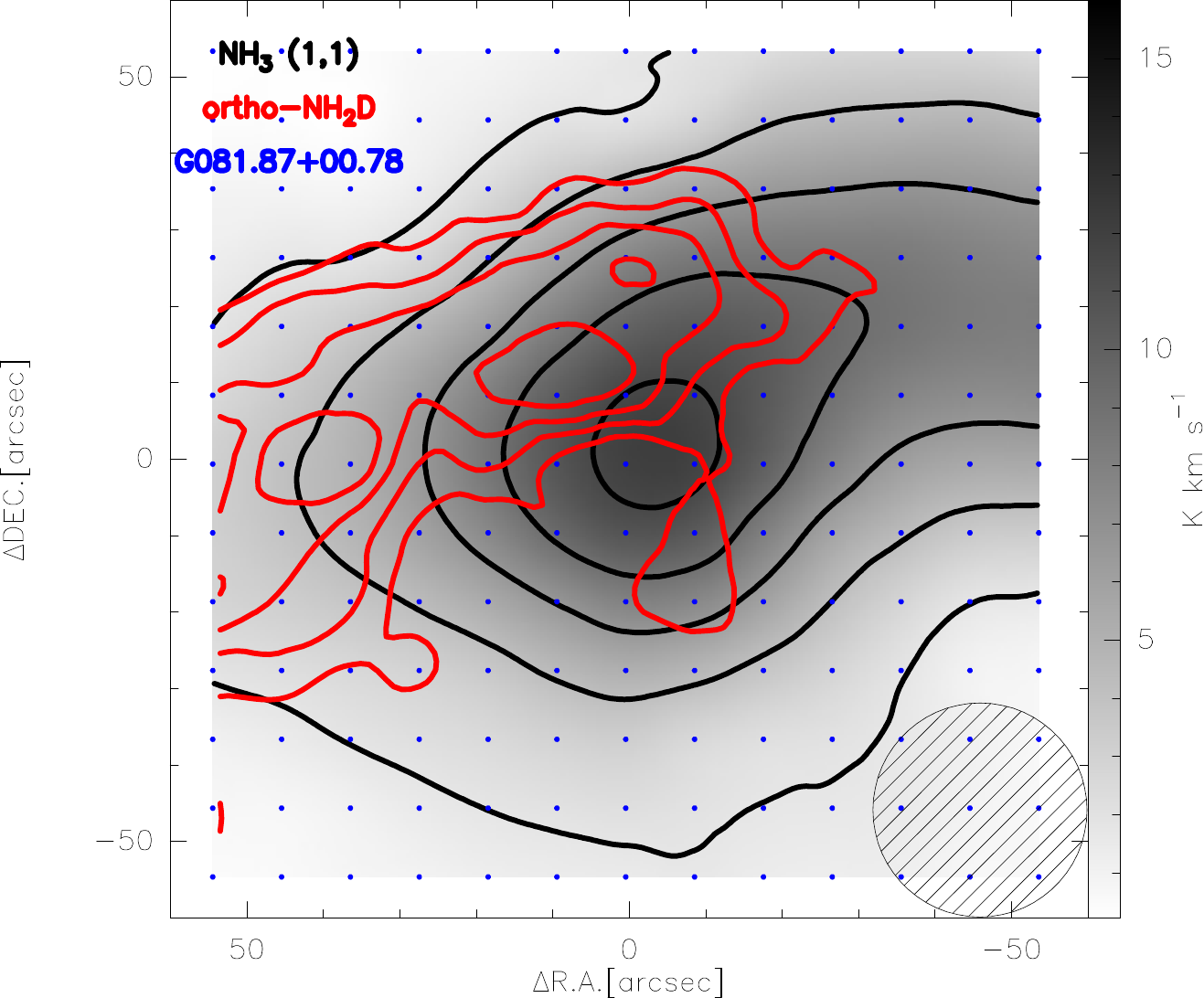}{0.5\textwidth}{(a)}
          \fig{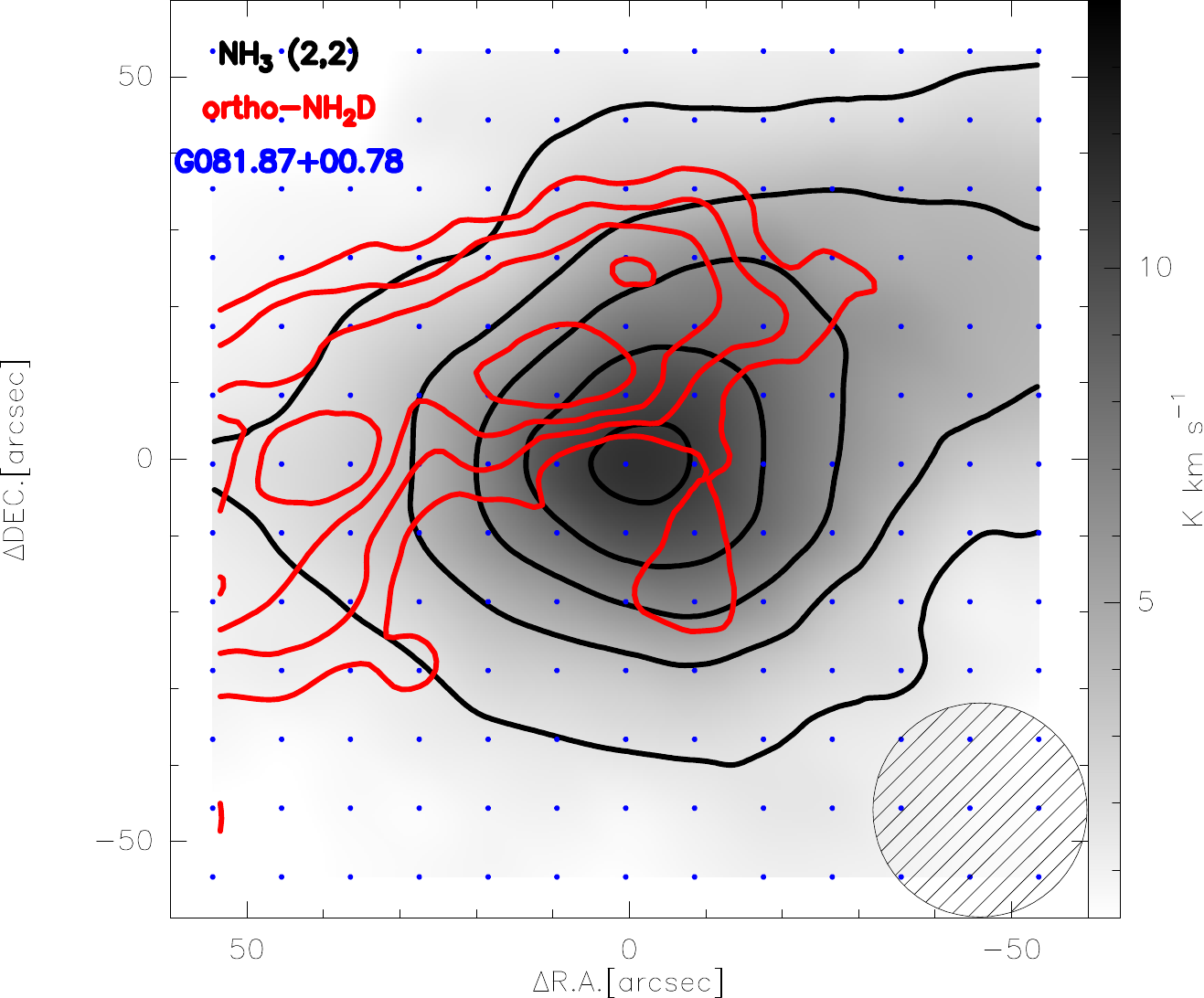}{0.5\textwidth}{(b)}}
\caption{Same as Figure \ref{G015_NH3}, but for source G081.87+00.78. (a) The red contour levels start at 3$\sigma$ in steps of 2$\sigma$, the black contour levels start at 15$\sigma$ in steps of 36$\sigma$, and the gray scale starts at 3$\sigma$. (b) The red contour levels start at 3$\sigma$ in steps of 2$\sigma$, the black contour levels start at 15$\sigma$ in steps of 24$\sigma$, and the gray scale starts at 3$\sigma$.
}
\label{G08187_NH3}
\end{figure}

\begin{figure}[h]
\gridline{\fig{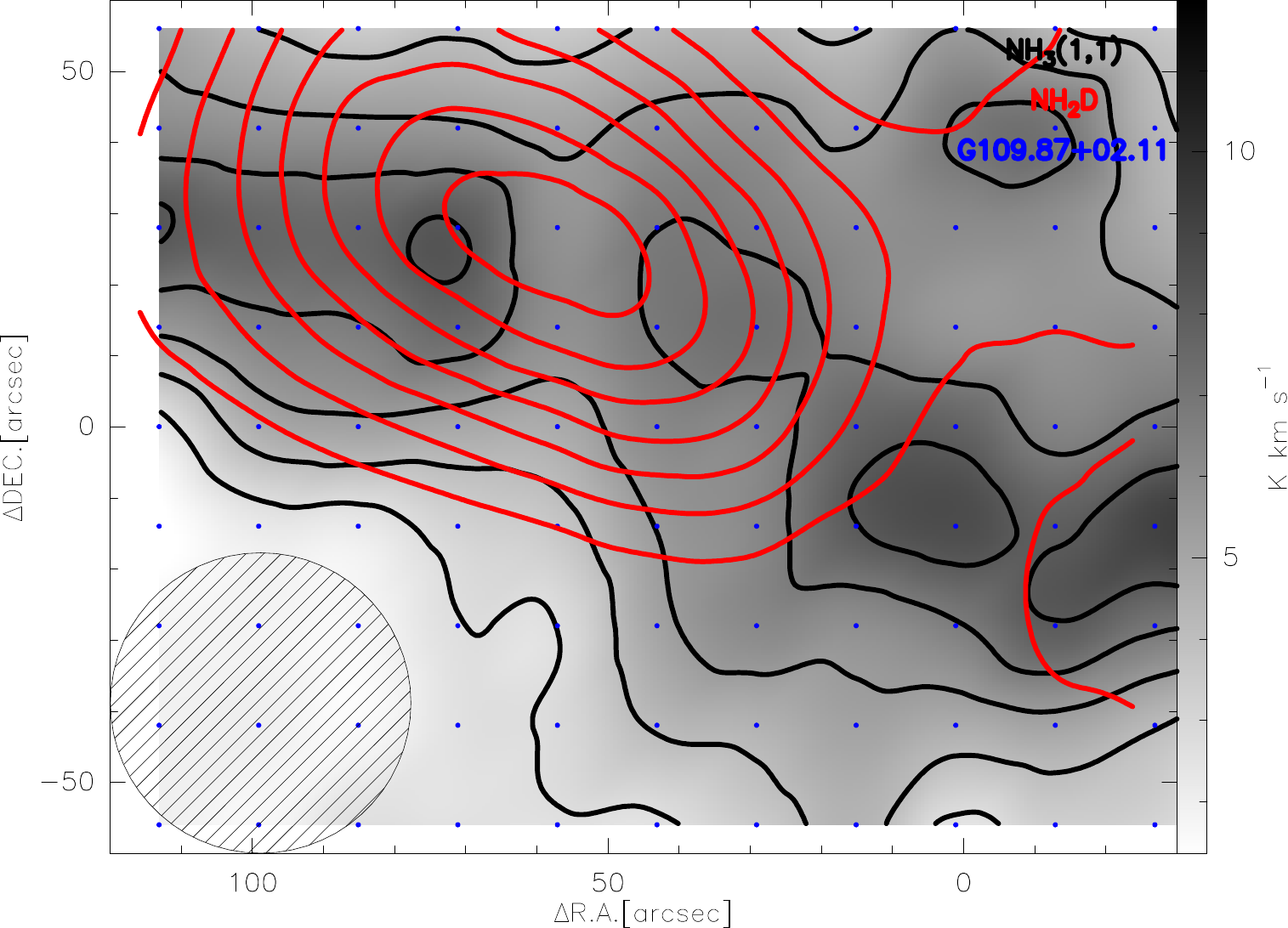}{0.5\textwidth}{(a)}
          \fig{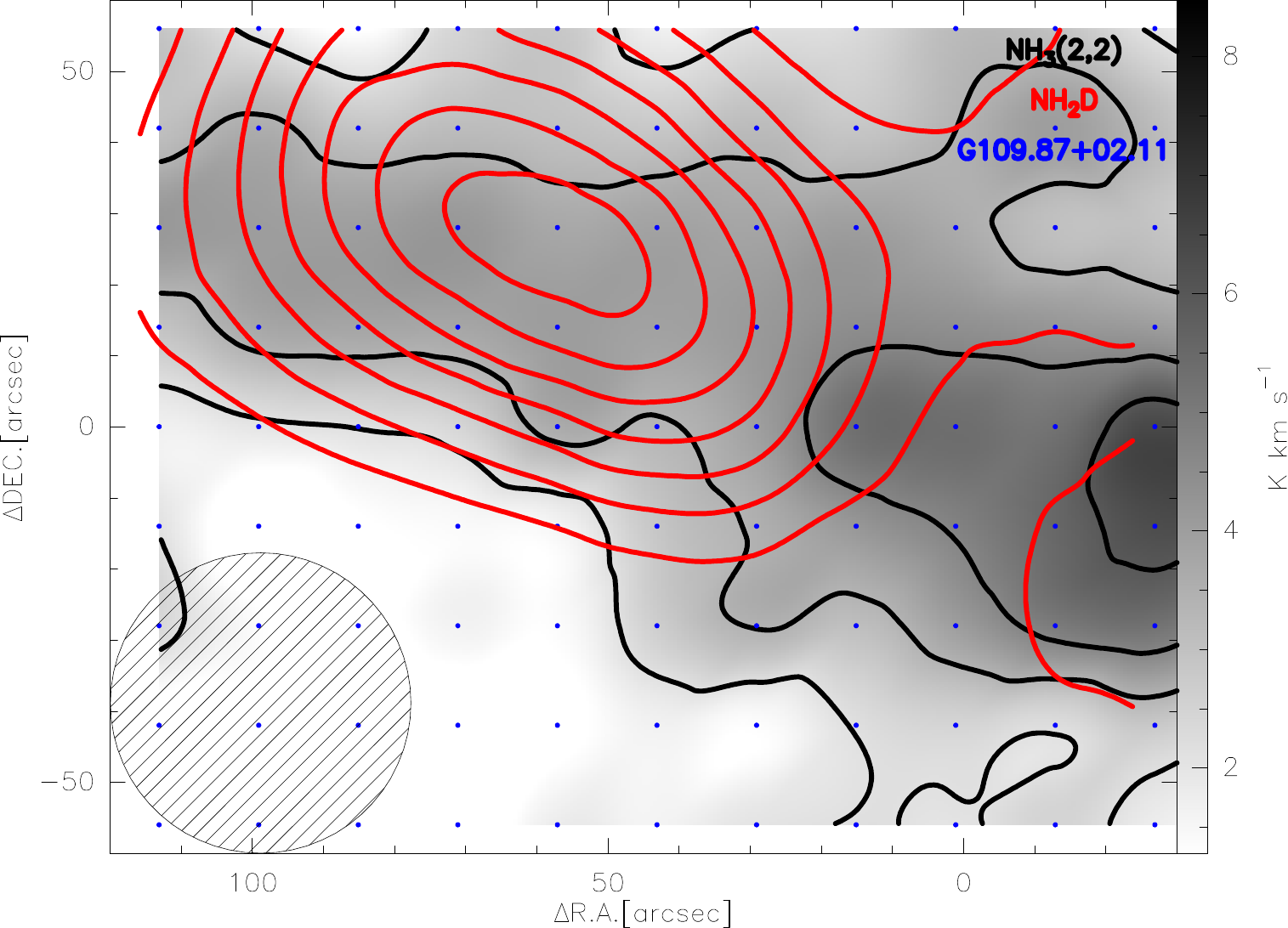}{0.5\textwidth}{(b)}}
\caption{Same as Figure \ref{G015_NH3}, but for source G109.87+02.11. (a) The red contour levels start at 5$\sigma$ in steps of 2$\sigma$, the black contour levels start at 6$\sigma$ in steps of 3$\sigma$, and the gray scale starts at 3$\sigma$. (b) The red contour levels start at 5$\sigma$ in steps of 2$\sigma$, the black contour levels start at 5$\sigma$ in steps of 3$\sigma$, and the gray scale starts at 3$\sigma$.
}
\label{G109_NH3}
\end{figure}

\begin{figure}[h]
\gridline{\fig{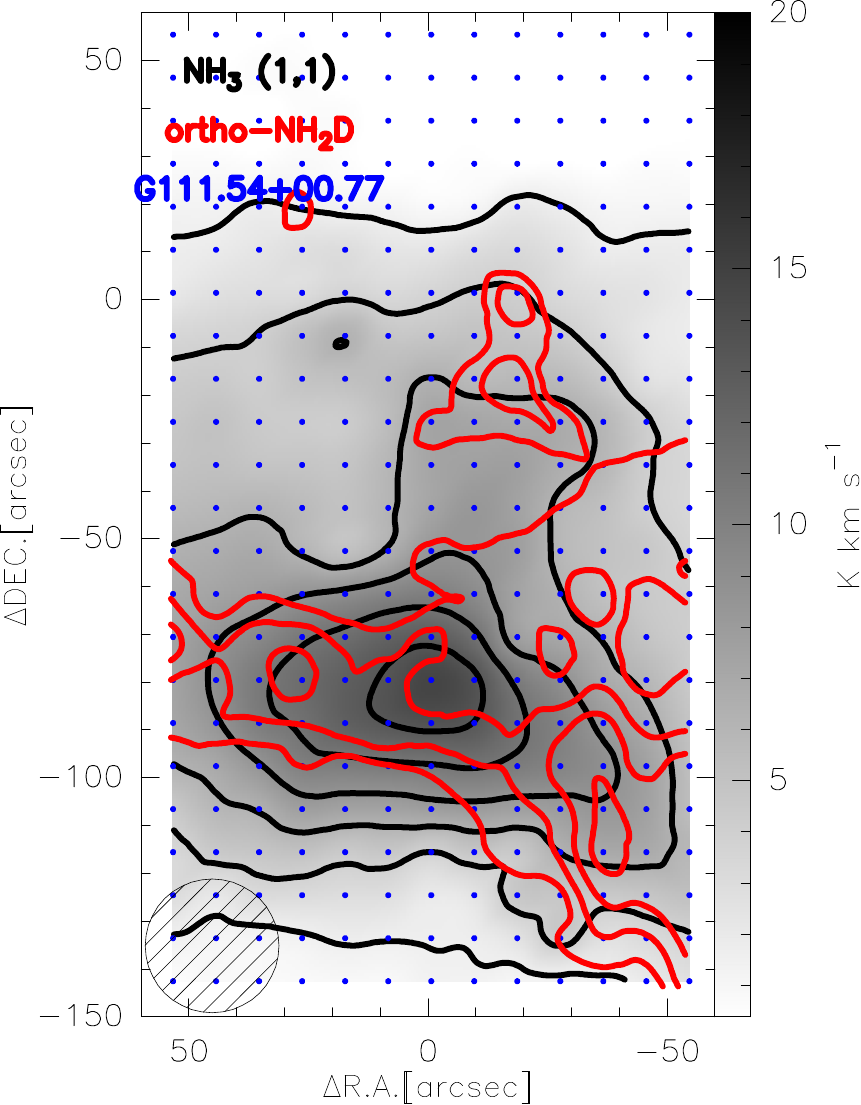}{0.4\textwidth}{(a)}
          \fig{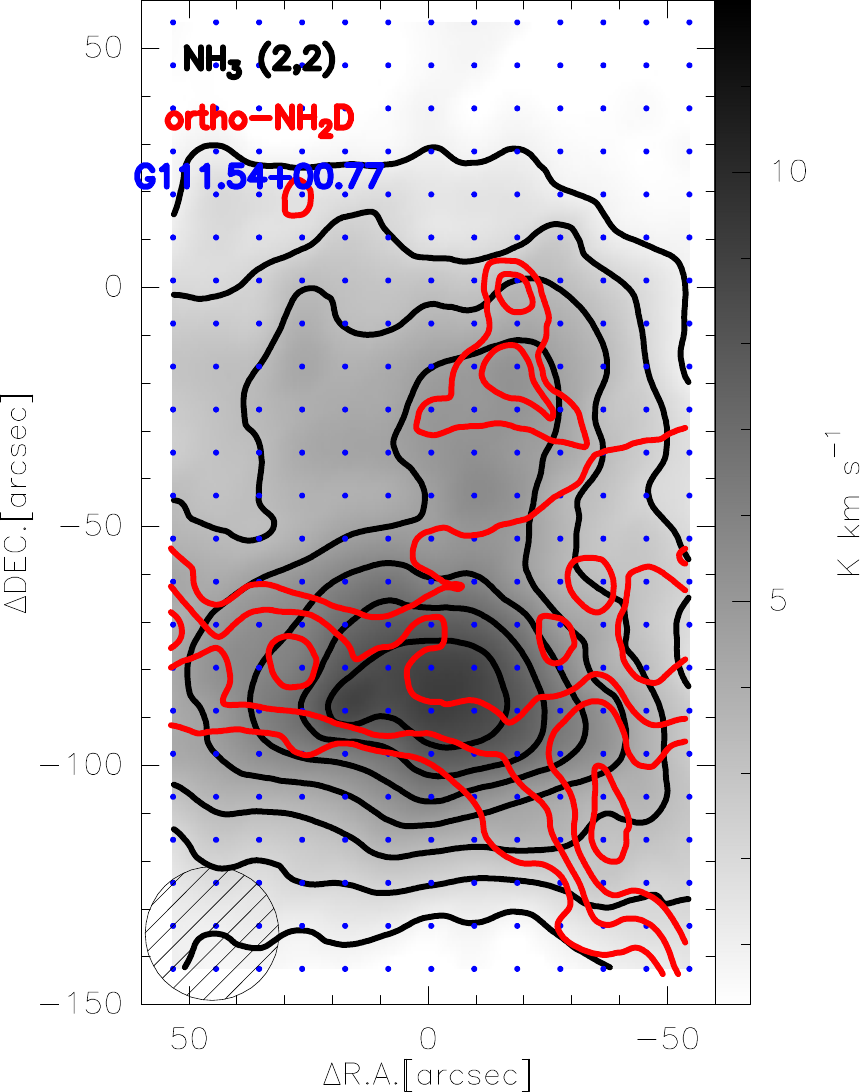}{0.4\textwidth}{(b)}}
\caption{Same as Figure \ref{G015_NH3}, but for source G111.54+00.77. (a) The red contour levels start at 3$\sigma$ in steps of 2$\sigma$, the black contour levels start at 12$\sigma$ in steps of 18$\sigma$, and the gray scale starts at 3$\sigma$. (b) The red contour levels start at 3$\sigma$ in steps of 2$\sigma$, the black contour levels start at 9$\sigma$ in steps of 12$\sigma$, and the gray scale starts at 3$\sigma$.
}
\label{G111_NH3}
\end{figure}

\begin{figure}[h]
\gridline{\fig{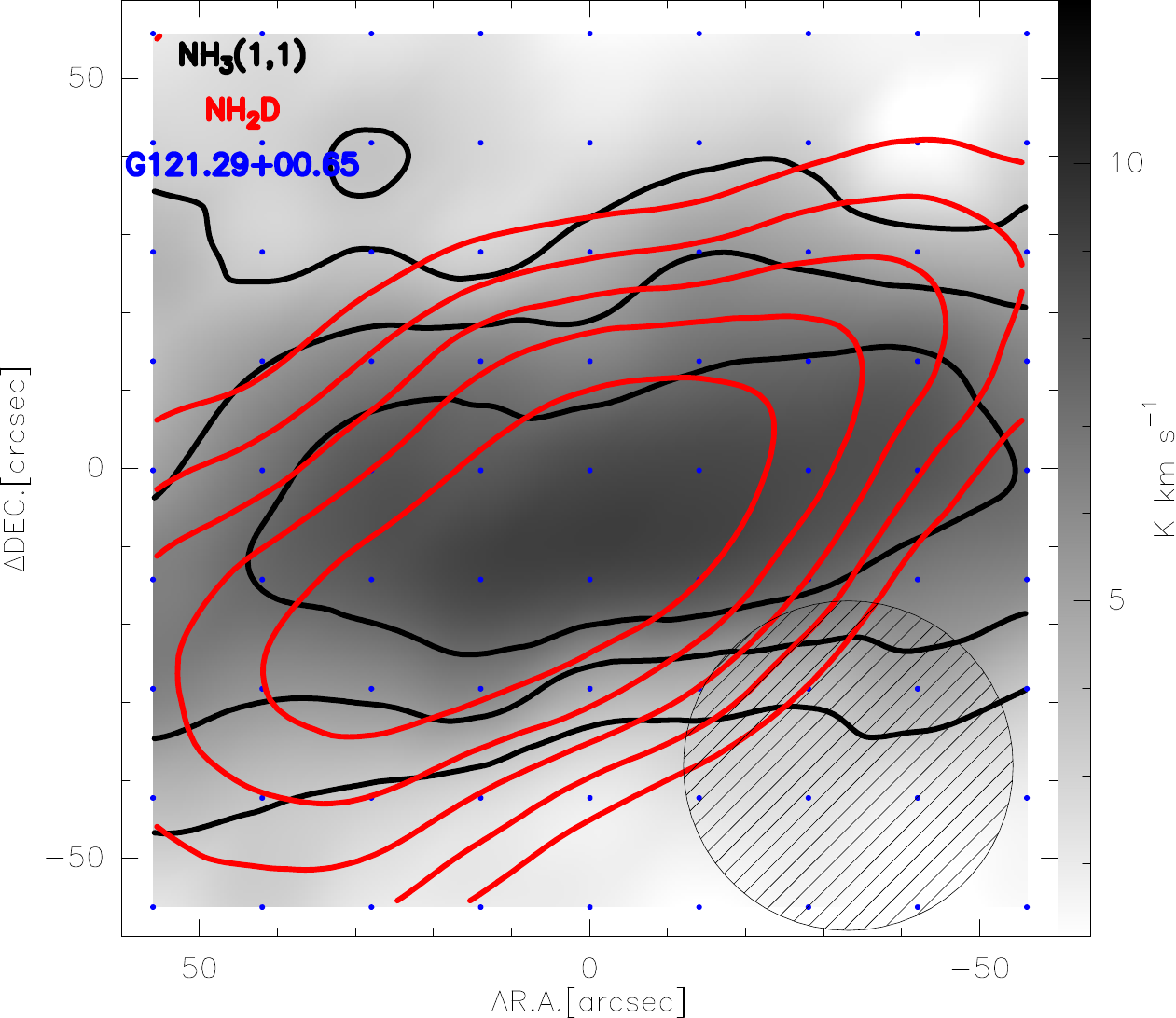}{0.5\textwidth}{(a)}
          \fig{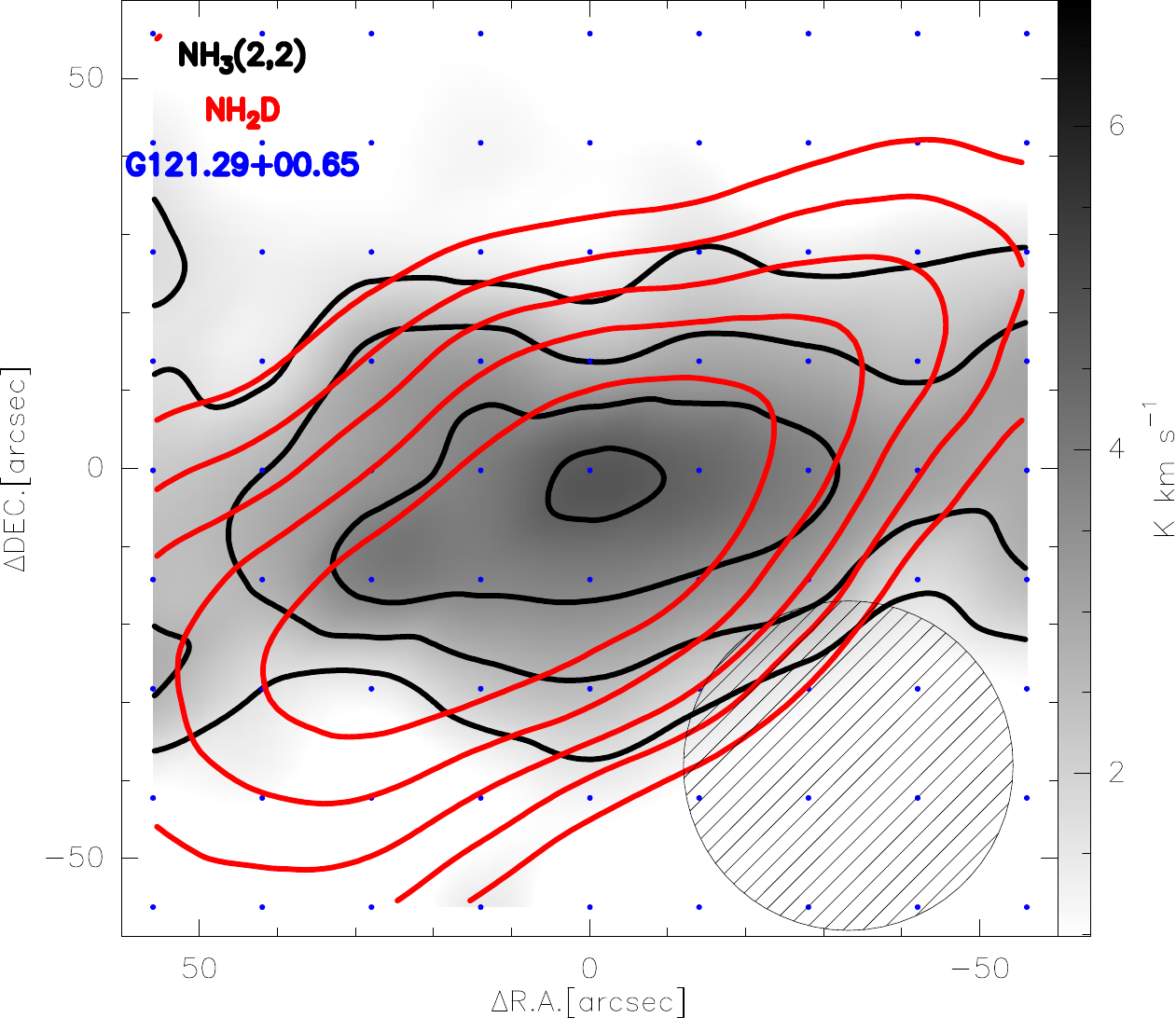}{0.5\textwidth}{(b)}}
\caption{Same as Figure \ref{G015_NH3}, but for source G121.29+00.65. (a) The red contour levels start at 7$\sigma$ in steps of 2$\sigma$, the black contour levels start at 7$\sigma$ in steps of 5$\sigma$, and the gray scale starts at 3$\sigma$. (b) The red contour levels start at 7$\sigma$ in steps of 2$\sigma$, the black contour levels start at 5$\sigma$ in steps of 3$\sigma$, and the gray scale starts at 3$\sigma$.
}
\label{G121_NH3}
\end{figure}

We present the spatial distribution maps NH$_3$(1,1), NH$_3$(2,2) and ortho-NH$_2$D $1_{11}^s-1_{01}^a$ in Figure \ref{G015_NH3}-\ref{G121_NH3}.

\begin{figure}[h]
\gridline{\fig{figure/NH3/G03519_NH3_1-1.pdf}{0.5\textwidth}{(a)}
          \fig{figure/NH3/G03519_NH3_2-2.pdf}{0.5\textwidth}{(b)}}
\caption{Same as Figure \ref{G015_NH3}, but for source G035.19-00.74. (a) The red contour levels start at 5$\sigma$ in steps of 5$\sigma$, the black contour levels start at 90$\sigma$ in steps of 54$\sigma$, and the gray scale starts at 3$\sigma$. (b) The red contour levels start at 5$\sigma$ in steps of 5$\sigma$, the black contour levels start at 54$\sigma$ in steps of 27$\sigma$, and the gray scale starts at 3$\sigma$.
}
\label{G03519_NH3}
\end{figure}

\clearpage

\section{Physical parameters of the calculation and the error}
\setcounter{table}{0}
The rotation temperature $T_{\rm rot}$(2,2;1,1), colunm denstity of NH$_3$ and NH$_2$D, and deuterium fractionation are list in Table \ref{D-fraction}.

\restartappendixnumbering
\startlongtable
\begin{deluxetable*}{ccccccc}
\tablecaption{Column density and deuterium fractionation}
\label{D-fraction}
\tablewidth{0pt}
\tablehead{
\colhead{Source} & \colhead{Offset} & \colhead{$T_{\rm rot}$(2,2;1,1)} &
 \colhead{$N$(NH$_3$)$\times$10$^{15}$} & \colhead{$N$(NH$_2$D)$\times$10$^{13}$} & \colhead{$D_{\rm frac}$(NH$_3$)$\times$100} \\
\colhead{} & \colhead{(arcsec, arcsec)} & \colhead{(K)} &
 \colhead{(cm$^{-2}$)} & \colhead{(cm$^{-2}$)} & \colhead{}
}
\startdata
G015.03-00.67 &(-9,-27) &24.8$\pm$0.7 &2.0$\pm$0.2 &1.2$\pm$0.1 &0.59$\pm$0.09 \\ 
 &(-18,-18) &24.4$\pm$0.6 &2.1$\pm$0.2 &1.2$\pm$0.1 &0.56$\pm$0.08 \\ 
 &(-27,-9) &24.1$\pm$0.5 &1.7$\pm$0.1 &1.1$\pm$0.1 &0.7$\pm$0.1 \\ 
 &(-36,-18) &23.3$\pm$0.5 &1.7$\pm$0.1 &1.2$\pm$0.1 &0.67$\pm$0.09 \\ 
 &(-45,-81) &21.6$\pm$0.9 &1.7$\pm$0.1 &0.9$\pm$0.1 &0.51$\pm$0.08 \\ 
 &(-54,-108) &19.5$\pm$0.8 &1.1$\pm$0.2 &1.3$\pm$0.1 &1.2$\pm$0.2 \\ \hline 
G023.44-00.18 &(9,9) &20.6$\pm$0.6 &0.46$\pm$0.04 &1.2$\pm$0.1 &2.5$\pm$0.3 \\ 
 &(-9,9) &20.8$\pm$0.6 &0.70$\pm$0.05 &1.5$\pm$0.1 &2.1$\pm$0.2 \\ 
 &(9,-9) &21.6$\pm$0.5 &0.79$\pm$0.07 &2.0$\pm$0.1 &2.6$\pm$0.3 \\ 
 &(-9,-9) &21.5$\pm$0.5 &0.94$\pm$0.07 &2.6$\pm$0.1 &2.8$\pm$0.2 \\ 
 &(-27,-9) &20.4$\pm$0.6 &1.02$\pm$0.05 &1.4$\pm$0.1 &1.3$\pm$0.1 \\ 
 &(0,-27) &21.2$\pm$0.5 &0.80$\pm$0.07 &1.8$\pm$0.1 &2.3$\pm$0.2 \\ 
 &(-18,-27) &20.3$\pm$0.5 &0.67$\pm$0.06 &1.8$\pm$0.1 &2.7$\pm$0.3 \\ 
 &(-36,-27) &17.8$\pm$0.7 &1.29$\pm$0.09 &1.1$\pm$0.1 &0.9$\pm$0.1 \\ 
 &(-27,36) &14.8$\pm$0.7 &0.31$\pm$0.05 &0.9$\pm$0.1 &3.1$\pm$0.6 \\ \hline 
G031.28+00.06 &(18,18) &21.7$\pm$0.5 &0.69$\pm$0.08 &0.81$\pm$0.09 &0.07$\pm$0.01 \\ 
 &(9,18) &22.1$\pm$0.5 &0.9$\pm$0.1 &0.81$\pm$0.09 &0.062$\pm$0.008 \\ 
 &(-18,-18) &20.3$\pm$0.5 &0.9$\pm$0.1 &0.87$\pm$0.09 &0.08$\pm$0.01 \\ 
 &(-18,-27) &18.9$\pm$0.5 &0.8$\pm$0.1 &0.94$\pm$0.09 &0.11$\pm$0.01 \\ 
 &(-18,-36) &17.6$\pm$0.4 &0.7$\pm$0.1 &0.94$\pm$0.08 &0.12$\pm$0.01 \\ 
 &(-18,-45) &16.7$\pm$0.5 &0.63$\pm$0.08 &0.84$\pm$0.09 &0.12$\pm$0.02 \\ 
 &(-27,-18) &19.5$\pm$0.5 &0.8$\pm$0.1 &0.87$\pm$0.09 &0.09$\pm$0.01 \\ 
 &(-27,-27) &18.2$\pm$0.5 &0.8$\pm$0.1 &0.96$\pm$0.09 &0.11$\pm$0.01 \\ 
 &(-27,-36) &17.1$\pm$0.4 &0.8$\pm$0.1 &0.96$\pm$0.08 &0.12$\pm$0.01 \\ 
 &(-27,-45) &16.4$\pm$0.4 &0.70$\pm$0.09 &0.94$\pm$0.09 &0.13$\pm$0.02 \\ \hline 
G034.39+00.22 &(-9,45) &19.5$\pm$0.4 &2.1$\pm$0.2 &3.5$\pm$0.2 &1.7$\pm$0.2 \\ 
 &(-9,36) &20.0$\pm$0.4 &2.2$\pm$0.2 &4.2$\pm$0.2 &1.9$\pm$0.2 \\ 
 &(-9,27) &20.2$\pm$0.4 &2.2$\pm$0.2 &4.9$\pm$0.2 &2.2$\pm$0.2 \\ 
 &(-9,18) &20.2$\pm$0.4 &2.2$\pm$0.2 &5.1$\pm$0.2 &2.3$\pm$0.2 \\ 
 &(-9,9) &20.1$\pm$0.4 &2.1$\pm$0.2 &4.9$\pm$0.2 &2.3$\pm$0.2 \\ 
 &(-9,0) &20.0$\pm$0.3 &2.1$\pm$0.2 &4.7$\pm$0.1 &2.2$\pm$0.2 \\ 
 &(-9,-9) &19.9$\pm$0.3 &2.0$\pm$0.2 &4.8$\pm$0.1 &2.4$\pm$0.2 \\ 
 &(-9,-18) &19.6$\pm$0.3 &2.0$\pm$0.2 &5.4$\pm$0.2 &2.7$\pm$0.2 \\ 
 &(-9,-27) &19.2$\pm$0.3 &1.9$\pm$0.2 &6.0$\pm$0.2 &3.1$\pm$0.3 \\ 
 &(-9,-36) &18.7$\pm$0.3 &1.8$\pm$0.1 &6.1$\pm$0.2 &3.4$\pm$0.3 \\ 
 &(-9,-45) &18.1$\pm$0.3 &1.6$\pm$0.1 &5.6$\pm$0.2 &3.5$\pm$0.3 \\ \hline 
G035.19-00.74 &(-45,45) &15.8$\pm$0.3 &0.78$\pm$0.07 &3.1$\pm$0.1 &4.0$\pm$0.4 \\ 
 &(-36,36) &17.1$\pm$0.3 &0.91$\pm$0.09 &3.4$\pm$0.1 &3.7$\pm$0.4 \\ 
 &(-27,27) &18.9$\pm$0.4 &1.1$\pm$0.1 &3.7$\pm$0.1 &3.3$\pm$0.3 \\ 
 &(-18,18) &20.5$\pm$0.3 &1.4$\pm$0.1 &3.6$\pm$0.1 &2.6$\pm$0.2 \\ 
 &(-9,9) &21.7$\pm$0.3 &1.5$\pm$0.1 &2.9$\pm$0.1 &2.0$\pm$0.2 \\ 
 &(0,0) &22.1$\pm$0.3 &1.4$\pm$0.1 &2.3$\pm$0.1 &1.7$\pm$0.2 \\ 
 &(9,-9) &21.7$\pm$0.3 &1.2$\pm$0.1 &1.92$\pm$0.09 &1.6$\pm$0.1 \\ 
 &(18,-18) &20.6$\pm$0.3 &1.02$\pm$0.09 &1.80$\pm$0.09 &1.8$\pm$0.2 \\ 
 &(27,-27) &18.9$\pm$0.4 &0.80$\pm$0.07 &1.7$\pm$0.1 &2.1$\pm$0.2 \\ 
 &(36,-36) &16.7$\pm$0.5 &0.71$\pm$0.06 &1.8$\pm$0.1 &2.5$\pm$0.3 \\ 
 &(45,-45) &14.5$\pm$0.6 &0.85$\pm$0.09 &1.8$\pm$0.1 &2.1$\pm$0.3 \\ \hline 
G035.20-01.73 &(-18,0) &22.4$\pm$0.6 &0.65$\pm$0.08 &2.0$\pm$0.1 &3.0$\pm$0.4 \\ 
 &(-27,0) &21.6$\pm$0.5 &0.89$\pm$0.09 &2.5$\pm$0.1 &2.8$\pm$0.3 \\ 
 &(-36,0) &21.0$\pm$0.4 &1.1$\pm$0.1 &3.4$\pm$0.1 &3.0$\pm$0.3 \\ 
 &(-45,0) &20.4$\pm$0.4 &1.3$\pm$0.1 &4.3$\pm$0.1 &3.2$\pm$0.3 \\ 
 &(-54,0) &19.7$\pm$0.4 &1.4$\pm$0.1 &4.9$\pm$0.1 &3.4$\pm$0.4 \\ 
 &(-63,0) &19.0$\pm$0.3 &1.3$\pm$0.1 &4.6$\pm$0.1 &3.5$\pm$0.4 \\ 
 &(-72,0) &18.3$\pm$0.3 &1.1$\pm$0.1 &3.7$\pm$0.1 &3.4$\pm$0.3 \\ 
 &(-54,18) &19.7$\pm$0.4 &0.75$\pm$0.07 &3.1$\pm$0.1 &4.1$\pm$0.4 \\ 
 &(-54,-18) &20.0$\pm$0.4 &1.0$\pm$0.1 &2.4$\pm$0.1 &2.4$\pm$0.3 \\ \hline 
G075.76+00.33 &(0,-27) &22.9$\pm$0.2 &0.54$\pm$0.04 &1.2$\pm$0.1 &2.3$\pm$0.3 \\ 
 &(-9,-27) &22.7$\pm$0.2 &0.62$\pm$0.04 &1.4$\pm$0.1 &2.2$\pm$0.2 \\ 
 &(-18,-27) &22.3$\pm$0.2 &0.68$\pm$0.05 &1.4$\pm$0.1 &2.0$\pm$0.2 \\ 
 &(-27,-27) &21.9$\pm$0.3 &0.69$\pm$0.05 &1.3$\pm$0.1 &1.8$\pm$0.2 \\ 
 &(-36,-27) &21.4$\pm$0.3 &0.66$\pm$0.05 &1.2$\pm$0.1 &1.8$\pm$0.2 \\ 
 &(-45,-27) &21.1$\pm$0.4 &0.57$\pm$0.04 &1.1$\pm$0.1 &2.0$\pm$0.3 \\ \hline 
G081.75+00.59 &(17,46) &18.1$\pm$0.4 &1.5$\pm$0.2 &6.1$\pm$0.2 &4.0$\pm$0.4 \\ 
 &(3,46) &18.2$\pm$0.4 &1.6$\pm$0.2 &6.7$\pm$0.2 &4.2$\pm$0.4 \\ 
 &(-11,46) &18.6$\pm$0.4 &1.3$\pm$0.1 &5.5$\pm$0.1 &4.3$\pm$0.4 \\ 
 &(17,32) &18.1$\pm$0.4 &1.5$\pm$0.2 &6.5$\pm$0.2 &4.3$\pm$0.4 \\ 
 &(3,32) &18.3$\pm$0.3 &1.7$\pm$0.2 &7.1$\pm$0.2 &4.2$\pm$0.4 \\ 
 &(-11,32) &18.6$\pm$0.3 &1.5$\pm$0.1 &6.1$\pm$0.1 &4.1$\pm$0.4 \\ 
 &(17,18) &18.6$\pm$0.4 &1.4$\pm$0.1 &6.9$\pm$0.2 &4.8$\pm$0.5 \\ 
 &(3,18) &18.7$\pm$0.3 &1.7$\pm$0.2 &7.6$\pm$0.2 &4.4$\pm$0.4 \\ 
 &(-11,18) &19.1$\pm$0.3 &1.6$\pm$0.1 &6.9$\pm$0.1 &4.3$\pm$0.4 \\ 
 &(17,4) &19.1$\pm$0.4 &1.2$\pm$0.1 &6.7$\pm$0.2 &5.6$\pm$0.6 \\ 
 &(3,4) &19.0$\pm$0.4 &1.5$\pm$0.1 &7.7$\pm$0.2 &5.0$\pm$0.5 \\ 
 &(-11,4) &19.3$\pm$0.4 &1.6$\pm$0.1 &7.3$\pm$0.2 &4.7$\pm$0.4 \\ 
 &(17,-10) &18.9$\pm$0.4 &0.93$\pm$0.09 &5.7$\pm$0.2 &6.2$\pm$0.6 \\ 
 &(3,-10) &19.1$\pm$0.4 &1.2$\pm$0.1 &6.9$\pm$0.2 &5.8$\pm$0.6 \\ 
 &(-11,-10) &19.3$\pm$0.4 &1.3$\pm$0.1 &6.8$\pm$0.2 &5.3$\pm$0.5 \\ 
 &(17,-24) &19.6$\pm$0.5 &0.57$\pm$0.06 &3.9$\pm$0.1 &6.8$\pm$0.8 \\ 
 &(3,-24) &19.6$\pm$0.5 &0.77$\pm$0.08 &5.0$\pm$0.1 &6.5$\pm$0.7 \\ 
 &(-11,-24) &19.7$\pm$0.4 &0.97$\pm$0.09 &5.4$\pm$0.1 &5.6$\pm$0.6 \\ 
 &(-23,-11) &19.2$\pm$0.4 &1.3$\pm$0.1 &6.0$\pm$0.1 &4.4$\pm$0.4 \\ 
 &(-37,-11) &18.4$\pm$0.3 &1.4$\pm$0.1 &4.8$\pm$0.1 &3.4$\pm$0.3 \\ 
 &(-51,-11) &18.0$\pm$0.4 &1.3$\pm$0.1 &3.7$\pm$0.1 &2.8$\pm$0.3 \\ 
 &(-23,-25) &19.6$\pm$0.4 &1.2$\pm$0.1 &5.4$\pm$0.1 &4.7$\pm$0.4 \\ 
 &(-37,-25) &18.6$\pm$0.3 &1.5$\pm$0.1 &4.9$\pm$0.1 &3.4$\pm$0.3 \\ 
 &(-51,-25) &18.0$\pm$0.4 &1.5$\pm$0.1 &4.0$\pm$0.1 &2.6$\pm$0.3 \\ 
 &(-23,-39) &19.9$\pm$0.4 &1.1$\pm$0.1 &4.8$\pm$0.1 &4.4$\pm$0.4 \\ 
 &(-37,-39) &19.1$\pm$0.4 &1.4$\pm$0.1 &4.8$\pm$0.1 &3.4$\pm$0.3 \\ 
 &(-51,-39) &18.3$\pm$0.4 &1.5$\pm$0.1 &3.7$\pm$0.1 &2.6$\pm$0.3 \\ 
 &(-23,-53) &19.1$\pm$0.4 &1.3$\pm$0.1 &4.6$\pm$0.1 &3.6$\pm$0.4 \\ 
 &(-37,-53) &19.2$\pm$0.3 &1.5$\pm$0.1 &4.7$\pm$0.1 &3.0$\pm$0.3 \\ 
 &(-51,-53) &18.7$\pm$0.4 &1.5$\pm$0.1 &3.6$\pm$0.1 &2.4$\pm$0.2 \\ 
 &(-23,-67) &19.3$\pm$0.4 &1.4$\pm$0.1 &5.1$\pm$0.1 &3.5$\pm$0.3 \\ 
 &(-37,-67) &19.5$\pm$0.3 &1.8$\pm$0.2 &5.2$\pm$0.1 &3.0$\pm$0.3 \\ 
 &(-51,-67) &19.2$\pm$0.4 &1.7$\pm$0.2 &3.9$\pm$0.1 &2.3$\pm$0.2 \\ 
 &(-23,-81) &20.6$\pm$0.5 &1.4$\pm$0.1 &5.6$\pm$0.2 &4.1$\pm$0.4 \\ 
 &(-37,-81) &20.1$\pm$0.4 &1.8$\pm$0.2 &5.9$\pm$0.2 &3.4$\pm$0.4 \\ 
 &(-51,-81) &19.8$\pm$0.5 &1.7$\pm$0.2 &4.6$\pm$0.1 &2.7$\pm$0.3 \\ 
 &(-23,-95) &21.9$\pm$0.6 &1.3$\pm$0.1 &5.6$\pm$0.2 &4.4$\pm$0.5 \\ 
 &(-37,-95) &21.0$\pm$0.6 &1.6$\pm$0.2 &6.2$\pm$0.2 &3.8$\pm$0.4 \\ 
 &(-51,-95) &20.1$\pm$0.6 &1.7$\pm$0.2 &4.9$\pm$0.2 &2.9$\pm$0.3 \\ 
 &(-23,-109) &23.0$\pm$0.6 &1.4$\pm$0.1 &5.2$\pm$0.2 &3.6$\pm$0.4 \\ 
 &(-37,-109) &21.9$\pm$0.7 &1.7$\pm$0.2 &5.8$\pm$0.2 &3.5$\pm$0.4 \\ 
 &(-51,-109) &20.5$\pm$0.8 &1.7$\pm$0.2 &4.6$\pm$0.2 &2.7$\pm$0.3 \\ 
 &(-23,-123) &23.9$\pm$0.6 &1.7$\pm$0.2 &5.1$\pm$0.2 &3.0$\pm$0.4 \\ 
 &(-37,-123) &22.8$\pm$0.7 &2.4$\pm$0.2 &5.4$\pm$0.2 &2.2$\pm$0.2 \\ 
 &(-51,-123) &21.6$\pm$0.8 &1.5$\pm$0.2 &4.1$\pm$0.2 &2.8$\pm$0.4 \\ \hline 
G081.87+00.78 &(45,-18) &20.2$\pm$0.3 &0.25$\pm$0.02 &1.7$\pm$0.1 &6.8$\pm$0.7 \\ 
 &(45,-9) &20.2$\pm$0.3 &0.30$\pm$0.02 &2.1$\pm$0.1 &6.8$\pm$0.6 \\ 
 &(45,0) &20.5$\pm$0.3 &0.30$\pm$0.02 &2.2$\pm$0.1 &7.2$\pm$0.6 \\ 
 &(45,9) &20.7$\pm$0.4 &0.25$\pm$0.02 &2.2$\pm$0.1 &8.6$\pm$0.7 \\ 
 &(36,9) &22.1$\pm$0.4 &0.31$\pm$0.02 &2.1$\pm$0.1 &6.8$\pm$0.6 \\ 
 &(27,18) &23.3$\pm$0.4 &0.30$\pm$0.02 &1.8$\pm$0.1 &6.1$\pm$0.6 \\ 
 &(18,18) &24.2$\pm$0.4 &0.43$\pm$0.03 &2.2$\pm$0.1 &5.1$\pm$0.4 \\ 
 &(9,18) &25.0$\pm$0.4 &0.60$\pm$0.04 &2.7$\pm$0.1 &4.5$\pm$0.4 \\ 
 &(0,18) &25.6$\pm$0.4 &0.76$\pm$0.05 &2.7$\pm$0.1 &3.6$\pm$0.3 \\ \hline 
G109.87+02.11 &(95,45) &21.6$\pm$1.3 &0.3$\pm$0.1 &1.19$\pm$0.09 &3.8$\pm$1.5 \\ 
 &(81,31) &21.4$\pm$0.8 &0.44$\pm$0.06 &1.80$\pm$0.09 &4.1$\pm$0.6 \\ 
 &(67,17) &21.9$\pm$0.8 &0.51$\pm$0.06 &2.0$\pm$0.1 &3.9$\pm$0.5 \\ 
 &(53,3) &22.5$\pm$1.0 &0.51$\pm$0.08 &1.7$\pm$0.1 &3.3$\pm$0.5 \\ 
 &(39,-11) &21.2$\pm$1.1 &0.6$\pm$0.1 &1.02$\pm$0.09 &1.6$\pm$0.3 \\ 
 &(81,45) &21.8$\pm$1.3 &0.3$\pm$0.1 &1.6$\pm$0.1 &5.5$\pm$2.4 \\ 
 &(67,31) &21.2$\pm$0.9 &0.51$\pm$0.06 &2.0$\pm$0.1 &4.0$\pm$0.5 \\ 
 &(53,17) &21.2$\pm$0.9 &0.61$\pm$0.08 &2.1$\pm$0.1 &3.4$\pm$0.5 \\ 
 &(39,3) &22.1$\pm$1.0 &0.56$\pm$0.08 &1.6$\pm$0.1 &2.9$\pm$0.5 \\ 
 &(25,-11) &23.1$\pm$0.9 &0.57$\pm$0.08 &0.86$\pm$0.09 &1.5$\pm$0.3 \\ 
 &(-3,30) &21.7$\pm$1.0 &0.45$\pm$0.07 &0.62$\pm$0.08 &1.4$\pm$0.3 \\ 
 &(-3,16) &23.3$\pm$1.0 &0.42$\pm$0.06 &0.64$\pm$0.08 &1.5$\pm$0.3 \\ 
 &(-17,-21) &24.6$\pm$0.8 &0.58$\pm$0.07 &0.6$\pm$0.1 &1.1$\pm$0.2 \\ \hline 
G111.54+00.77 &(45,-81) &21.3$\pm$0.3 &0.75$\pm$0.07 &1.5$\pm$0.1 &2.0$\pm$0.2 \\ 
 &(36,-81) &21.5$\pm$0.2 &0.84$\pm$0.07 &1.6$\pm$0.1 &1.9$\pm$0.2 \\ 
 &(27,-81) &22.1$\pm$0.2 &0.89$\pm$0.08 &1.7$\pm$0.1 &1.9$\pm$0.2 \\ 
 &(18,-81) &22.9$\pm$0.2 &0.93$\pm$0.08 &1.7$\pm$0.1 &1.8$\pm$0.2 \\ 
 &(9,-81) &23.5$\pm$0.2 &1.02$\pm$0.09 &1.6$\pm$0.1 &1.6$\pm$0.2 \\ 
 &(0,-81) &23.9$\pm$0.3 &1.1$\pm$0.1 &1.5$\pm$0.1 &1.4$\pm$0.2 \\ 
 &(-9,-81) &24.7$\pm$0.3 &0.94$\pm$0.08 &1.4$\pm$0.1 &1.5$\pm$0.2 \\ 
 &(-18,-81) &24.9$\pm$0.4 &0.74$\pm$0.07 &1.3$\pm$0.1 &1.8$\pm$0.2 \\ 
 &(-27,-81) &24.2$\pm$0.4 &0.56$\pm$0.05 &1.5$\pm$0.2 &2.7$\pm$0.4 \\ 
 &(-36,-81) &22.6$\pm$0.4 &0.46$\pm$0.04 &1.3$\pm$0.1 &2.8$\pm$0.4 \\ 
 &(-45,-45) &21.9$\pm$0.5 &0.27$\pm$0.03 &1.2$\pm$0.1 &4.3$\pm$0.6 \\ 
 &(-45,-54) &21.8$\pm$0.5 &0.29$\pm$0.03 &1.1$\pm$0.1 &3.9$\pm$0.6 \\ 
 &(-45,-63) &21.6$\pm$0.5 &0.33$\pm$0.03 &1.0$\pm$0.1 &2.9$\pm$0.5 \\ 
 &(-45,-72) &21.4$\pm$0.4 &0.34$\pm$0.03 &0.9$\pm$0.1 &2.8$\pm$0.5 \\ 
 &(-45,-99) &21.0$\pm$0.5 &0.49$\pm$0.06 &2.2$\pm$0.1 &4.4$\pm$0.6 \\ 
 &(-45,-108) &20.3$\pm$0.5 &0.52$\pm$0.06 &2.4$\pm$0.1 &4.6$\pm$0.6 \\ 
 &(-45,-117) &20.0$\pm$0.5 &0.47$\pm$0.05 &2.3$\pm$0.1 &4.9$\pm$0.6 \\ 
 &(-45,-126) &20.1$\pm$0.6 &0.36$\pm$0.04 &2.0$\pm$0.1 &5.6$\pm$0.7 \\ 
 &(-45,-135) &21.1$\pm$0.7 &0.22$\pm$0.03 &1.5$\pm$0.1 &6.7$\pm$1.0 \\ 
 &(-18,-18) &26.5$\pm$0.5 &0.33$\pm$0.02 &1.3$\pm$0.2 &3.9$\pm$0.6 \\ \hline 
G121.29+00.65 &(49,-49) &13.6$\pm$1.4 &0.4$\pm$0.1 &1.3$\pm$0.3 &3.6$\pm$1.2 \\ 
 &(35,-35) &15.4$\pm$0.8 &0.51$\pm$0.09 &1.7$\pm$0.2 &3.3$\pm$0.7 \\ 
 &(21,-21) &17.3$\pm$0.6 &0.77$\pm$0.09 &2.0$\pm$0.1 &2.5$\pm$0.3 \\ 
 &(7,-7) &19.0$\pm$0.6 &0.9$\pm$0.1 &2.1$\pm$0.2 &2.2$\pm$0.3 \\ 
 &(-7,7) &18.6$\pm$0.6 &0.9$\pm$0.1 &2.0$\pm$0.2 &2.2$\pm$0.3 \\ 
 &(-21,21) &16.9$\pm$0.7 &0.7$\pm$0.1 &1.5$\pm$0.1 &2.3$\pm$0.4 \\ 
 &(-35,35) &16.3$\pm$1.2 &0.27$\pm$0.06 &1.1$\pm$0.2 &3.9$\pm$1.1 \\ 
 &(49,-30) &16.1$\pm$0.9 &0.48$\pm$0.08 &1.5$\pm$0.2 &3.1$\pm$0.6 \\ 
 &(35,-16) &17.1$\pm$0.6 &0.71$\pm$0.09 &1.8$\pm$0.1 &2.6$\pm$0.4 \\ 
 &(21,-2) &18.2$\pm$0.5 &0.9$\pm$0.1 &1.9$\pm$0.1 &2.2$\pm$0.3 \\ 
 &(7,12) &18.5$\pm$0.6 &0.9$\pm$0.1 &1.9$\pm$0.2 &2.1$\pm$0.3 \\ 
 &(-7,26) &17.3$\pm$0.8 &0.6$\pm$0.1 &1.4$\pm$0.1 &2.4$\pm$0.4 \\ 
 &(-21,40) &15.9$\pm$1.1 &0.36$\pm$0.08 &0.9$\pm$0.2 &2.6$\pm$0.7 \\ 
 &(30,-49) &13.8$\pm$1.4 &0.32$\pm$0.09 &1.3$\pm$0.2 &4.1$\pm$1.4 \\ 
 &(16,-35) &15.0$\pm$0.8 &0.6$\pm$0.1 &1.6$\pm$0.2 &2.7$\pm$0.5 \\ 
 &(2,-21) &18.2$\pm$0.6 &0.7$\pm$0.1 &1.9$\pm$0.1 &2.7$\pm$0.4 \\ 
 &(-12,-7) &18.9$\pm$0.6 &1.0$\pm$0.1 &2.0$\pm$0.2 &2.0$\pm$0.3 \\ 
 &(-26,7) &17.5$\pm$0.5 &0.9$\pm$0.1 &1.8$\pm$0.1 &2.0$\pm$0.3 \\ 
 &(-40,21) &16.0$\pm$0.6 &0.72$\pm$0.09 &1.5$\pm$0.1 &2.0$\pm$0.3 \\ 
 &(-54,35) &16.2$\pm$1.2 &0.32$\pm$0.07 &1.1$\pm$0.2 &3.4$\pm$1.0 \\ 
\enddata
\end{deluxetable*}

\clearpage

\section{The method of pixel averaging In Each Source }
\setcounter{figure}{0}
\renewcommand{\thefigure}{D\arabic{figure}}

\begin{figure}[h]
\centering
\includegraphics[width=0.25\textwidth]{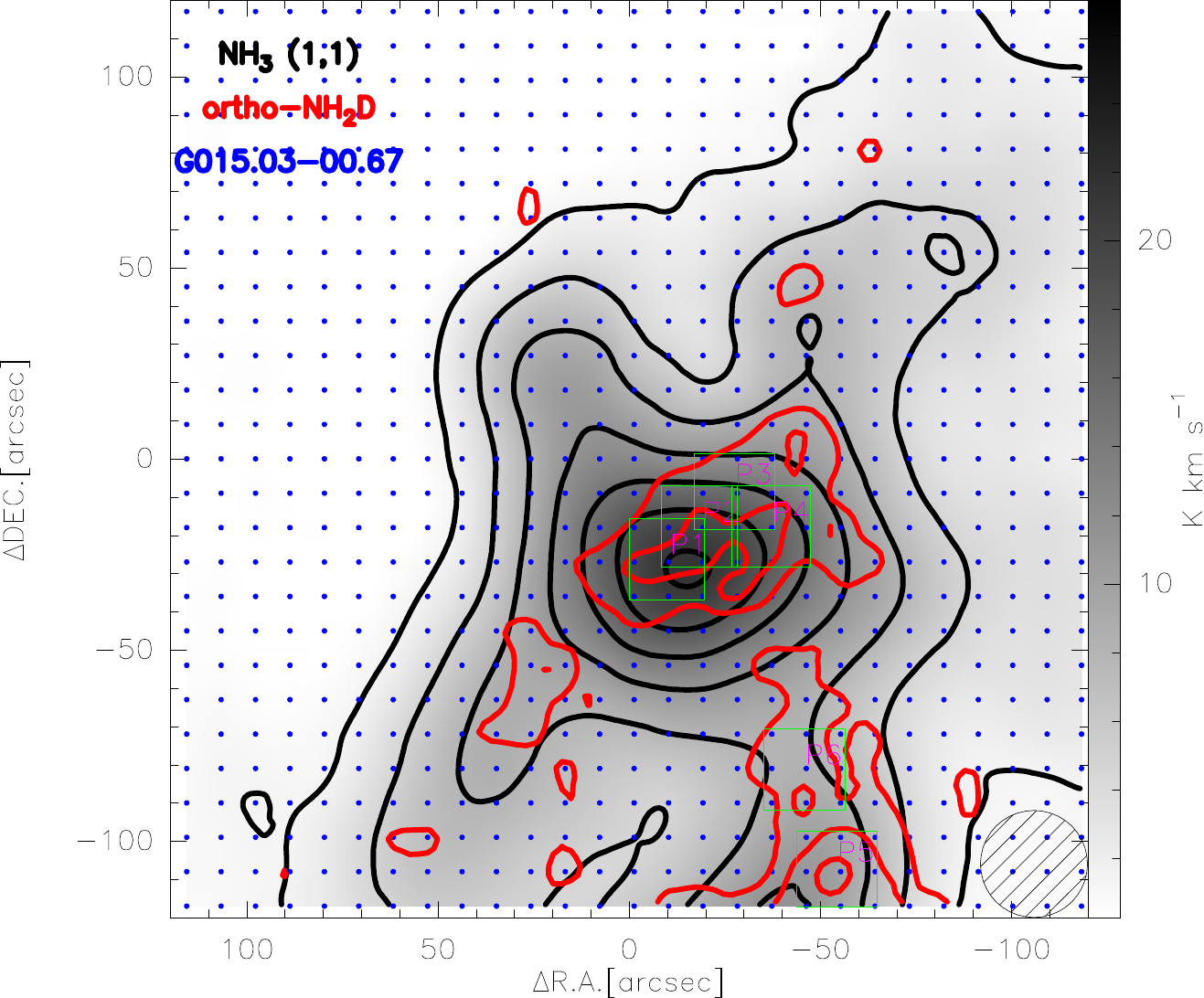}\includegraphics[width=0.25\textwidth]{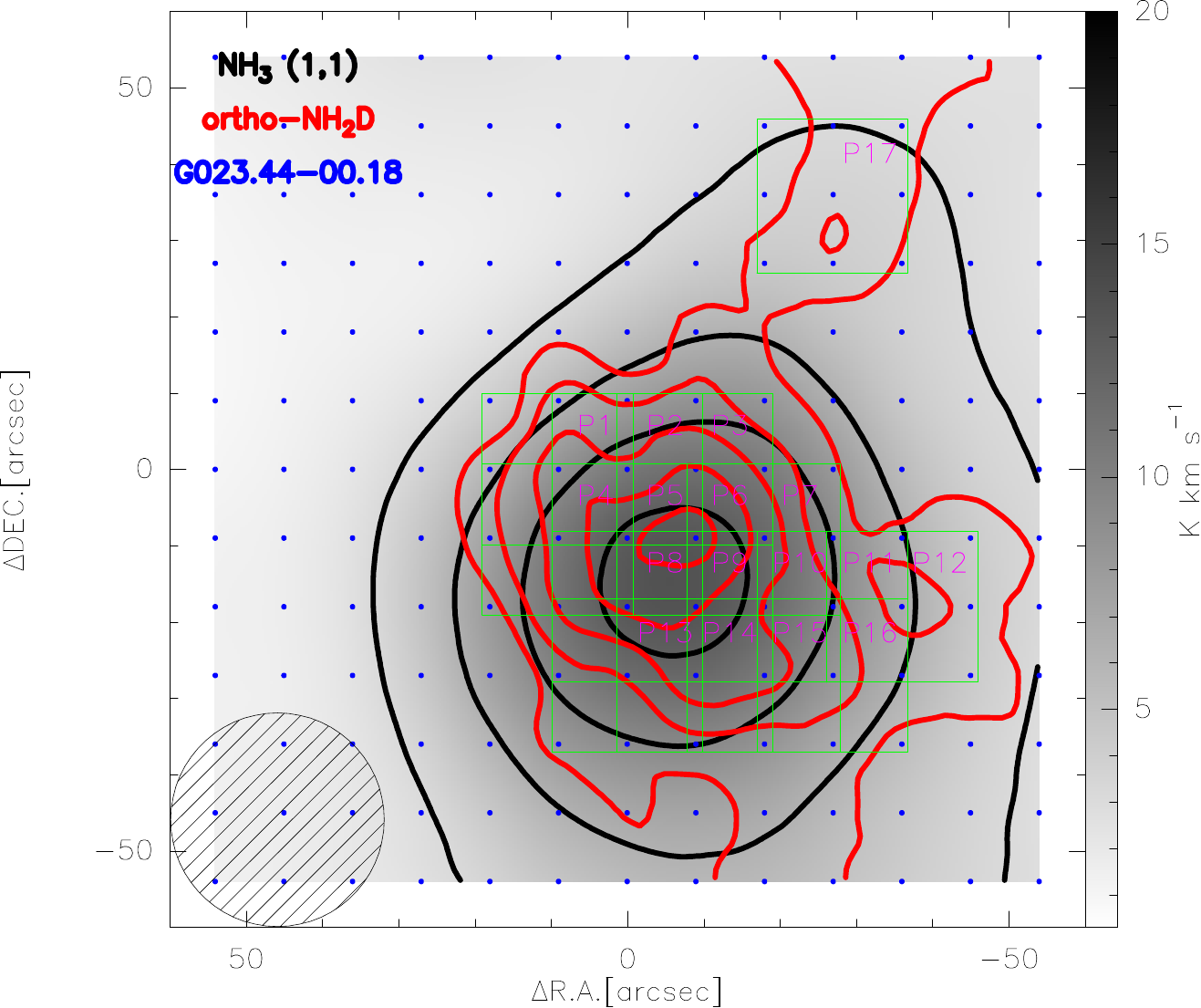}\includegraphics[width=0.25\textwidth]{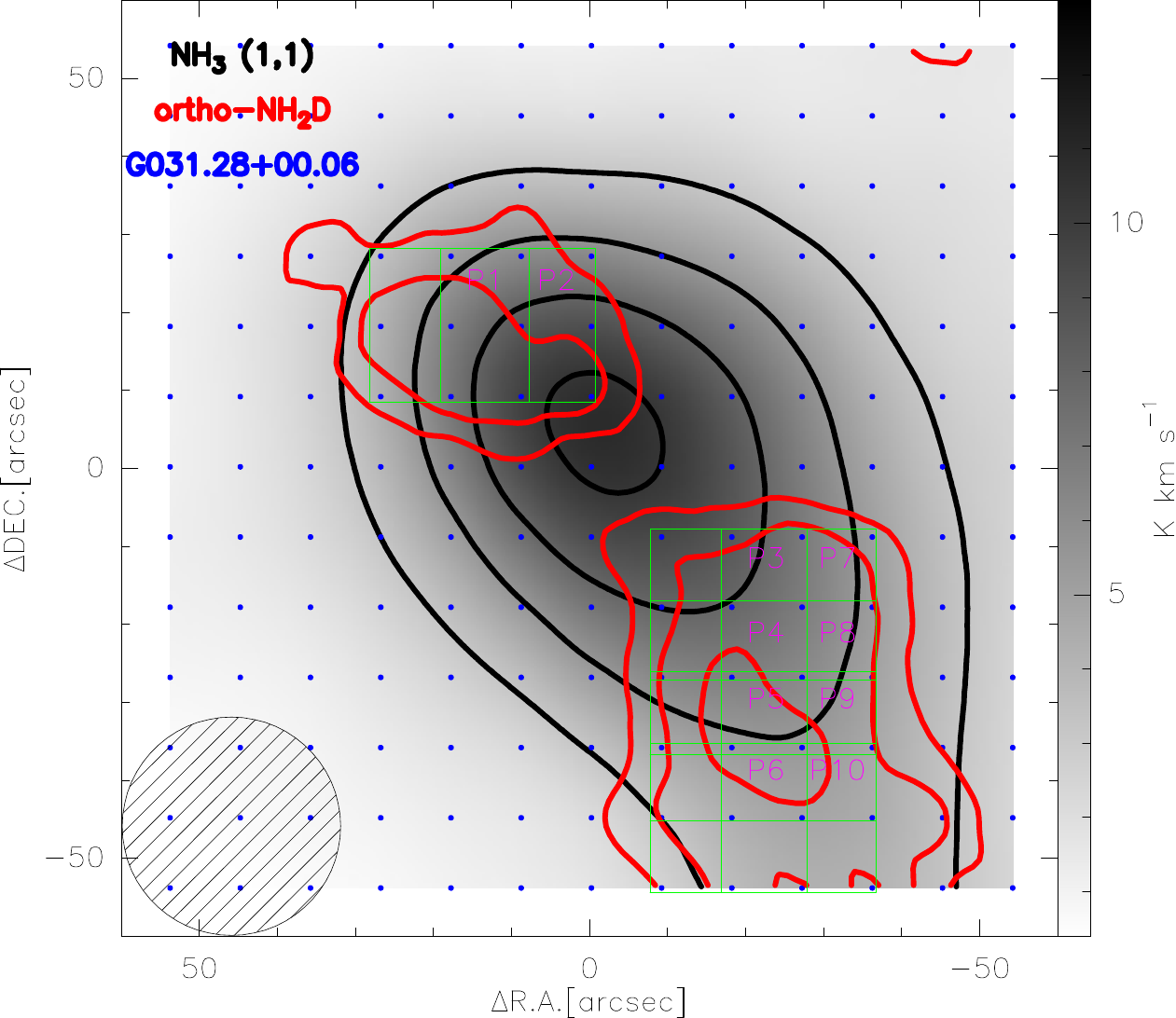}\\
\includegraphics[width=0.25\textwidth]{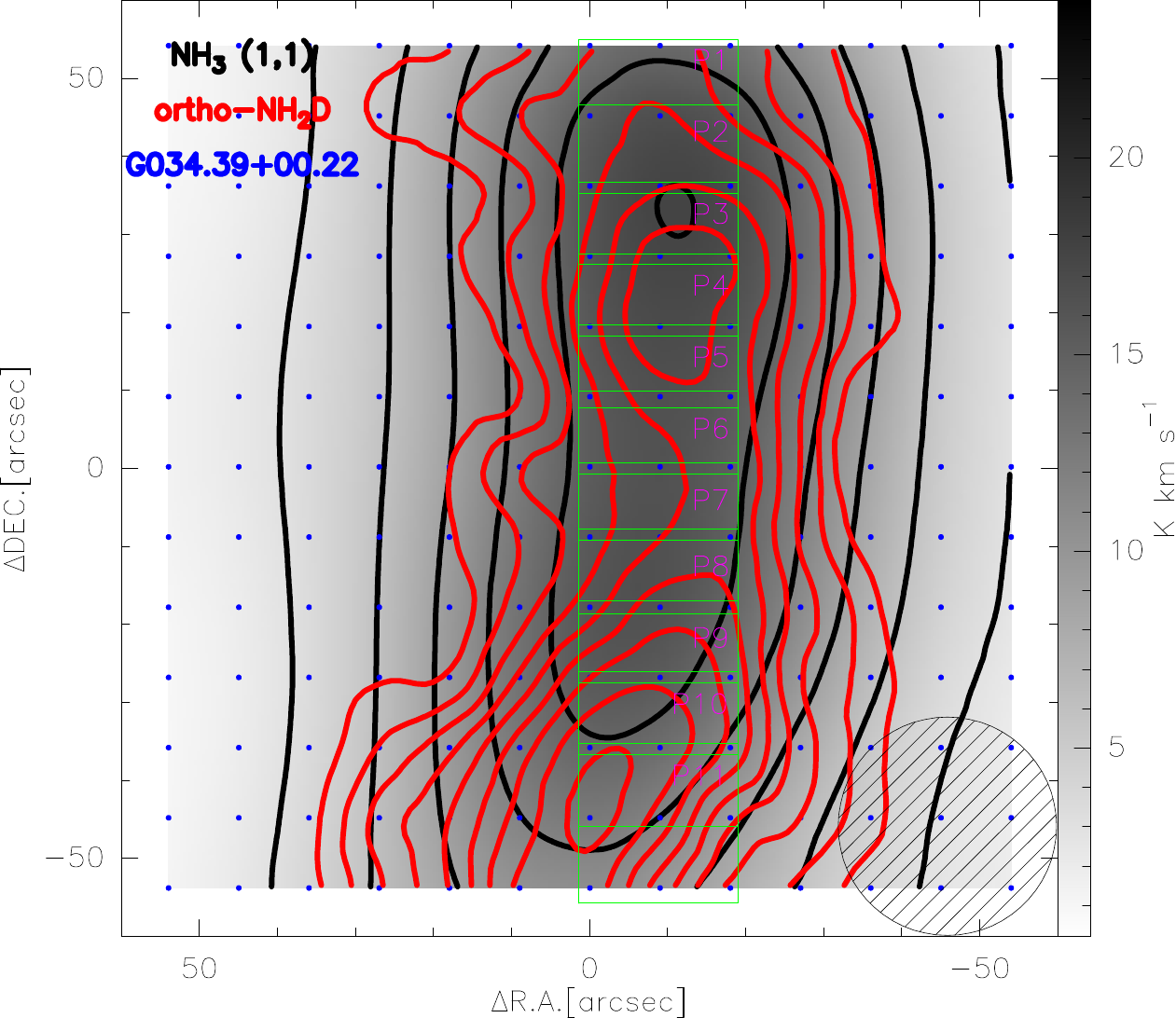}\includegraphics[width=0.25\textwidth]{figure/slice/G03519_NH3_1-1_slice.pdf}\includegraphics[width=0.25\textwidth]{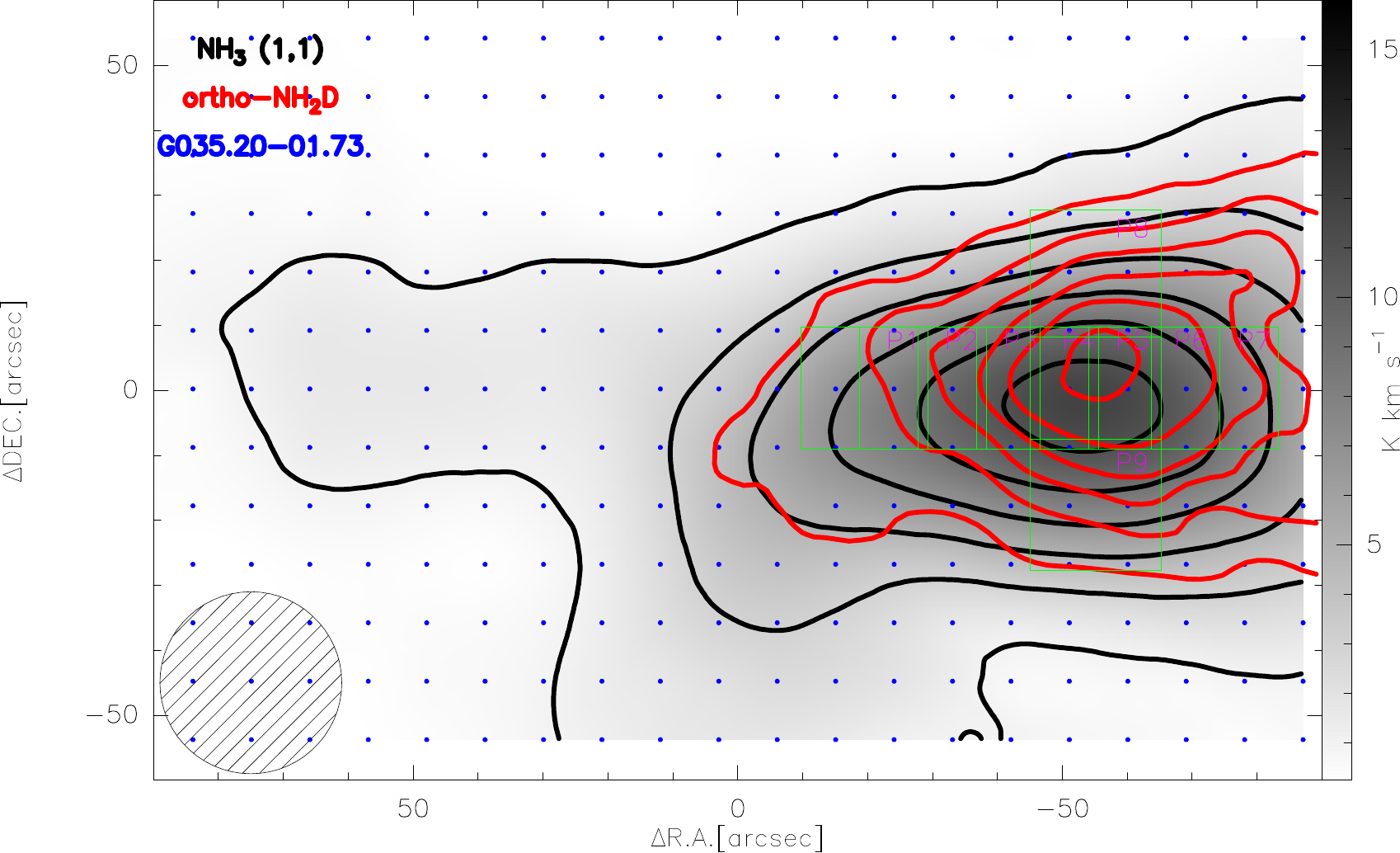}\\
\includegraphics[width=0.25\textwidth]{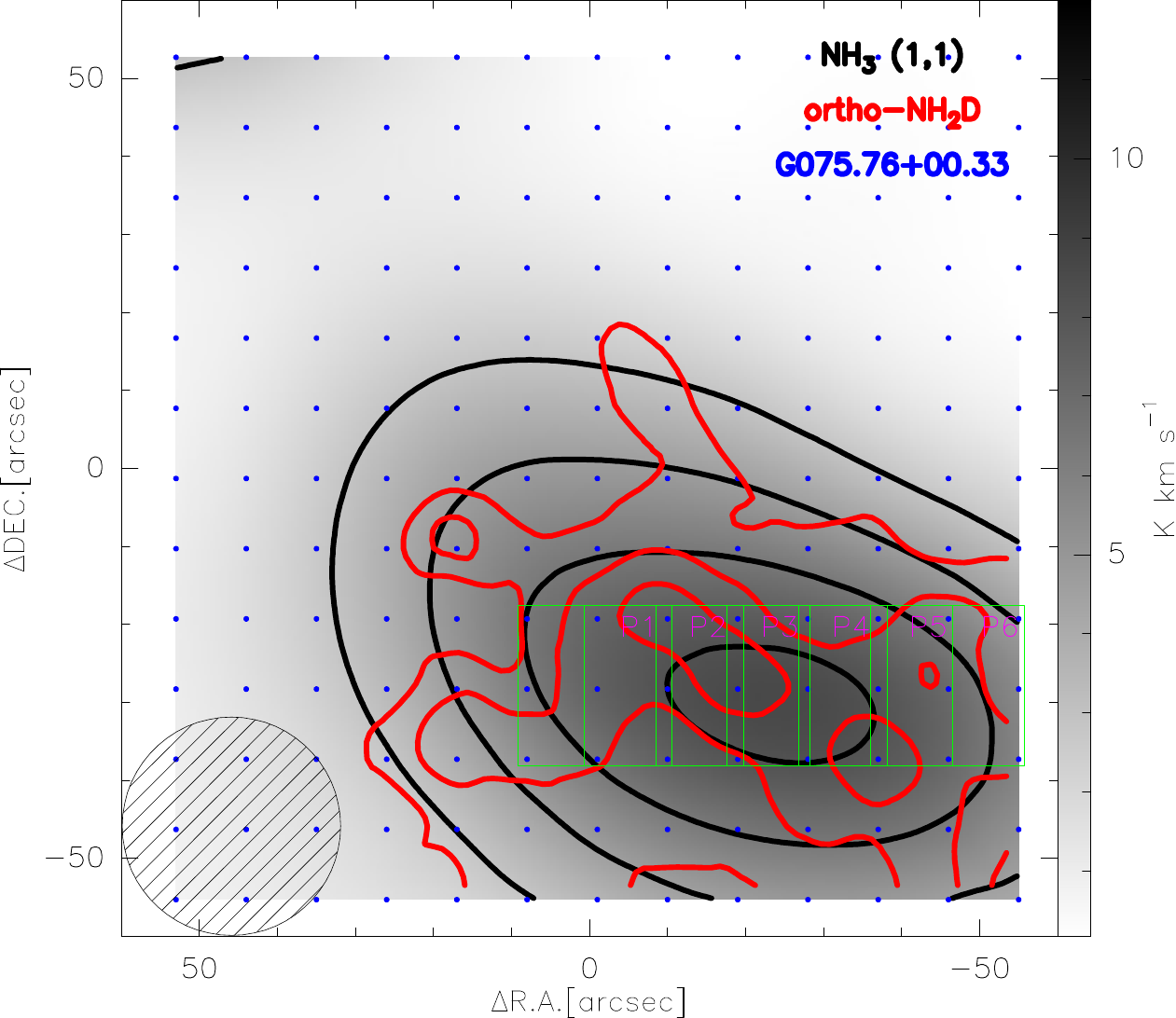}\includegraphics[width=0.25\textwidth]{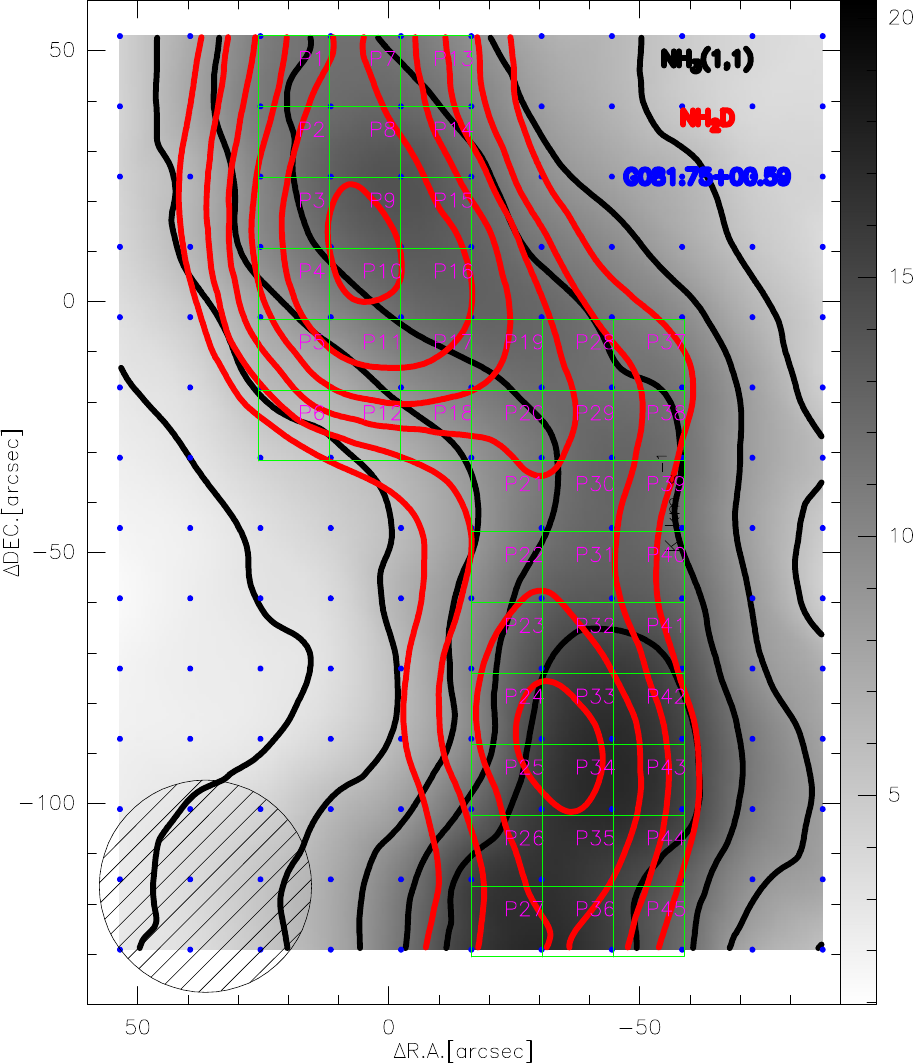}\includegraphics[width=0.25\textwidth]{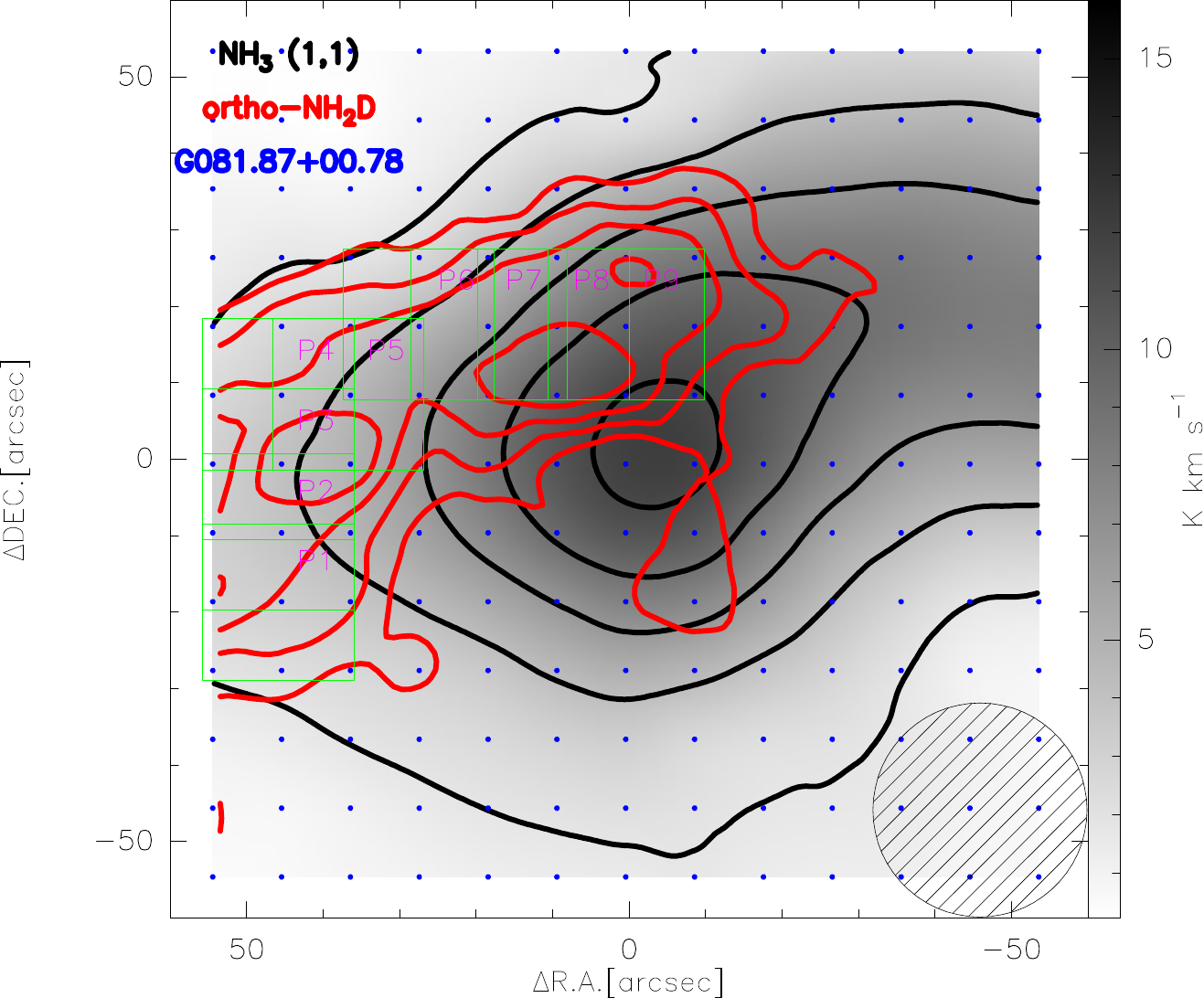}\\
\includegraphics[width=0.25\textwidth]{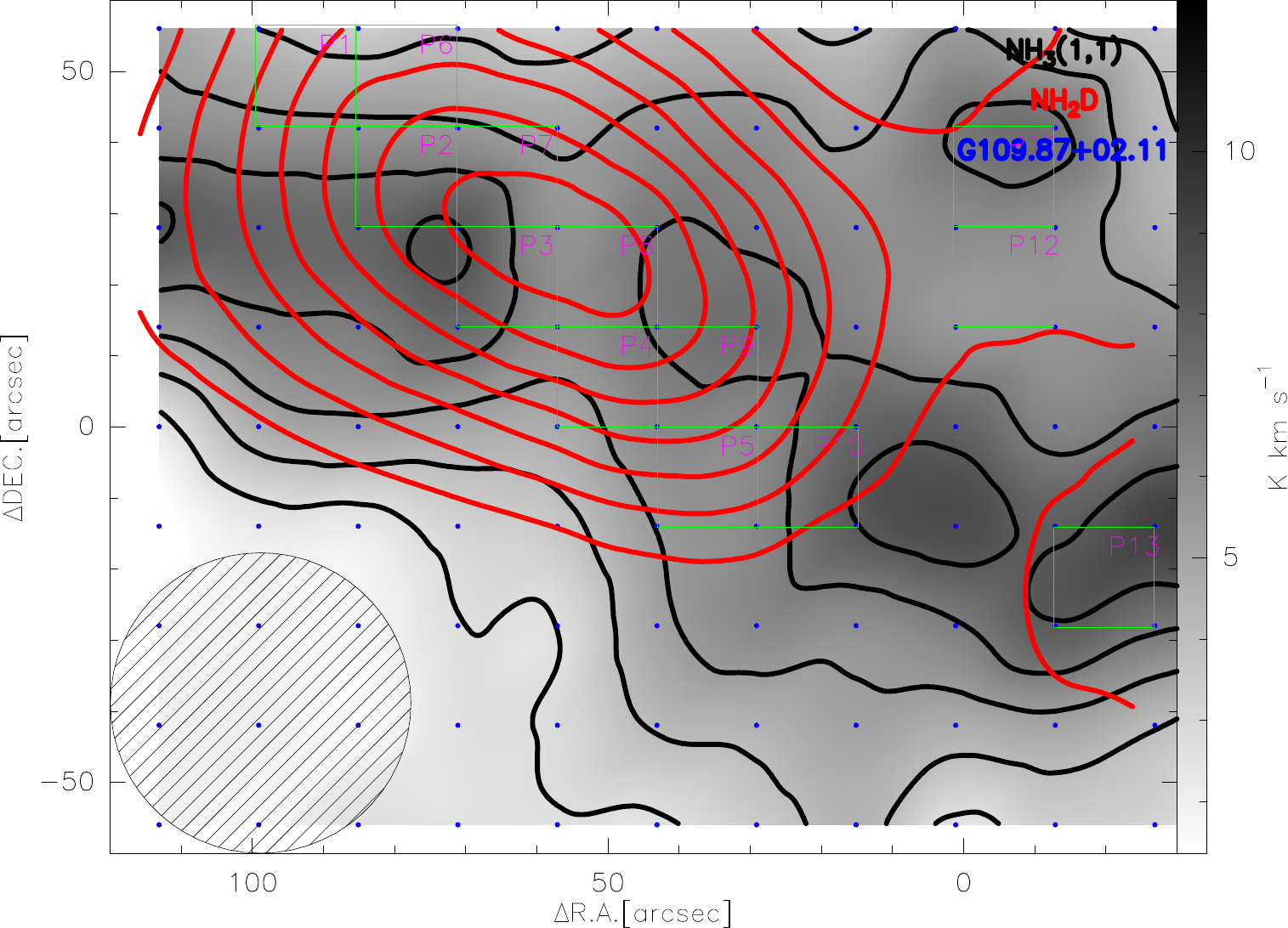}\includegraphics[width=0.25\textwidth]{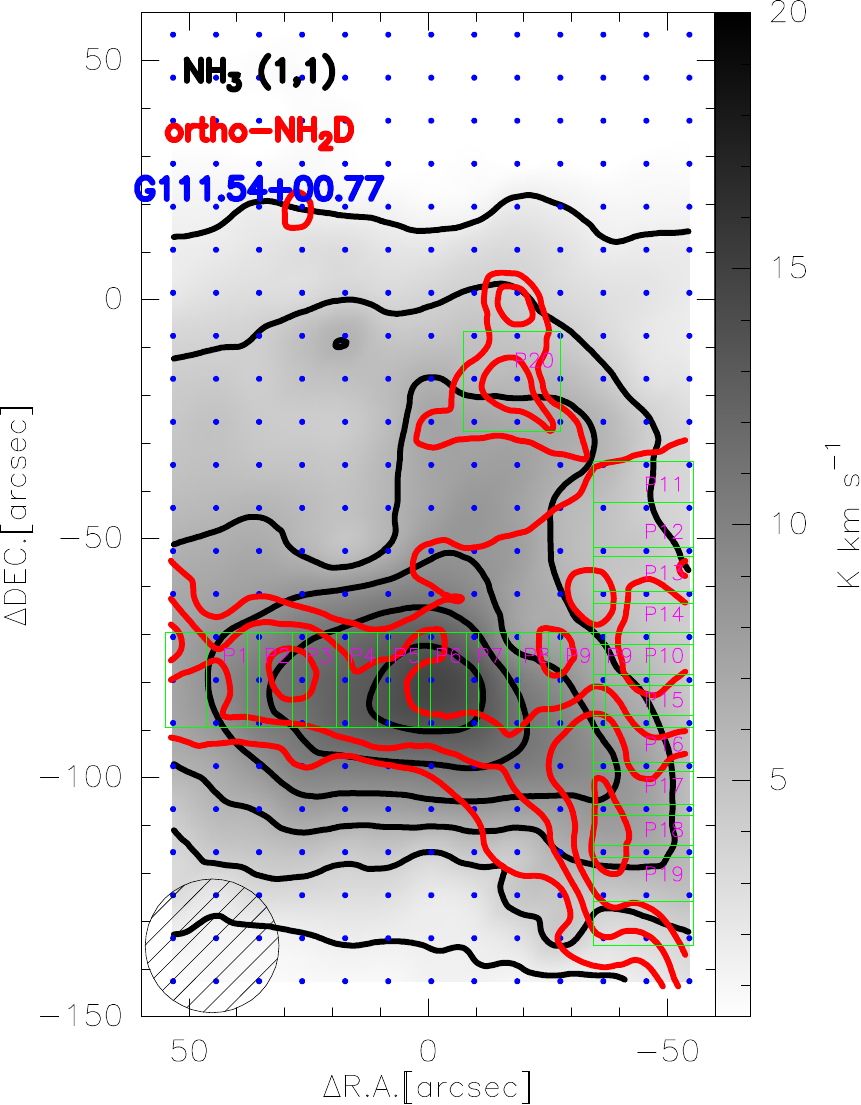}\includegraphics[width=0.25\textwidth]{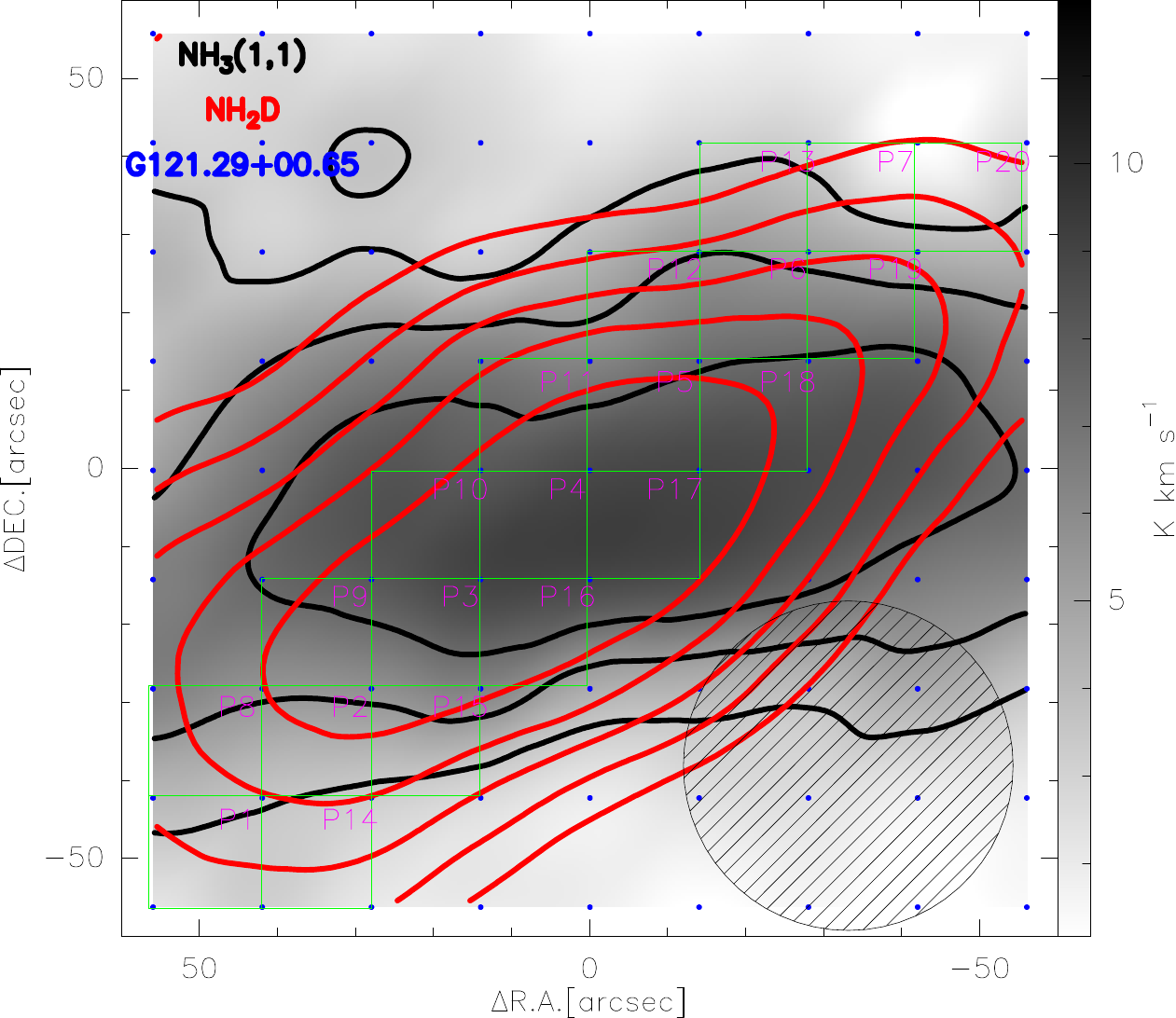}
\caption{The pixels in the box are the average area. Within each source, the order of the region numbers is the order in Table \ref{D-fraction}.}
\label{slice}
\end{figure}

The method of pixel averaging for each source is shown in Figure \ref{slice}.

\newpage

\section{The relationship between deuterium fractionation of NH$_3$ and distance of NH$_3$ Peak}
\setcounter{figure}{0}
\renewcommand{\thefigure}{E\arabic{figure}}

\begin{figure}[h]
\includegraphics[width=0.33\textwidth]{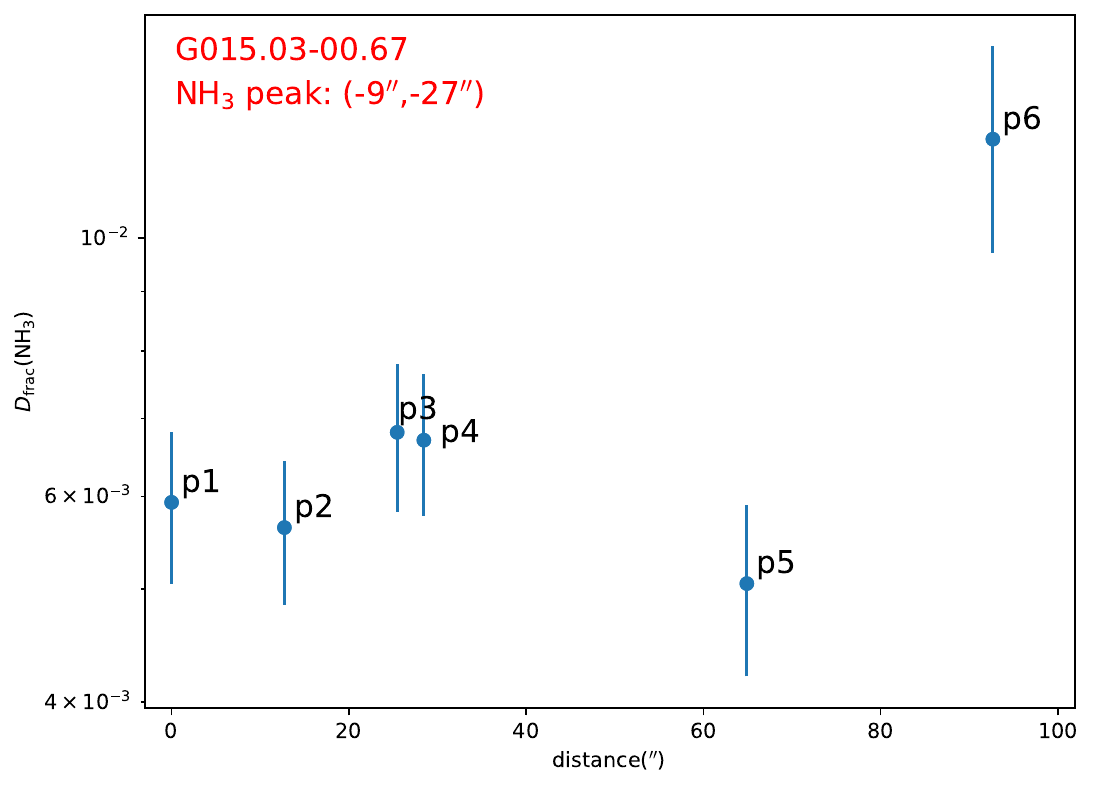}\includegraphics[width=0.33\textwidth]{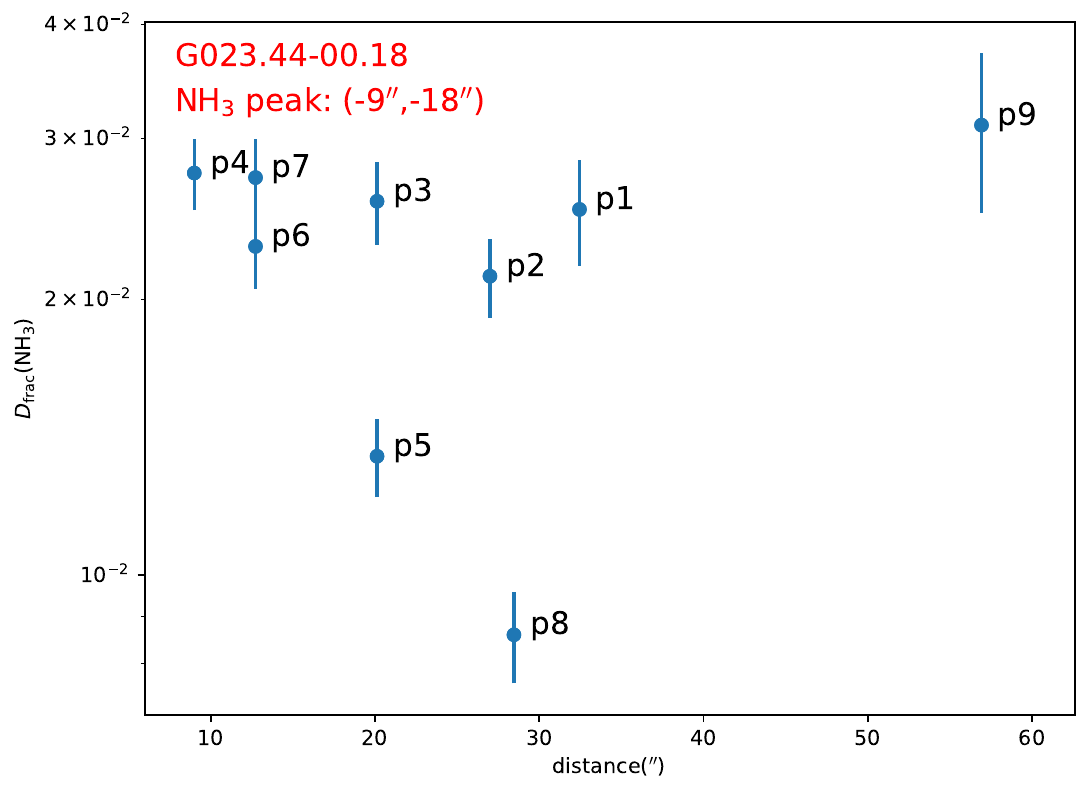}\includegraphics[width=0.33\textwidth]{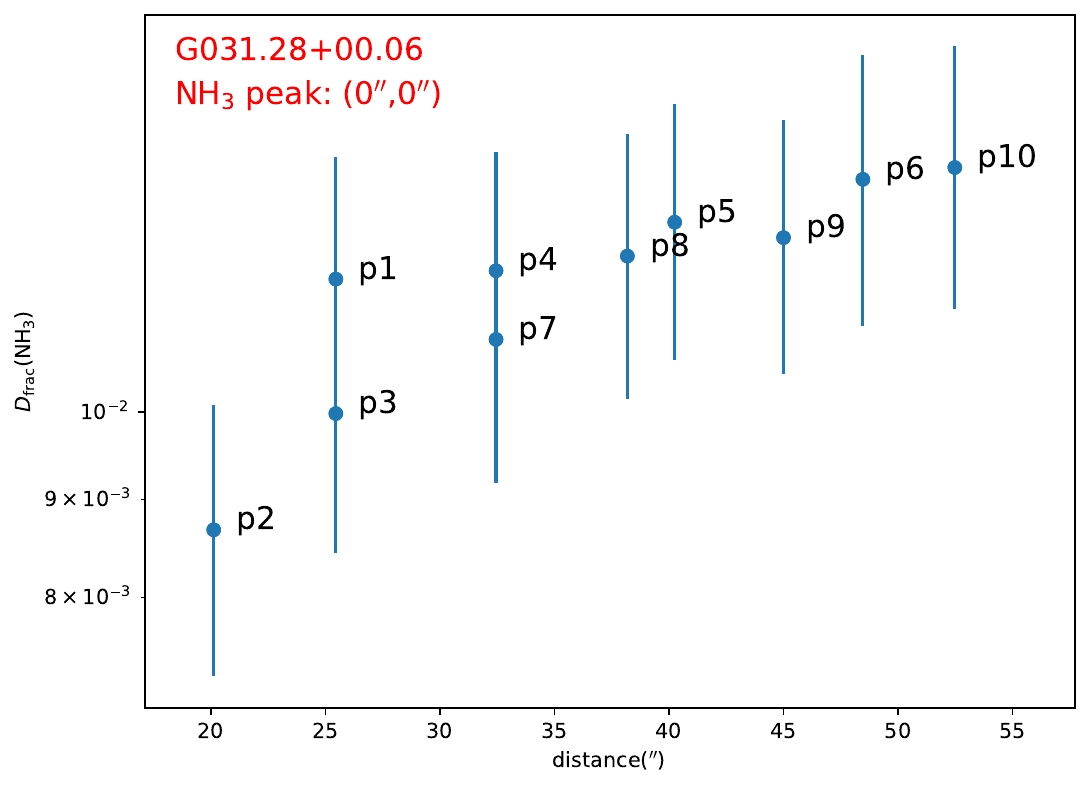}\\
\includegraphics[width=0.33\textwidth]{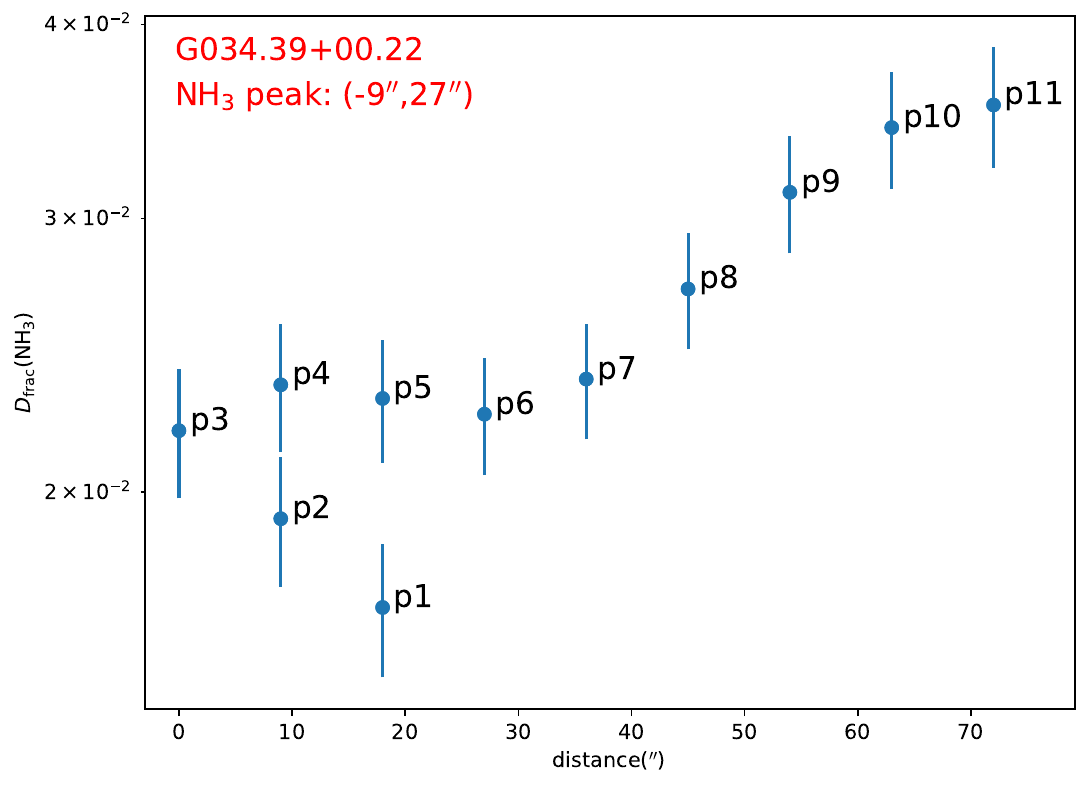}\includegraphics[width=0.33\textwidth]{figure/distance/G035190074_distance.pdf}\includegraphics[width=0.33\textwidth]{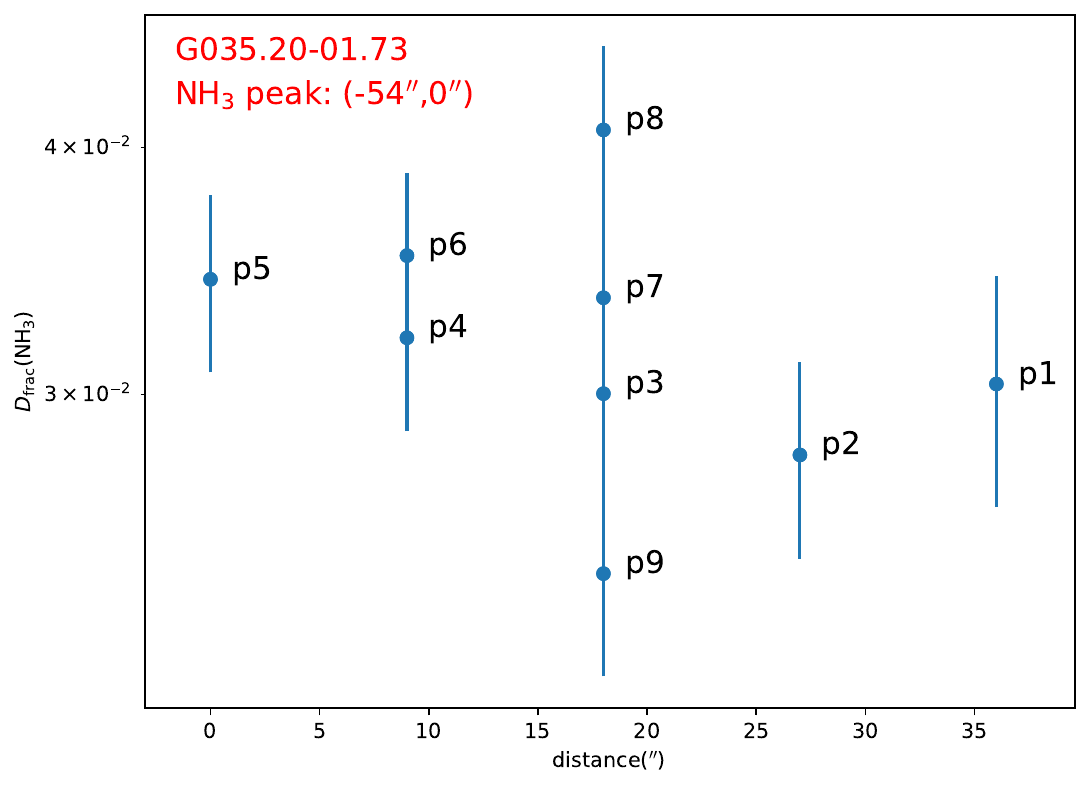}\\
\includegraphics[width=0.33\textwidth]{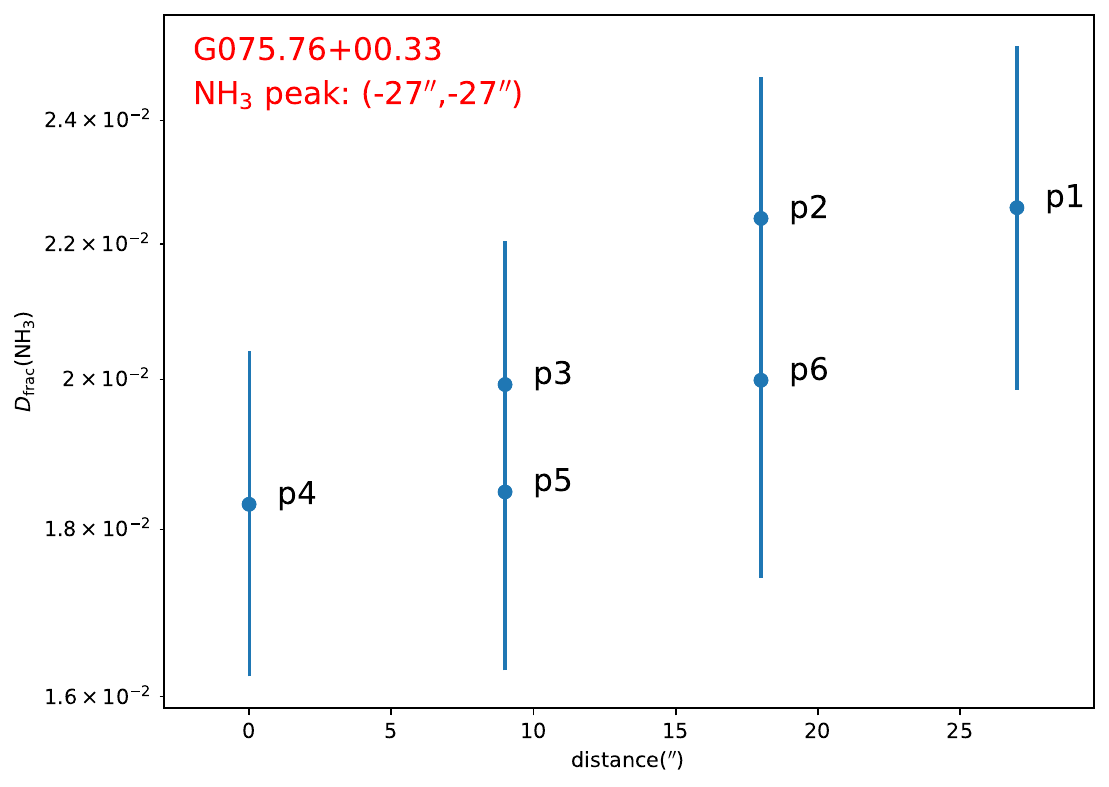}\includegraphics[width=0.33\textwidth]{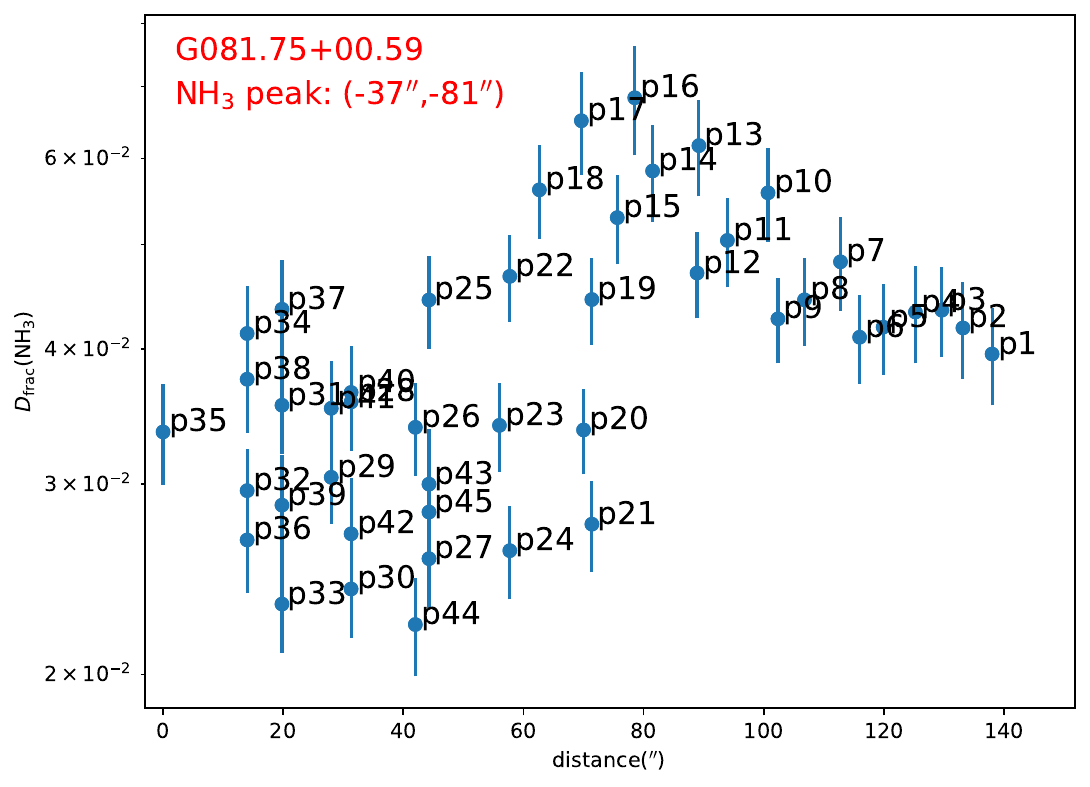}\includegraphics[width=0.33\textwidth]{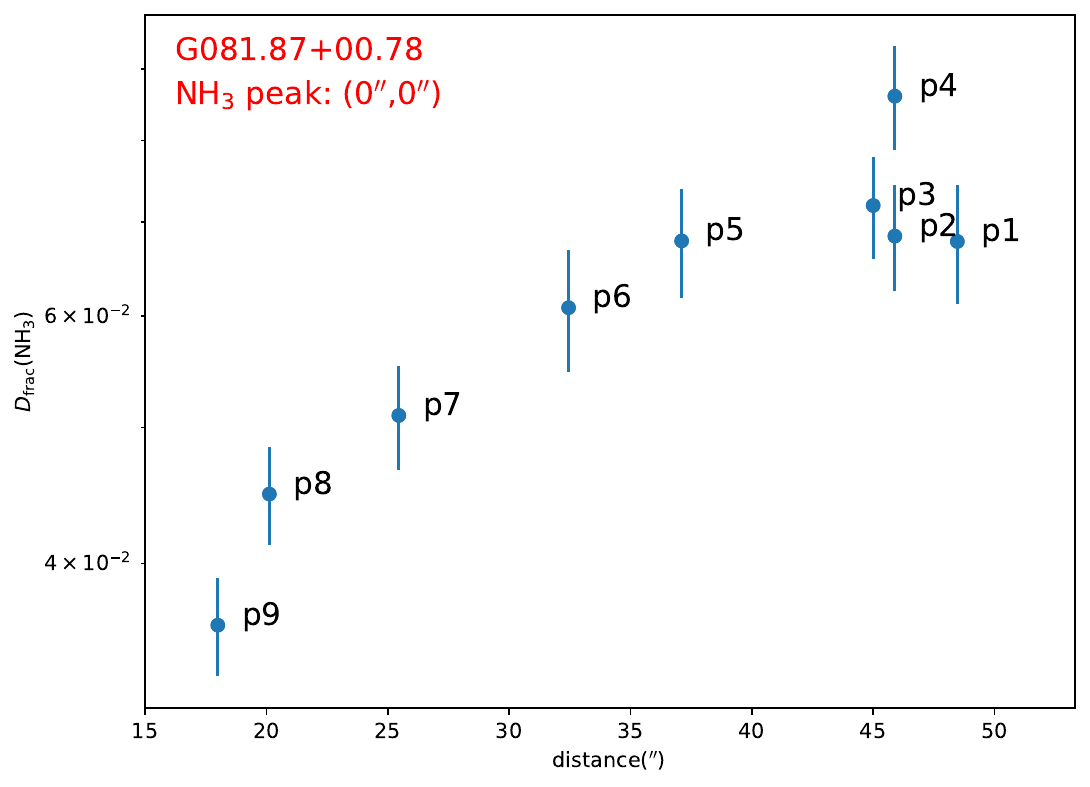}\\
\includegraphics[width=0.33\textwidth]{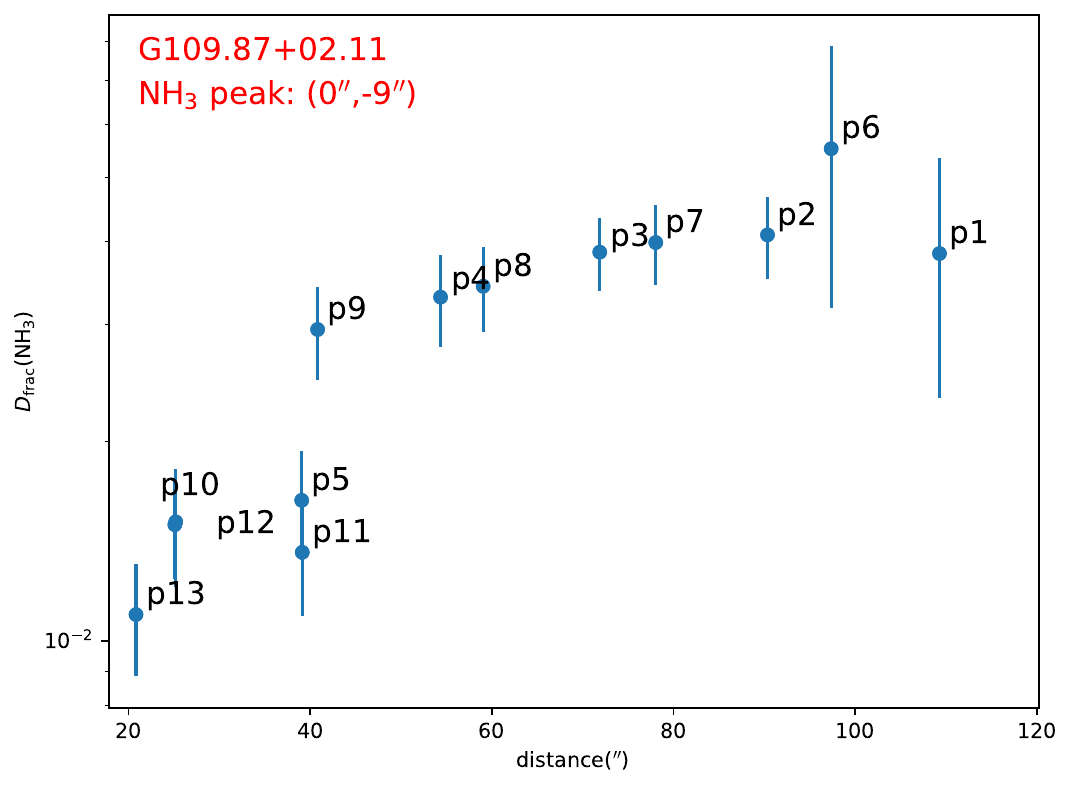}\includegraphics[width=0.33\textwidth]{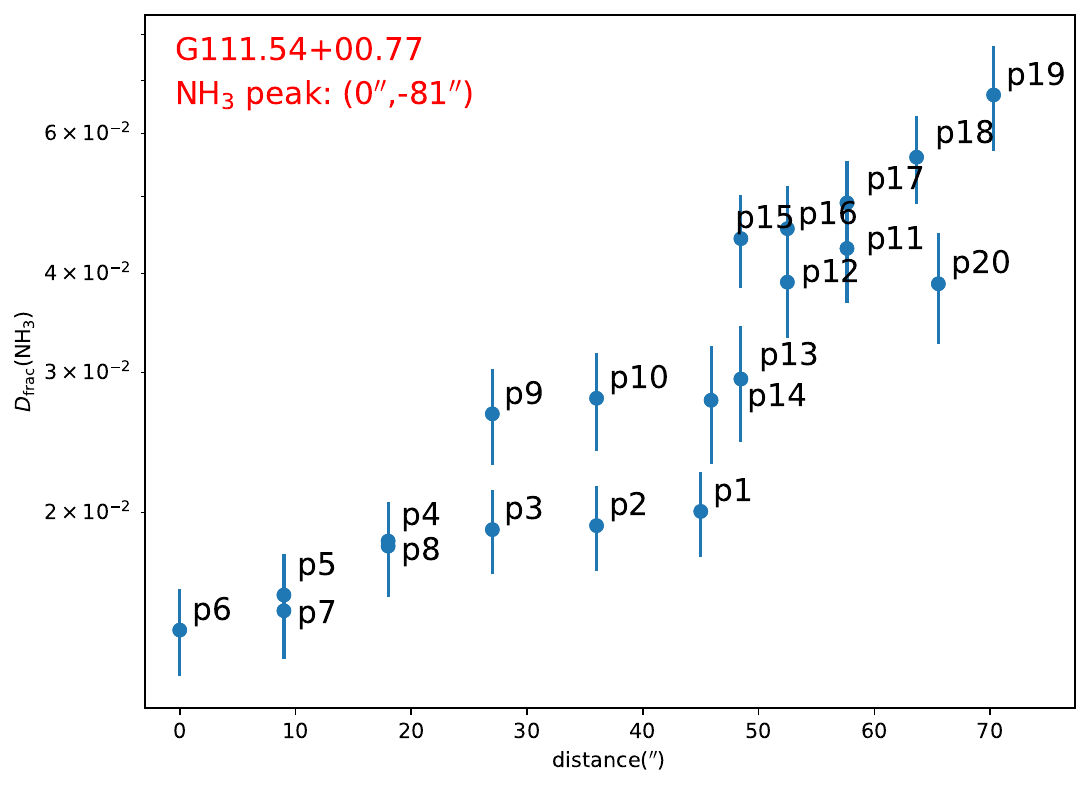}\includegraphics[width=0.33\textwidth]{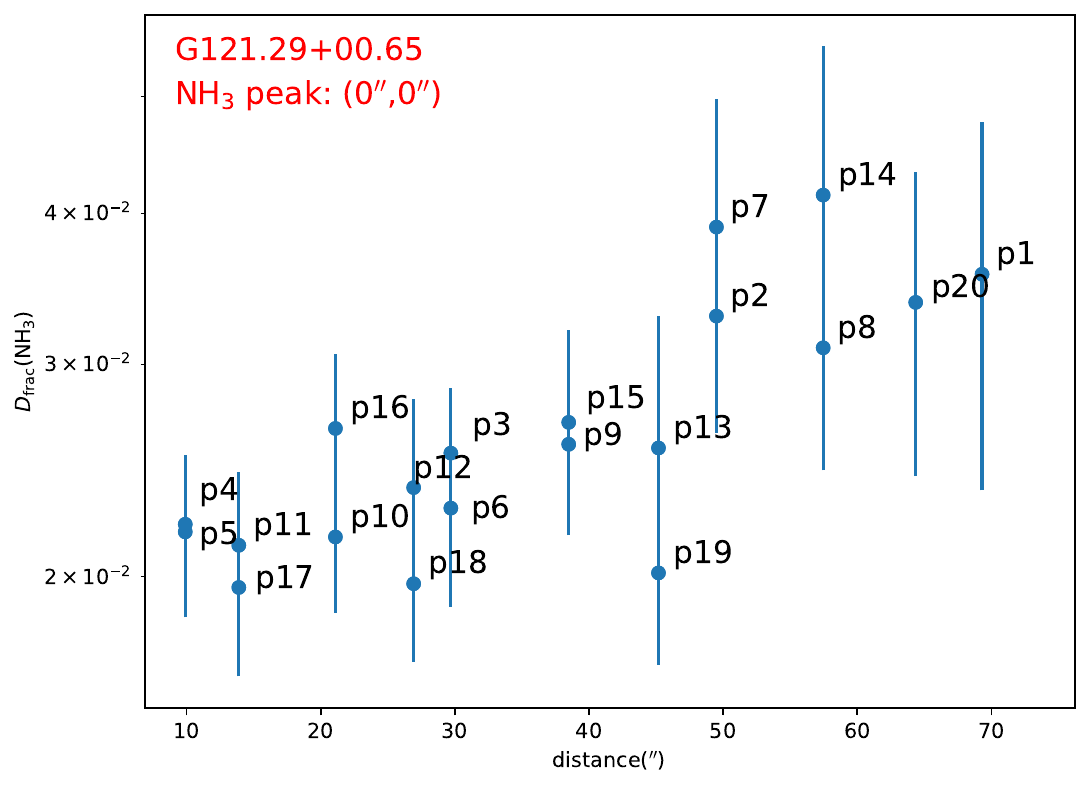}
\caption{The relationship between deuterium fractionation of NH$_3$ and distance of NH$_3$ peak.}
\label{Dfrac_distance}
\end{figure}

\begin{figure}[h]
\includegraphics[width=0.33\textwidth]{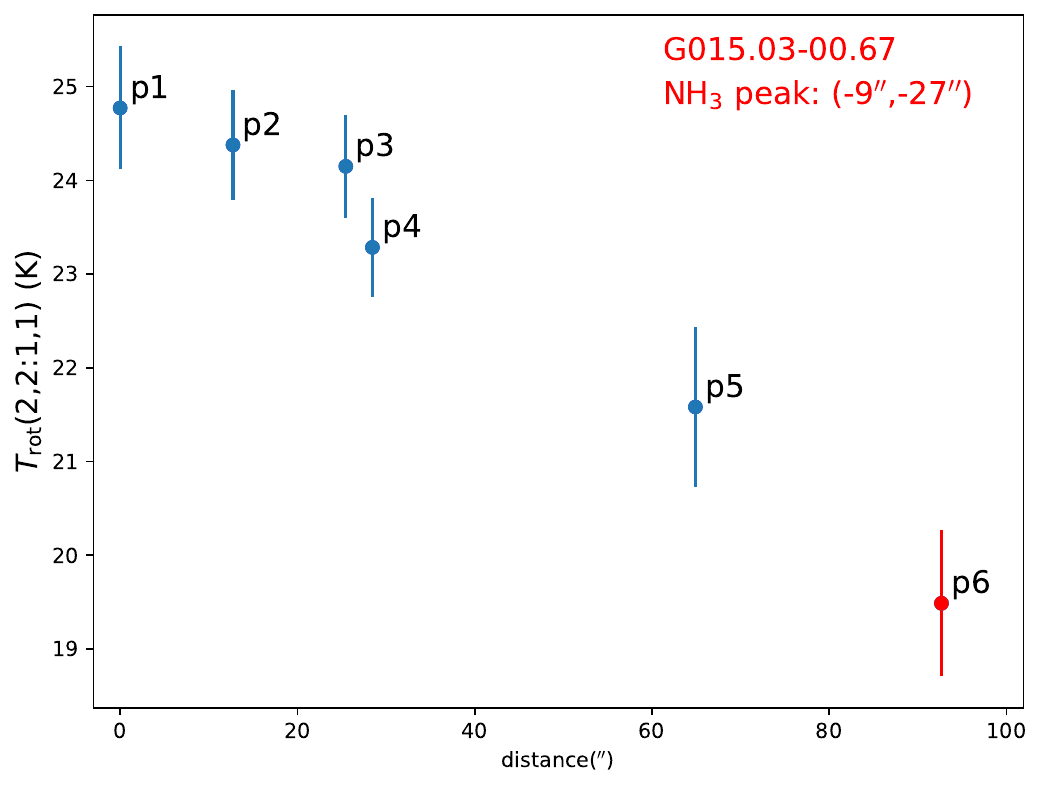}\includegraphics[width=0.33\textwidth]{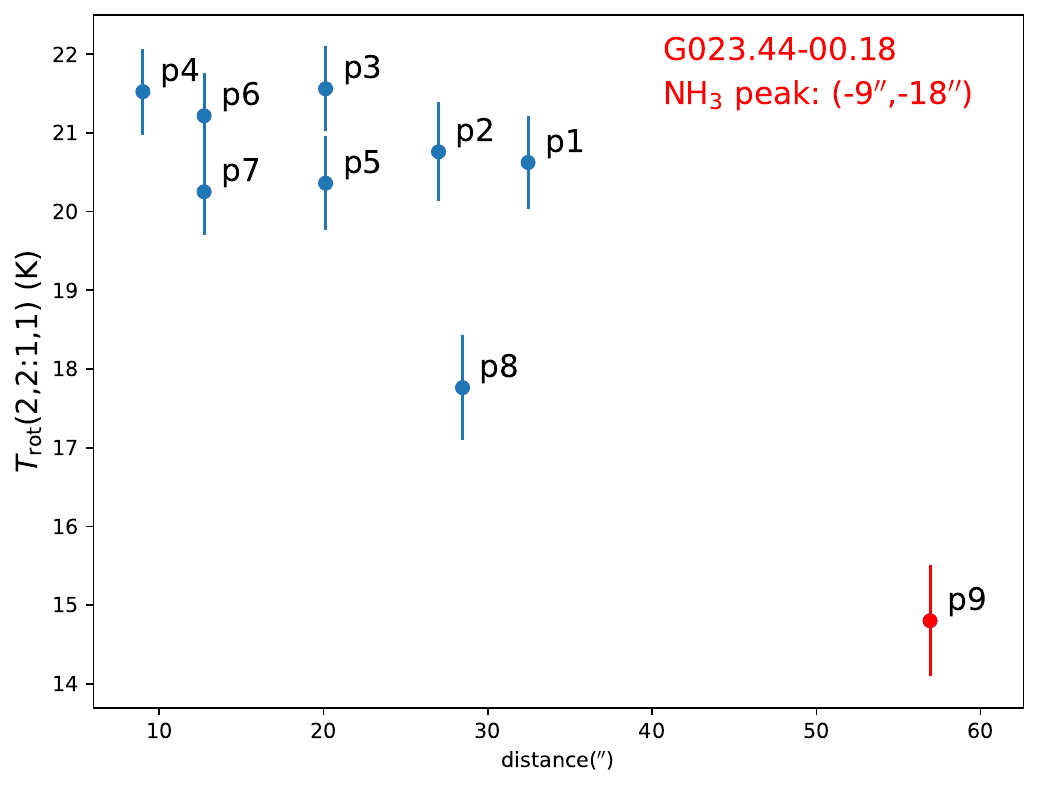}\includegraphics[width=0.33\textwidth]{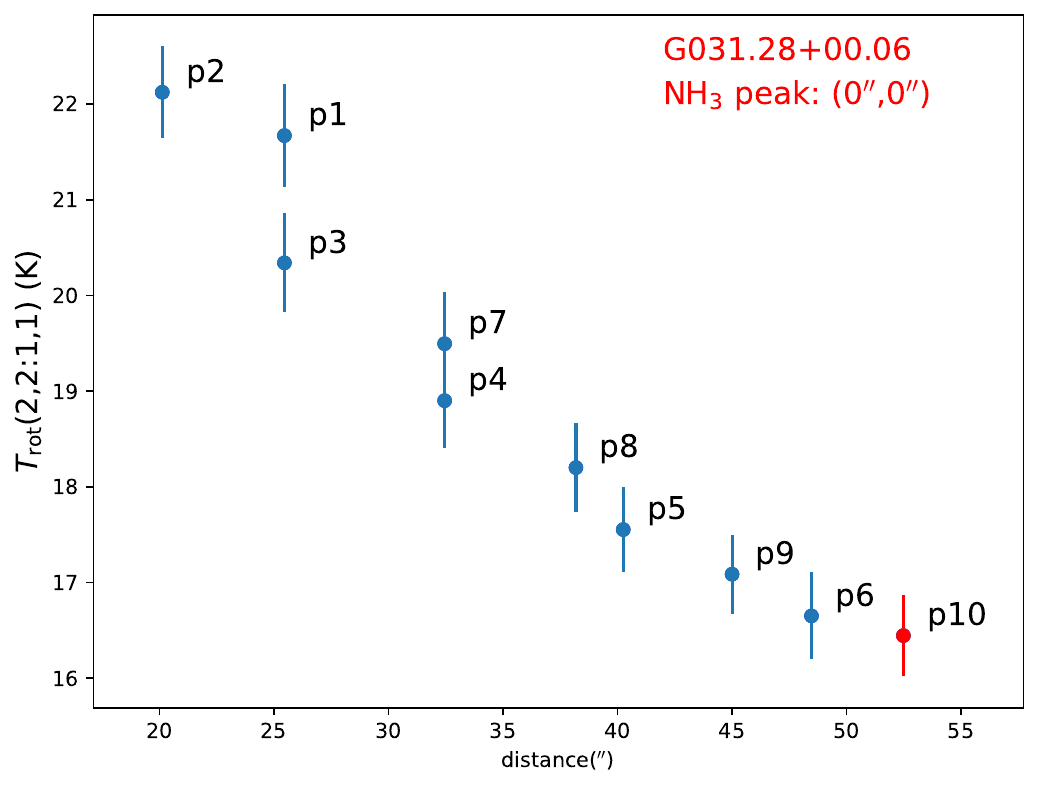}\\
\includegraphics[width=0.33\textwidth]{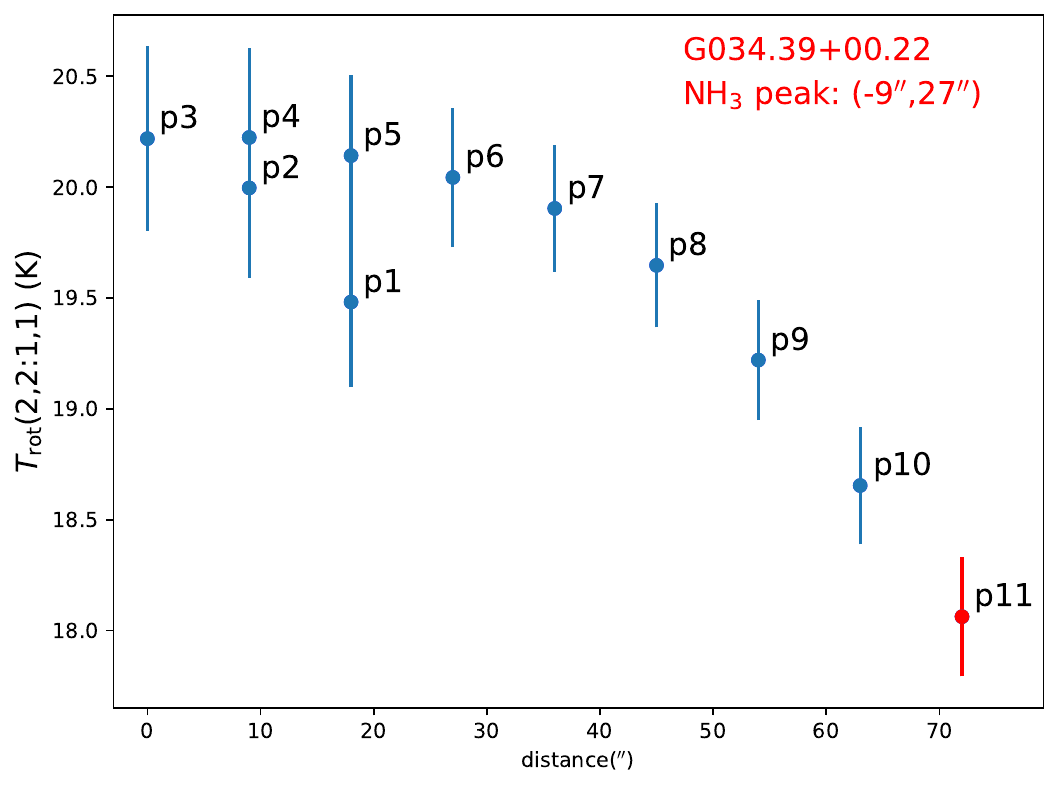}\includegraphics[width=0.33\textwidth]{figure/distance/G035190074_distance_temperature.pdf}\includegraphics[width=0.33\textwidth]{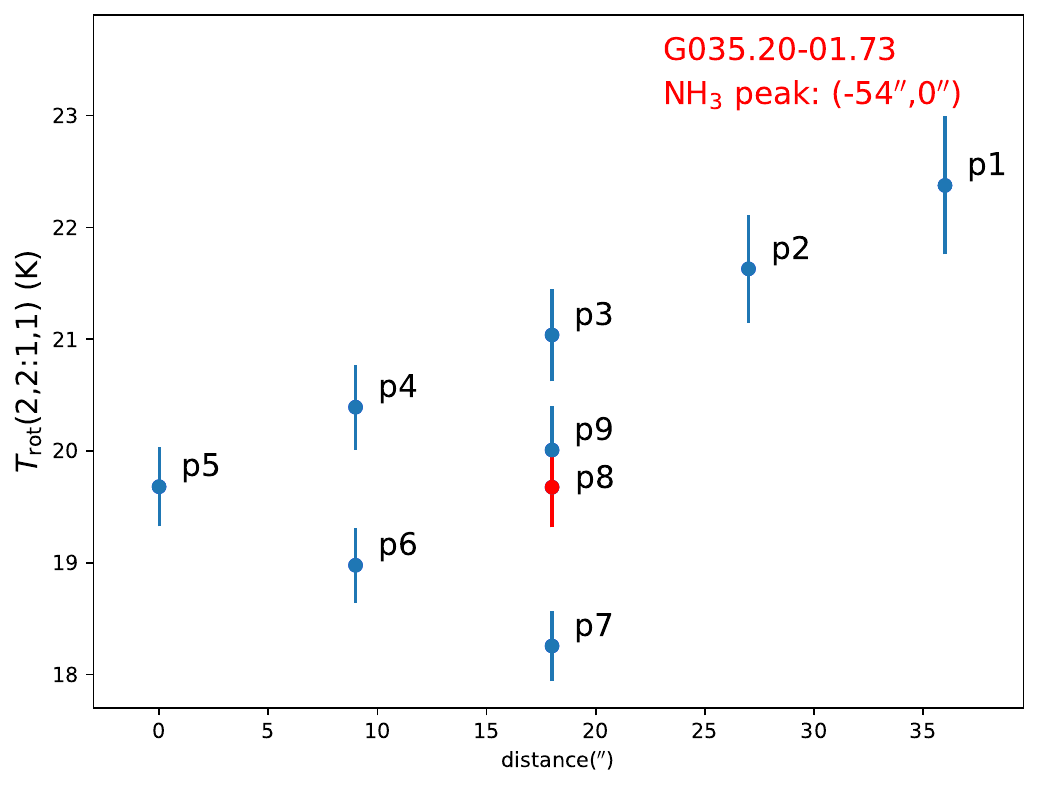}\\
\includegraphics[width=0.33\textwidth]{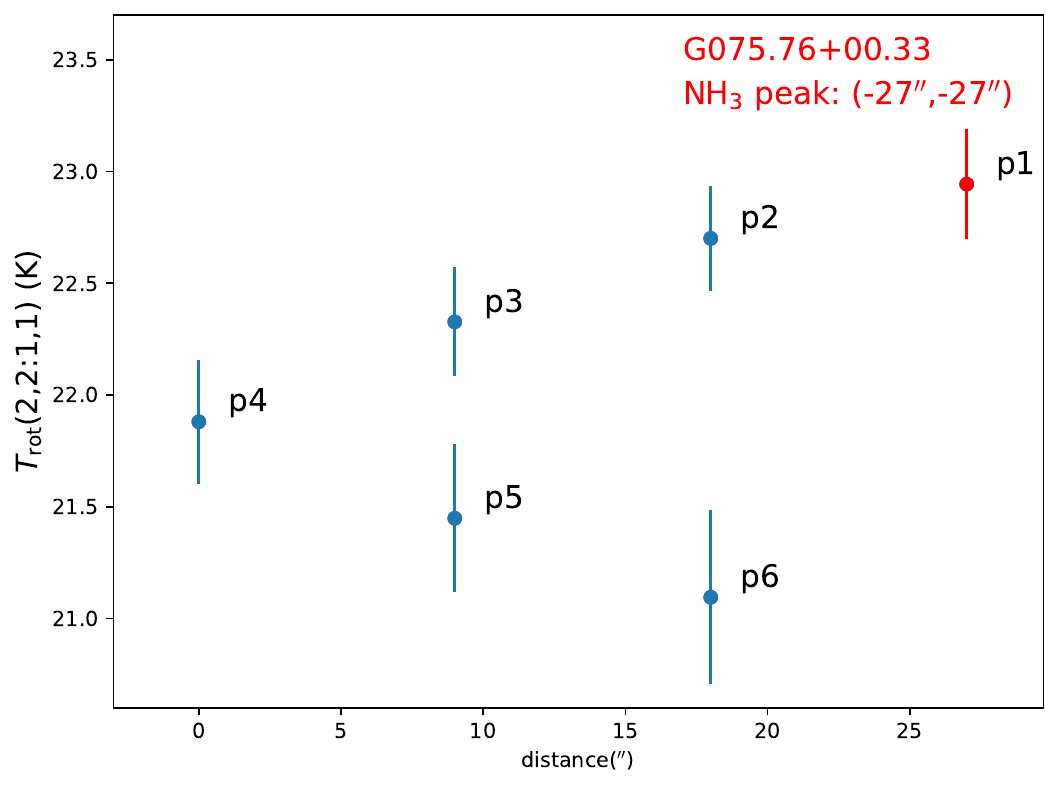}\includegraphics[width=0.33\textwidth]{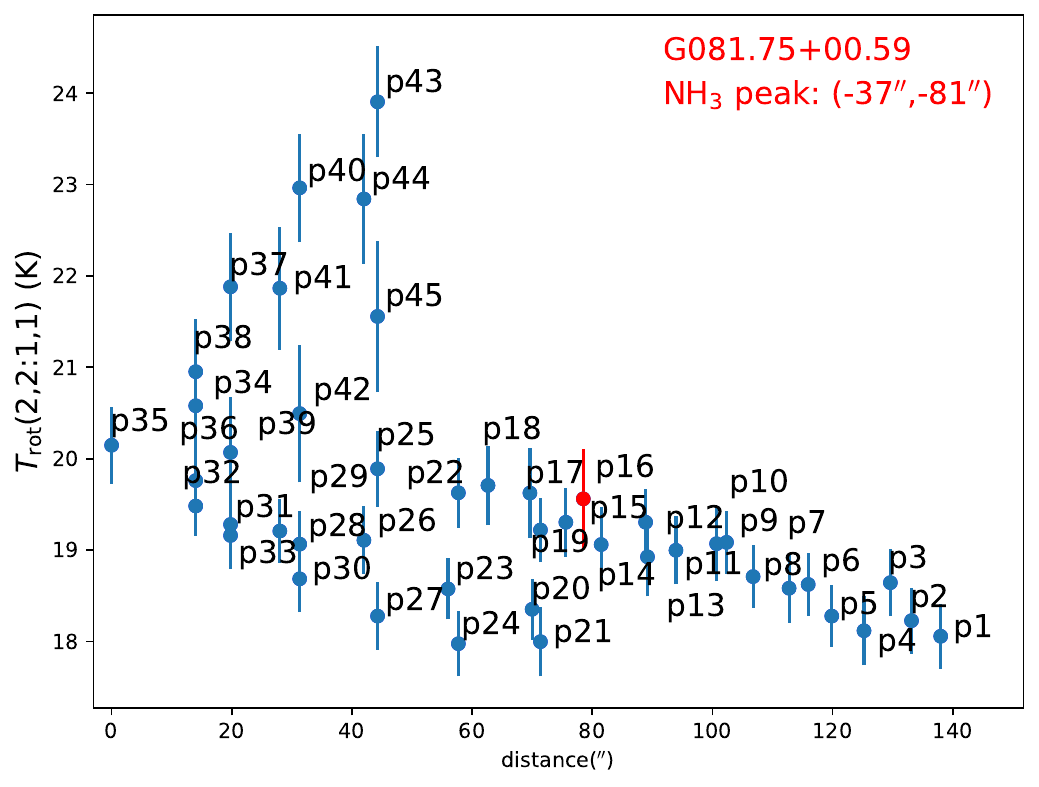}\includegraphics[width=0.33\textwidth]{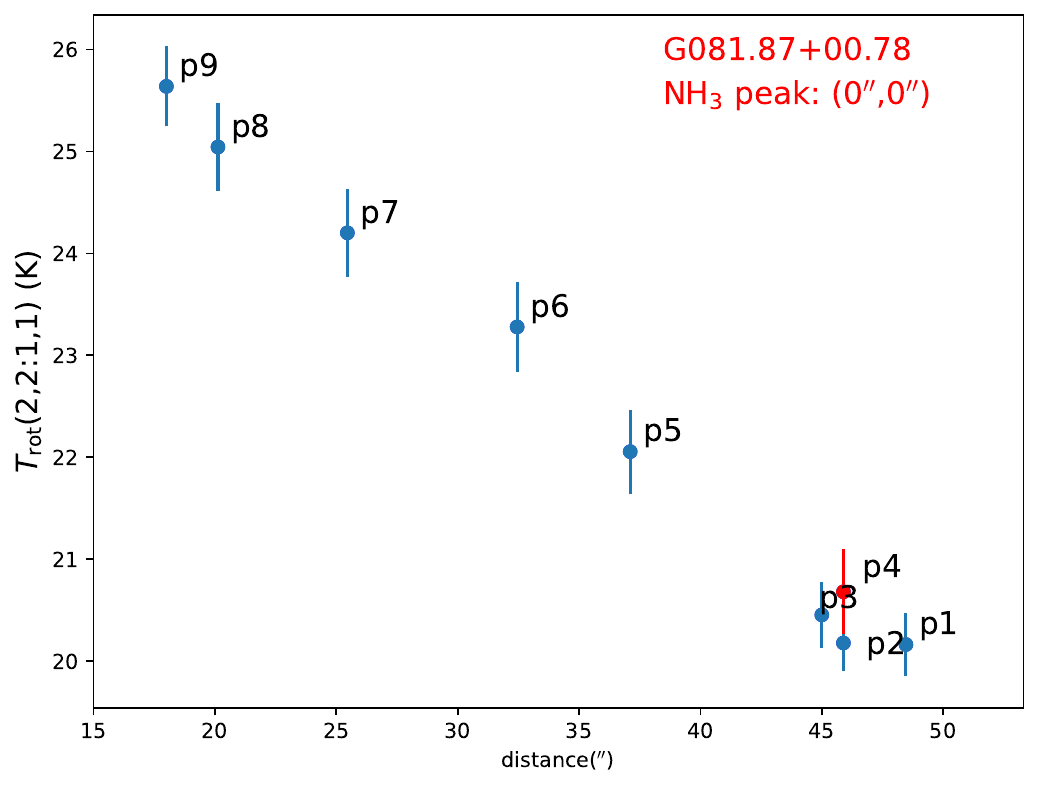}\\
\includegraphics[width=0.33\textwidth]{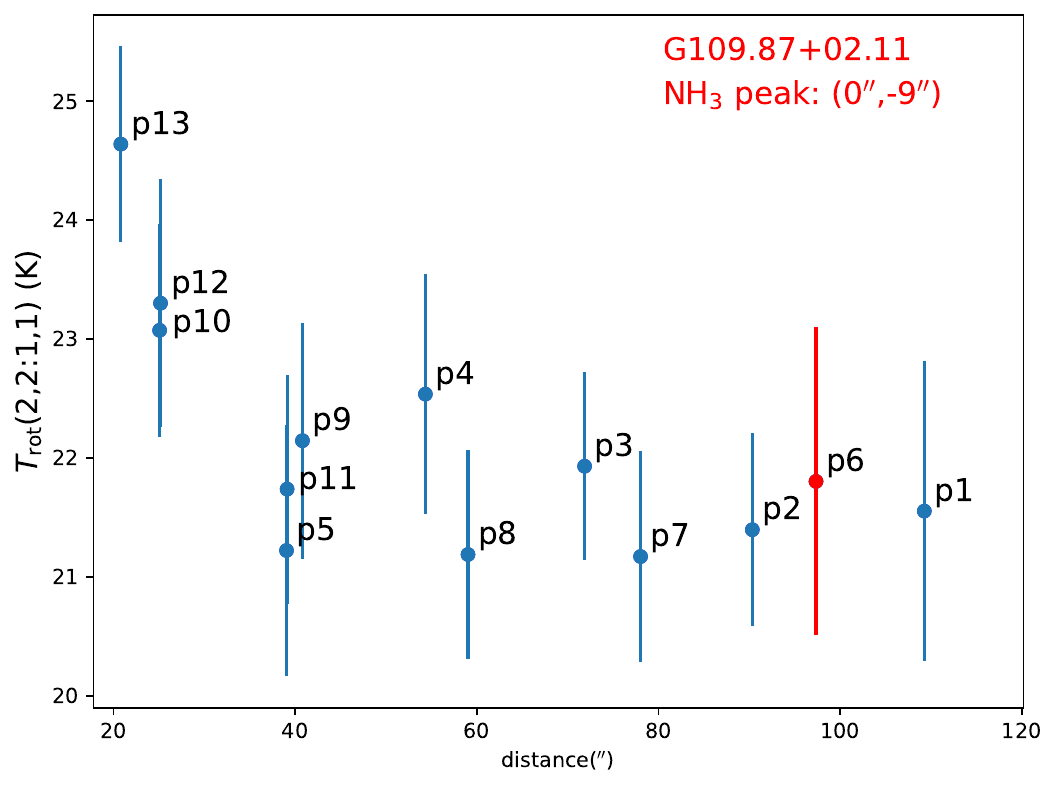}\includegraphics[width=0.33\textwidth]{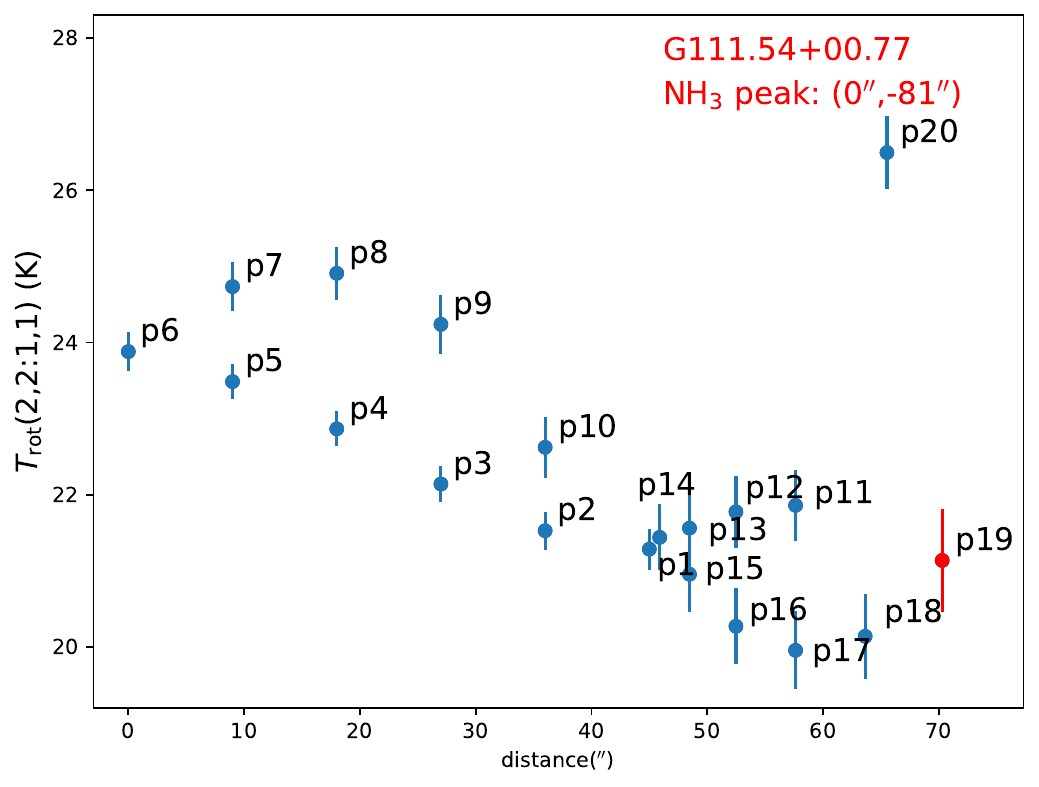}\includegraphics[width=0.33\textwidth]{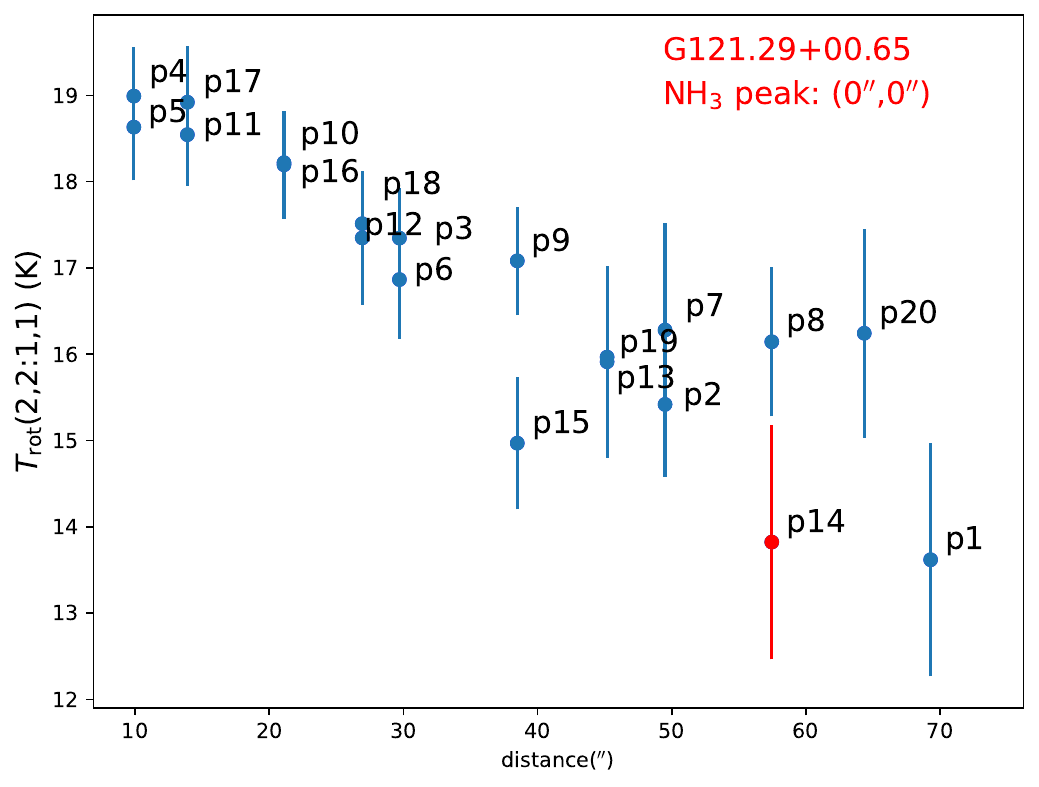}
\caption{The relationship between rotation temperature of NH$_3$ and distance of NH$_3$ peak.  The red marker is the position with the highest deuterium fractionation of NH$_3$.}
\label{distance_temperature}
\end{figure}

The relationship between deuterium fractionation of NH$_3$ and distance of NH$_3$ peak are 
shown in Figure \ref{Dfrac_distance}. The relationship between rotation temperature of NH$_3$ and distance of NH$_3$ peak are shown in Figure \ref{distance_temperature}.

\end{document}